\documentclass[12pt]{article}
\usepackage{amstext,amsmath,amsthm,amssymb,amsfonts,bbm,enumerate}
\usepackage[pagebackref, colorlinks = true, linkcolor = blue, urlcolor  = blue, citecolor = red]{hyperref}
\usepackage[latin1]{inputenc}
\usepackage{hyperref}
\usepackage{amsthm}
\usepackage{xcolor}
\usepackage{geometry}
\usepackage{graphicx}
\usepackage{framed}
\usepackage{mathbbol}
\usepackage{framed}
\usepackage{atbegshi, picture}
\AtBeginShipoutNext{\AtBeginShipoutUpperLeft{
 \put(\dimexpr\paperwidth-0.5cm\relax,-0.8cm){\makebox[0pt][r]{\small YITP-18-110}}
}}
 
\newtheorem{definition}{Definition}
\newtheorem{theorem}{Theorem}
\newtheorem{lemma}{Lemma}
\newtheorem{proposition}{Proposition}

 \theoremstyle{remark} \newtheorem{remark}{Remark}
 
\newcommand{\un}{\mathbb{1}}
\newcommand{\cI}{{\mathcal I}}
\newcommand{\cC}{{\mathcal C}}
\newcommand{\cD}{{\mathcal D}}
\newcommand{\cA}{{\mathcal A}}
\newcommand{\cB}{{\mathcal B}}
\newcommand{\cM}{\mathcal{M}}
\newcommand{\A}{{A}}
\newcommand{\R}{{r}}
\newcommand{\cV}{{\mathcal V}}

\newcommand{\cN}{{\mathcal N}}
\newcommand{\cT}{{\mathcal T}}
\newcommand{\cL}{{\mathcal L}}

\newcommand{\cW}{{\mathcal W}}

\newcommand{\cE}{{\mathcal{E}}}
\newcommand{\bN}{{\mathbb{N}}}

\newcommand{\bee}{\begin{equation}}
\newcommand{\be}{\begin{equation}}
\newcommand{\ee}{\end{equation}}
\newcommand{\Li}{\mathrm{L}}

\newcommand{\fR}{f^{(R)}}
\newcommand{\fI}{f^{(I)}}
\newcommand{\fr}{f_0}

\def \Ceil#1{\lceil{#1}\rceil}

\newcommand{\I}{\mathtt{I}}
\newcommand{\II}{\mathtt{II}}
\newcommand{\III}{\mathtt{III}}

\newcommand{\bea}{\begin{eqnarray}}
\newcommand{\eea}{\end{eqnarray}}

\definecolor{brown}{rgb}{0.59, 0.29, 0.0}
\definecolor{indigo}{rgb}{0.29, 0.0, 0.51}
\definecolor{armygreen}{rgb}{0.29, 0.33, 0.13}
\definecolor{flame}{rgb}{0.89, 0.35, 0.13}

\begin{document}

\title{\Huge\bf Melonic Turbulence}

\author{\rule{0pt}{10mm}St\'ephane Dartois,$^{a}$ Oleg Evnin,$^{b,c}$  Luca Lionni,$^{d,b}$\vspace{2mm}\\ Vincent Rivasseau$^{e}$ and Guillaume Valette$^{f}$\vspace{7mm}\\
${}^a${\small  School of Mathematics and Statistics, The University of Melbourne,}\\
{ \small  Victoria 3010, \textsc{Australia}}\vspace{1mm}\\
${}^b${\small Department of Physics,  Faculty of Science, Chulalongkorn University,}\\
{\small  Bangkok 10330,  \textsc{Thailand}}\vspace{1mm}\\
${}^c${\small Theoretische Natuurkunde, Vrije Universiteit Brussel (VUB) and}\\
{\small The International Solvay Institutes, B-1050 Brussel,  \textsc{Belgium}}\vspace{1mm}\\
${}^d${\small Yukawa Institute for Theoretical Physics, Kyoto University,}\\
{\small Kitashirakawa, Sakyo-ku, Kyoto 606-8502,  \textsc{Japan}}\vspace{1mm}\\
${}^e${\small Laboratoire de Physique Th\'eorique, CNRS UMR 8627,}\\
{\small  Universit\'e Paris-Sud, 91405 Orsay Cedex,  \textsc{France}}\vspace{1mm}\\
${}^f${ \small Service de Physique Th\'eorique et Math\'ematique,}\\
{\small  Universit\'e Libre de Bruxelles (ULB) and The International Solvay Institutes}\\
{\small Campus de la Plaine, CP 231,  B-1050 Bruxelles,  \textsc{Belgium}}\vspace{2mm}}
\date{\vspace{-5ex}}

\maketitle
\vfill
\begin{abstract}
\noindent We propose a new application of random tensor theory to studies of non-linear random flows in many variables. Our focus is on  
non-linear resonant  systems which often emerge as weakly non-linear approximations to problems whose linearized perturbations possess highly resonant spectra of frequencies (non-linear Schr\"odinger equations for Bose-Einstein condensates in harmonic traps, dynamics in Anti-de Sitter spacetimes, etc). 
We perform Gaussian averaging both for the tensor coupling between modes and for the initial conditions.
In the limit when the initial configuration has \emph{many modes} excited, we prove that there is 
a leading regime of perturbation theory governed by the \emph{melonic graphs} of random tensor theory.
Restricting the flow equation to the corresponding melonic approximation, we show that 
at least during a finite  time interval, the initial excitation spreads over 
more modes, as expected in a turbulent cascade. 
We call this phenomenon \emph{melonic turbulence}.

\end{abstract}


\setcounter{footnote}{0}
\setcounter{lemma}{0}
\setcounter{theorem}{0}

\tableofcontents
\section{Introduction}

Random (rectangular) matrices were first introduced by Wishart \cite{Wishart}. 
The Hermitian case was developed in physics to understand the quantum mechanics 
of large systems, with nuclear physics as an intended application
\cite{Wigner}. The Wigner-Dyson laws for the eigenvalues of a Gaussian independently and identically distributed (i.i.d.) Hermitian matrix show a remarkable eigenvalue
\emph{repulsion} \cite{Dyson}, due to the Vandermonde determinant coming from integration over the unitary group. It is also 
the source of the Tracy-Widom law for their extreme eigenvalue statistics \cite{MajumdarSchehr1}. 
A main later development was 't~Hooft's $1/N$ expansion for \emph{interacting} matrix models,
which builds a remarkable bridge between topology and physics \cite{Hooft}, connecting two-dimensional 
quantum gravity and random matrices \cite{2dgrav}.

The application of random matrix theory to \emph{linear} random flows, pioneered
in the paper of May \cite{May} on stability of large ecological systems, had immense influence, 
from ecological theory \cite{Allesina} to superstring theory \cite{Dine}. For our illustrative purposes, it is sufficient to inspect the following simplified version of May's setup: consider a flow  for a variable $X$ in 
${\mathbb R}^N$ with $N$ large. Near any equilibrium  we can approximate it with a linear flow,
which is integrable and complete:
\bee \dot X = MX \ => \ X = e^{Mt} X_0 .   \label{linearflow}
\ee
Suppose the matrix $M$ is random. If $M$ is assumed diagonal with \emph{independent} parity-symmetric random eigenvalues,
the probability of stability at positive times (which means all eigenvalues are negative) 
is obviously $2^{-N} = e^{-N \log 2}$. But it is more interesting to consider systems in which all modes are coupled together,
so that all the entries of the matrix $M$ are independent and identically distributed.
In this case, random matrix theory applies and the (fermion-like) repulsion between eigenvalues makes this probability 
of stability (all eigenvalues negative) much smaller, of the form $e^{-K N^2}$
at large $N$.\footnote{The constant $K$ can be exactly computed see, e.g., \cite{MajumdarSchehr1}.}
Hence, such a random linear flow is almost never stable. 

This lack of stability discovered by May means that there are typically many unstable directions in which the variables grow. 
To understand the generic behavior of random flows in many variables therefore requires
going beyond the linear regime. However, only linear flows are integrable and complete in time;
non-linear flows can diverge in finite time. This entails subtleties with defining averaged quantities for non-linear random flows, even for a finite time interval.  At a more basic level, non-linear flows are much more difficult to analyze, as their coefficients are no longer matrices, but \emph{tensors}. 
As efficient random tensor theory has not been developed until recently, these complications have hampered the study of non-linear random flows
in the past.

Random tensors, which generalize random matrices, were initially 
introduced for studies of discrete random geometry and quantum gravity \cite{Ambjorn}. 
Their theory has been given a boost with the discovery of specific $1/N$ expansions \cite{1/N} for tensors with a large number of dimensions given by $N$,
dominated by a very simple family of so-called melonic graphs \cite{Bonzom:2011zz},
which provides an analytic tool to investigate the large $N$ limit \cite{Gurauetal}.
This melonic dominance is now well understood as a robust universal property of random tensors \cite{TensUniv,MelonUniv}. Besides, the possible existence of enhanced scalings for tensor interactions, which leads to a richer dominant sector, has also been studied \cite{EnhScal}.
In parallel developments, tensor field analogs of non-commutative field theories were introduced and renormalized \cite{TensRen} and their renormalization 
group flows were investigated \cite{SigmaF}.
Non-perturbative or constructive aspects are also actively studied \cite{SigmaC}.
More recently, random tensor models were connected to the interesting holographic and quantum 
gravitational properties of the Sachdev-Ye-Kitaev (SYK) model and related models \cite{Kitaevetal,wittenetal} on the one hand, and of matrix models in the large $D$ limit \cite{LargeD} on the other hand. This class of  time-dependent models displays an interesting mix of maximally chaotic behavior 
and solvability in a certain limit governed by the melonic graphs. The SYK model has in fact originally appeared in the nuclear physics context starting with \cite{BohigasFlores, FrenchWong}, for a textbook treatment see \cite{Kota} -- while a recent treatment along similar lines of the quantum version of the resonant systems we shall be focusing on below can be found in \cite{quantres}.

In view of the above, it seems timely to apply these recent developments
to study non-linear random flows in many variables. In this paper we focus on
the question of energy cascades characteristic of turbulent flows, and restrict ourselves to a specific 
resonant non-linear equation, namely 
\bee  
i\, \frac{d\alpha_j }{dt}(t)  = \sum_{\substack{{j',k,k'=0}\\ {j+j'= k+k'}}}^\infty C_{jj'kk'} \bar \alpha_{j'}  (t)\alpha_k (t) \alpha_{k'} (t).  
\label{basicequa}
\ee
Here, $\alpha_n$ with $n\ge 0$ are an infinite sequence of complex-valued functions of time (whose physical origin is in complex amplitudes of linear normal modes of a weakly non-linear system). Such equations naturally emerge in weakly non-linear analysis of PDEs whose frequency spectra of linearized perturbations are highly resonant (more specifically, differences of any two frequencies of the linearized normal modes are integer in appropriate units). Two (related) applications of these ideas to concrete PDEs in recent literature, motivated by physical problems, are studies of weakly non-linear dynamics in Anti-de Sitter spacetime \cite{AdSres} (inspired, in particular, by its conjectured gravitational instability \cite{BR}), as well as applications to the Gross-Pitaevskii equation (which can also be called a non-linear Schr\"odinger equation) for Bose-Einstein condensates in harmonic traps \cite{GPres}.

To illustrate how resonant systems of the form \eqref{basicequa} arise from weakly non-linear PDE analysis, we briefly focus on the one-dimensional non-linear Schr\"odinger equation in a harmonic trap,
\begin{equation}
i\,\frac{\partial \Psi}{\partial t}=\frac12\left(-\frac{\partial^2}{\partial x^2}+x^2\right)\Psi +g|\Psi|^2\Psi,
\label{NLS1d}
\end{equation}
which provides a particularly straightforward setting. The linearized problem ($g=0$) is simply the harmonic oscillator Schr\"odinger equation, and its general solution is written as
\begin{equation}
\Psi=\sum_{n=0}^\infty \alpha_n \psi_n(x) e^{-iE_n t},\qquad E_n=n+\frac12,\qquad \frac12\left(-\frac{\partial^2}{\partial x^2}+x^2\right)\psi_n=E_n\psi_n,
\label{NLS1dlin}
\end{equation}
with constant $\alpha_n$. When a small non-zero coupling $g$ is turned on, $\alpha_n$ cease being constant and acquire slow drifts. One can of course derive an exact equation describing these slow drifts by substituting \eqref{NLS1dlin} into \eqref{NLS1d} and projecting on $\psi_j(x)$, which yields
\begin{equation} 
 i \frac{d}{dt}\alpha_j (t)  = g
  \sum_{\substack{{j',k,k'=0}}}^\infty C_{jj'kk'} \bar \alpha_{j'}  (t)\alpha_k (t) \alpha_{k'} (t)
 \,e^{i(E_j+E_{j'}-E_k-E_{k'})t}, 
\end{equation}
\noindent where $C_{jj'kk'}=\int dx \,\bar\psi_j  \bar\psi_{j'} \psi_k \psi_{k'}$, which in this $1$ dimensional problem reduces to $C_{jj'kk'}=\int dx \,\psi_j  \psi_{j'} \psi_k \psi_{k'}$ as the eigenstate wavefunctions are real in this case. This equation is still exactly identical to the original PDE, however in the weakly non-linear regime $g\ll 1$ simplifications occur. In this regime, 
$\alpha_j$ vary extremely slowly (on time scales of order $1/g$), while most of the terms on the right hand side oscillate fast (on time scales of order 1) due to the last exponential factor. Mathematical results on time-averaging (for a textbook exposition, see \cite{murdock}, and for a rigorous mathematical discussion specifically adapted to non-linear Schr\"odinger equations, see \cite{KM}) guarantee that one can simply discard any such oscillatory terms while still providing a uniformly accurate approximation on time scales of order $1/g$ for small $g$. In implementing this procedure, only terms satisfying $E_j+E_{j'}-E_k-E_{k'}\equiv j+j'-k-k'=0$ are retained (this is known as the ``resonance condition"), after which the evolution is conveniently re-expressed in terms of the ``slow time" $gt$, giving an equation of the form \eqref{basicequa}. Notice that in \eqref{basicequa} the sum is taken only over modes that satisfy the resonance condition.

Similar weakly non-linear analysis can be applied to other (often much more complicated) PDEs \cite{AdSres,GPres}, resulting again in effective \emph{resonant systems} of the form \eqref{basicequa}. The only difference is in the values of the interaction coefficients $C_{jj'kk'}$, which depend on the physics of the problem through the structure of the linearized normal modes and the specific form of the non-linearity. Resonant systems of the form \eqref{basicequa} are also interesting enough in their own right to be studied from a non-linear dynamics perspective. A particularly simple choice is $C_{jj'kk'}=1$, which results in a Lax-integrable system that has been proposed and examined in a series of works by G\'erard and Grellier with rigorous and far-reaching results for its solutions \cite{GG}. Furthermore, a very large class of partially solvable resonant systems generalizing some of the properties observed in the physically motivated examples of \cite{AdSres,GPres} has been constructed in \cite{AO}. Such profusion of different systems of the form \eqref{basicequa} naturally invites the question of typicality: which properties of solutions of \eqref{basicequa} hold \emph{on average}, in large ensembles of resonant systems defined by random $C_{jj'kk'}$  drawn from some distribution? Such statistical approach may often be more viable than analyzing extremely complicated non-linear dynamics of concrete individual resonant systems.

A key question for solutions of resonant systems is the emergence of \emph{turbulence}, which means excitation of modes $\alpha_n$ with $n\gg 1$ starting from initial data in which such modes are strongly suppressed. This question underlies a number of investigations of non-linear Schr\"odinger equations, and it is of pivotal importance for the conjectured weakly non-linear instability of Anti-de Sitter spacetime. Similarly, the Lax-integrable model of \cite{GG} has been explicitly designed as a tractable setting in which the question of turbulence can be thoroughly analyzed. The main topic of our investigation will be the turbulent properties of solutions of \eqref{basicequa} averaged over ensembles of resonant systems.
We can solve the equation as a power series in time.
When, in the spirit of May, we randomize this power series (over both couplings and initial conditions), Feynman-like perturbation theory emerges for averaged Sobolev norms that quantify the strength of the turbulent cascade.
In the limit of \emph{many initially excited low-lying modes},
we prove that the dominant terms at each order of this perturbation theory are given by the very specific \emph{melonic graphs} which generically dominate the perturbation theory of tensor models 
of large size \cite{Bonzom:2011zz,Gurauetal,MelonUniv}. 
This melonic approximation has been known in turbulence theory under the name of \emph{Direct Interaction Approximation} since the pioneering work of Kraichnan \cite{Kraichnan}. It also goes under the name
of \emph{Mode Coupling Approximation} (MCA) for critical dynamics or liquids and disordered systems such as spin glasses, see the review
 \cite{Bouchaud:1995wx}. However to our knowledge it has not been applied to
 the strongly resonant systems considered below. In this paper we \emph{prove} that for our models the \emph{melonic approximation} 
\emph{displays an energy cascade}, in the sense of Sobolev norm growth, at least within a certain initial time interval. This main result
of our paper hopefully should lead to more detailed study of the critical regime at which analyticity 
of our melonic approximation breaks down, and to
other applications of random tensor theory in the area of non-linear dynamics.

\section*{Acknowledgements}

We thank Peter Grassberger for a stimulating discussion at the beginning of this project. The work of S.D.~was supported by the Australian Research Council grant DP170102028. O.E. has been funded by CUniverse research promotion project (CUAASC)
at Chulalongkorn University. L.L.~is a JSPS International Research Fellow. G.V.~ is a Research Fellow at the Belgian F.R.S.-FNRS. This project has been initiated during the \emph{2nd Bangkok workshop on discrete geometry and statistics} at Chulalongkorn University.


 \section{Resonant systems}

\subsection{The model} 

Consider the resonant equation \eqref{basicequa},
in which the coefficients $C_{jj'kk'} $ are real and symmetric under the exchange of $j$ and $j'$, of $k$ and $k'$, and of the pairs $(jj')$ and $(kk')$. 
Under such conditions, the non-linear evolution equation~\eqref{basicequa} and its complex conjugate
can be considered as the canonical equations for the Hamiltonian
\bee 
H = \frac12 \sum_{\substack{{j,j',k,k'=0}\\ {j+j'= k+k'}}}^\infty C_{jj'kk'} \bar \alpha_{j}  (t)\bar \alpha_{j'}  (t)\alpha_k (t) \alpha_{k'} (t) \label{Ham}
\ee
with the symplectic form $i\sum_n d\bar\alpha_n\wedge d\alpha_n$.

An elegant way to take into account the resonance condition $j+j'= k+k'$ 
is to change from the variables $\{j,j',k,k'\}$ to the variables $\{S,j,k\}$ defined as 
$S=j+j'=k+k'$. 
In this new set of variables, the tensor couplings $C$ can be rewritten as an infinite family of real symmetric  $(S+1)\times(S+1)$ matrices
$C^S_{jk}$ (where $S$ runs over non-negative integers, that is, $S \in {\mathbb N}$) and the previously mentioned symmetries of C transform into
\bee C^S_{jk}=C^S_{kj}=C^S_{S-j, k }=C^S_{j, S-k}. \label{symcons}
\ee
The Hamiltonian becomes
\bee H = \frac12 \sum_{S=0}^\infty \sum_{j,k=0}^S C^S_{jk}\bar \alpha_j  (t)\bar \alpha_{S-j}  (t)\alpha_k (t) \alpha_{S-k} (t), \label{Ham1}
\ee
and the equations of motion become
\bee  i\, \frac{d\alpha_j }{dt}=  \sum_{S = j}^\infty \sum_{k=0}^S C^S_{jk} \bar \alpha_{S-j}  (t)\alpha_k (t) \alpha_{S-k} (t). \label{basicequa1}
\ee

In order to study the spread of energy over modes, we introduce the Sobolev norms 
\begin{equation}
S_\gamma ( t)= \sum_{\R \ge 0} \R^\gamma \bar{\alpha}_\R (t)\alpha_\R (t)  = \sum_{n\ge 0}   s_{\gamma,n} t^n,
\end{equation}
and study how they evolve over time. We have indicated the expansion of $S_\gamma ( t)$ in powers of $t$, which
shall be employed in our derivations.
Two specific cases, $S_0 (t)$ and $S_1 (t)$, are in fact independent of $t$, and correspond to known conserved quantities of \eqref{basicequa}. For $S_0$, it can be checked trivially, and 
for $S_1$ it requires the resonant condition. For resonant systems emerging from weakly nonlinear analysis of PDEs, $S_0$ can be thought of as a ``particle number'' quantifying excitations of the linearized modes (it becomes literally that if the model is quantized, as in \cite{quantres}), while $S_1$ is the total energy of the normal modes in the linearized theory.
On the other hand, $S_\gamma (t)$ for $\gamma>1$  are generically not conserved, and can be used to quantify the transfer of energy from the long wavelength modes to those with shorter wavelengths. The growth of these quantities indicates that the excitation of higher modes is getting stronger. Note that, evidently, $S_{\gamma'}>S_\gamma$ for $\gamma'>\gamma>0$. Hence the growth of $S_\gamma$ provides a lower bound on the growth of $S_{\gamma'}$.

Returning to \eqref{basicequa} and applying the same strategy as May, we can consider the symmetric coefficients $C_{jj'kk'}$ as Gaussian i.i.d.\ variables
satisfying the resonance condition or, equivalently, the family of symmetric matrices $C^S_{jk}$
as Gaussian i.i.d.\ variables. In the simpler $\{S,j,k\}$ parametrization, it corresponds to
imposing the covariance
\begin{align}
\label{eq: CovarianceNew}
\langle C^S_{jk}C^{S'}_{j'k'}\rangle_C =\frac{\delta_{SS'}}{8} 
 & \Bigl( \delta_{jj'}\delta_{kk'} + \delta_{j,S-j'}\delta_{kk'} + \delta_{jj'}\delta_{k,S-k'}+ \delta_{j,S-j'}\delta_{k,S-k'} \nonumber \\
& + \delta_{jk'}\delta_{kj'} + \delta_{j,S-k'}\delta_{kj'} + \delta_{jk'}\delta_{k,S-j'}+ \delta_{j,S-k'}\delta_{k,S-j'} \Bigr).
\end{align}
on the infinite family of real $(S+1)\times(S+1)$ matrices $\{C^S_{jk}\}_{S\in\bN}$
with \emph{no symmetries}. Indeed, the necessary symmetries are automatically implemented
by the eight terms in \eqref{eq: CovarianceNew}.\footnote{This point, although fundamental, is often confusing. As a clarification, the reader may be reminded that the Gaussian measure with covariance
0 on ${\mathbb R}$ \emph{is} the Dirac measure $\delta (x)$ implementing the constraint $x=0$; 
the Gaussian measure on ${\mathbb R^2}$ with covariance $\begin{pmatrix}1&1\\1&1\end{pmatrix}$ 
is proportional to $e^{-x^2/2}\delta (x-y)$  implementing the constraint $x=y$; so  Gaussian measures can implement \emph{constraints}. The choice of the Gaussian measure 
with covariance \eqref{eq: CovarianceNew} implements the desired symmetry constraints \eqref{symcons} on $C$.}

We choose initial conditions for the modes $\alpha_j$ in which the higher modes are suppressed. 
More precisely, we draw the initial conditions  from a random Gaussian ensemble, in which they are
independently but \emph{not} identically distributed with respect to $j$, and they spread over a large number $N \gg 1$ of low-lying modes. This is expressed by the following covariance:
\bee 
\langle  \alpha_j (0) \bar \alpha_{j'} (0) \rangle_\alpha  =\frac{ \delta_{jj'}}{N} \chi_N(j),  \quad
\langle  \alpha_j (0) \alpha_{j'} (0) \rangle_\alpha   =  \langle  \bar \alpha_j (0) \bar \alpha_{j'} (0) \rangle_\alpha =0,
\label{initN}
\ee
where the function $\chi_N(j)$ is such that $\sum_{j\ge 0 }\chi_N(j) = N$, so that we have the normalization condition 
\bee \sum_{j=0}^\infty   \langle  \vert \alpha_j (0) \vert^2  \rangle_\alpha =1 .\label{normalized}
\ee
In practice, the distribution that we use throughout this paper 
decays exponentially over $j$, so that 
\be 
\chi_N(j) =p^j \quad \text{ and } \quad N = \frac{1}{1-p},
\ee
where $0<p<1$ is fixed. The limit $N \to \infty$ corresponds to the limit\footnote{We could also consider an equidistribution with cutoff $N$, hence
where $\chi_N (j) =  1 $ if $0 \le j \le N-1$ and  $\chi_N (j) =  0 $  if $j>N$. This is the simplest distribution for a $1/N$ expansion.
The precise form of the $\chi$ function is not important for what will follow; however it is important 
to consider the $N \to \infty$ regime in which many modes are excited at $t=0$.} $p \to 1$.

\

The main quantities we want to study are the averaged Sobolev norms 
\bee \bar S_\gamma ( t):= \langle S_\gamma (t) \rangle_{C, \alpha} = \sum_{n \ {\rm even}}  \bar s_{\gamma,n} t^{n} .
\ee
Note that only even integers contribute in the sum, since the Gaussian distribution for $C$ is even.
Defining $\bar S_\gamma ( t)$ may be subtle, even though the individual coefficients of its time-series expansion
are perfectly well-defined and algorithmically computable. For instance, if it so happens that $S_\gamma$ blows up in finite time for some solutions, the fact that this blow-up time decreases when scaling up the mode couplings $C$ means that the ensemble-averaged $S_\gamma$ blows up for all finite times, and its time series necessarily has a zero radius of convergence. While examples where there is numerical evidence for finite-time blow up are known \cite{AdSres}, they involve the mode couplings $C$ growing without bound for large mode numbers, which is outside our ensemble of resonant systems.
Whether such complications actually occur in our context, is an open, interesting and complicated mathematical question (if they do, extra care will have to be taken in extracting meaningful information from our averaged quantities). Be it as it may, the dominant melonic part we shall extract from the expansion is always convergent, and should convey some information on the dynamics of initial configurations with a large spread over energies (a large number of initially excited low-lying modes).

When $N \to \infty$ in \eqref{initN}, hence $p \to 1$, \emph{more and more} low-lying modes are excited by the initial conditions. 
A well-established theoretical physics practice of $1/N$ expansions is to identify in $\bar s_{\gamma, n}$ the amplitudes that scale in the leading way as $N \to \infty$ 
at each fixed order $n$ of the perturbation series, and to \emph{restrict} the perturbation theory for $\bar S_\gamma (t)$ to these contributions. 
This leads to the cactus approximation of random vector models, to the planar approximation of random
matrix models and to the melonic approximation of random tensor models. 

At any fixed order  in $t$, we shall establish in Section~\ref{melodomi} below that
\bee
\bar s_{\gamma, n} =  s^{melo}_{\gamma, n}   + o_{\gamma, n} (1/N) .  \label{domiscaling1}
\ee
Accordingly, the melonic approximation 
\bee S^{melo}_\gamma (t) := \sum_{n \in 2{\mathbb N}} s^{melo}_{\gamma, n} t^{n} \label{domiscaling2}
\ee 
to the averaged Sobolev norm $\bar S_\gamma$ includes only graphs of the melonic type and we have
\begin{equation}
\bar S_\gamma (t)=  S^{melo}_\gamma (t)+ o_{\gamma, t} (1/N)  .\label{domiscaling3}
\end{equation}
We prove that, irrespectively of the possible complications for $\bar S_\gamma$, the time series for $S^{melo}_\gamma (t)$ 
has a finite, non-zero radius of convergence, hence it is well-defined and in fact \emph{analytic} at least for a finite time interval (in essence, this is because the melonic family has far \emph{fewer} graphs than the general family). In this sense, \eqref{domiscaling3} provides a heuristic motivation for our work, while our technical focus is on $S^{melo}_\gamma (t)$ defined through \eqref{domiscaling1} and \eqref{domiscaling2}. We can summarize the $1/N$ analysis as

\begin{theorem}\label{theo1}
The dominant graphs as $N \to \infty $ for the averaged Sobolev norm $\bar S_\gamma (t)$ are exactly the melonic graphs and the corresponding approximation
$S^{melo}_\gamma (t)$ is an analytic function of time in a disk $\vert t\vert < \rho$ of finite radius $\rho >0$.
\end{theorem}

We also check by a very simple explicit computation of $s^{melo}_{\gamma, 2} $ that
\begin{theorem}\label{theo2}
For any $\gamma >1$ there exists a constant $\delta$ such that  $S^{melo}_\gamma (t)$ grows monotonically in time for $t \in [0,\delta]$.
\end{theorem}

These are the main results of this paper: they mean that, \emph{in the melonic approximation},
energy spreads at least for a while from the low modes to the higher modes, as expected in a turbulent cascade. 
We call this phenomenon \emph{melonic turbulence}. (Note that technically, we employ the term `turbulence' in the
manner typical of studies of conservative deterministic turbulence in
PDEs, see, e.g., the classic paper \cite{NLStorus}, rather than in the
sense of dissipative hydrodynamic turbulence. In this setting,
turbulent phenomena are by definition characterized by various forms
of growth of Sobolev norms.)

According to the usual pattern of $1/N$ investigations, the next steps in the analysis could be
\begin{itemize}

\item  to identify the critical time at which the melonic approximation blows up and the corresponding critical behavior, 
then to incorporate more graphs near criticality. This corresponds to the single and double scaling limits
in matrix and tensor models \cite{2dgrav,TensScal};

\item to bound the non-melonic effects (possibly for a particular subclass of systems). This corresponds
to constructive studies \cite{RivConstruct} in quantum field theory.

\item to investigate with computer algebra the energy cascade and the speed of convergence to the melonic limit when $N \to \infty$.

\end{itemize}
Such developments are deferred to future work.

 \subsection{Tree expansion of non-linear flows in one dimension}
 
 We recall the combinatorial tree representation for the perturbative solution of any deterministic flow equation. This is best understood first in 
 the rather trivial case of a single real mode, for which variables separate and time evolution is solved by a single quadrature.
 Consider a function $x$ of time $t$ with subscript  notation $x_t$ instead of $x(t)$, hence initial condition $x_0$. The \emph{non-linear} flow equation 
\bee 
 \label{trivialq}
\dot x = \lambda x^{q} 
\ee
for $q \in {\mathbb N} $, $q \ge 2$, is trivially solved as 
\bee x_t = x_0 [ 1 - (q-1) \lambda t x_0^{q-1} ] ^{-\frac{1}{q-1}} = x_0 \sum_{h \in {\mathbb N}} \frac{ (\lambda t x_0^{q-1})^h}{h!} \prod _{k=1}^{h-1} [k(q-1)+1], \label{trivialsol}
\ee
and it converges for $ \vert  t  \vert < \frac{1}{(q-1) \lambda  x_0^{q-1} }$, but diverges after that, say, for positive $\lambda$ and $x_0$.

\

We now give the precise graphical representation of the Taylor series \eqref{trivialsol} in terms of \emph{trees}. The idea is to compute the $h^{\text{th}}$ derivative $x^{(h)}$ recursively using \eqref{trivialq}. We start with a particular vertex (the root)
and connect it with an edge to a first vertex of valency $q+1$. In this way we get a tree with one root, one vertex of valency $q+1$,  and $q$ leaves. To the vertex is associated a factor $\lambda$  and to each leaf a factor $x$, so that this first tree corresponds to $\dot x$ as given in \eqref{trivialq}.  

To compute $\ddot x$ we connect any of these leaves to a second vertex with valency $q+1$. In this way, we get $q$ possible trees with $2q-1$ leaves, corresponding to the computation of $\ddot x =  q\lambda \dot x x^{q-1} =  q \lambda^2 x^{2q -1} $. Note that the ordering of the edges around a vertex matters: it is responsible for the factor $q$. In the same way, we can then compute recursively $x^{(h)}$.

\begin{figure}[!ht]
	\centering
	\includegraphics[scale=0.75]{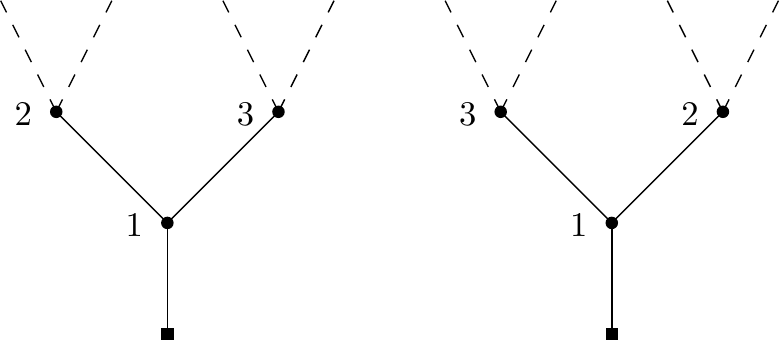}
	\caption{There are two heap-orderings for this binary 1-rooted tree. The root is the black square, the three true (i.e.\ 3-valent) vertices are shown as black disks
	and the four leaves are amputated.}
	\label{fig:Ordered-trees1}
\end{figure}

By looking at the examples with three vertices of valency $q+1$, corresponding to the case $h=3$, we see that the order in which the vertices of valency $q+1$ have been added in the recursion matters as well. For instance in the case $q=2$, the two trees of Fig.~\ref{fig:Ordered-trees1}  should be distinguished, where the labels indicate the order in which the vertices have been added. 

\

\begin{figure}[!ht]
	\centering
	\includegraphics[scale=0.75]{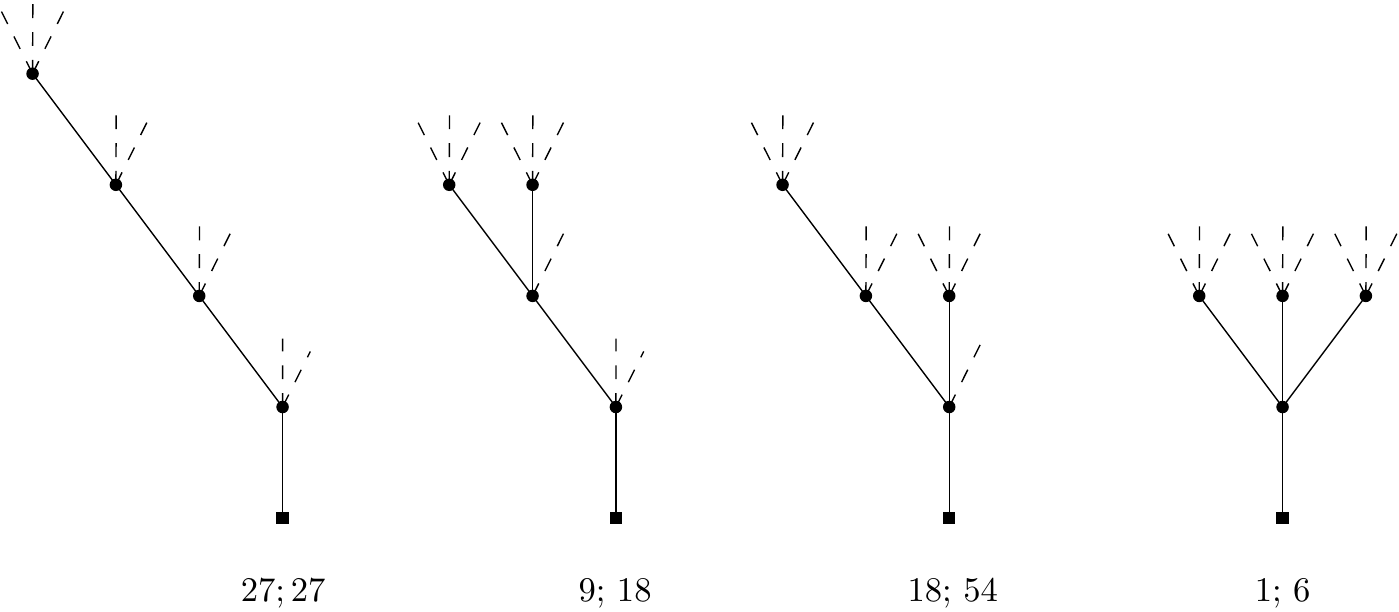}
	\caption{Ternary (heap-ordered) 1-rooted trees at order 4: for any of the four tree shapes, the first number gives the number of 1-rooted trees and the second
	the number of heap-ordered 1-rooted trees. They add up to 55 and 105 respectively.}
	\label{fig:Ordered-trees2}
\end{figure}

\noindent{\bf Ordered trees.} The trees arising in the iteration of this process  are \emph{heap-ordered, $q$-ary, 1-rooted trees}, which we now introduce. 
In this paper, a \emph{$1$-rooted tree} is a tree drawn on the plane, i.e.~a tree together with an ordering of the edges around each vertex, which in addition has a distinguished  vertex of valency one, called the root. The leaves are the vertices of valency 1 distinct from the root. It will be convenient in what follows to consider amputated leaves, hence to represent leaves simply as dashed half-edges hooked at another vertex but with no vertex at the end (see Figures \ref{fig:Ordered-trees1}-\ref{fig:Ordered-trees2}), and to represent the root as a black square of valency one. 

A rooted tree is said to be \emph{$q$-ary} if its vertices are all of valency $q+1$ (we also call these the ``true" vertices), except the root
and the leaves. The 1-rooted $q$-ary trees with $h$ true vertices are counted by $q$th Fuss-Catalan numbers $C_{h}^{q} := \frac{1}{qh+1 }{{qh+1}\choose{h}}$ 
\cite{FussCatalan}. 

Around each vertex $v$ of a rooted tree distinct from the root, there is a unique edge which belongs to the only path connecting this vertex to the root. We call this edge the parent-edge of $v$. The other edges incident to $v$ are the children-edges. This provides a kinship among vertices as well.

We define a \emph{heap-ordered tree} 
as a rooted tree together with a
labeling $\sigma$ of its $h$ true vertices from 1 to $h$, which respects the kinship of the vertices, that is  $\sigma (v) < \sigma (v') $ whenever $v$ is the parent of $v'$.

We denote $\cT_{1,q}^h$ the set of $q$-ary 1-rooted heap-ordered trees with $h$ true vertices.
They also have 
$h(q-1)+1$ leaves and $hq +1$ edges.  
An easy induction shows that $\vert \cT_{1,q}^h \vert = \prod _{k=1}^{h-1} [k(q-1)+1]$ (compare with \eqref{trivialsol}).
For instance, when $q=2$ there are five binary 1-rooted trees at order 3 and 14 at order 4 (the ordinary Catalan numbers) but there are 6 binary heap-ordered 1-rooted trees at order 3 and 24 at order 4. For $q=3$, hence ternary trees, these numbers become respectively 12  and 55 for the ordinary rooted trees (Fuss-Catalan numbers) and 15 and 105 for the heap-ordered ones (see Figure \ref{fig:Ordered-trees2}).

When iterating the computation of derivatives $x^{(h)}$ for the flow \eqref{trivialq}, one obtains a representation of $x^{(h)}$ as a sum over 
trees in $\cT_{1,q}^h$, with a factor $\lambda$  at each true vertex and a factor $x$ at each leaf. Indeed the labeling exactly keeps track  of the order at which the true vertices appear in the recursive process.

To compute $x_t$ from its Taylor expansion $x_t = \sum_{h \in {\mathbb N}} \frac{t^h}{h!} x_0^{(h)}$ as a power series in $t$, one simply needs to sum over $h$ the amplitudes $A(T)= \lambda^h x_0^{h(q-1) + 1}$ for  $T$  running over $\cT_{1,q}^h$ (where the amplitudes are now defined with a factor $x_0$ instead of $x$ at each leaf),  with an overall factor  $\frac{t^h}{h!}$:
\bee 
x_t = \sum_{h \in {\mathbb N}} \frac{t^h}{h!}\sum_{T \in \cT_{1,q}^h}  \A (T) .
\ee
Note that consistently, the case $h=0$ corresponds to a single leaf attached to the root, whose amplitude is $x_0$.
This is in agreement with \eqref{trivialsol} and gives the combinatorial tree representation of the perturbative solution to the non-linear flow. 
A generalization to flows in ${\mathbb R}^N$ or ${\mathbb C}^N$ is similar but we need labels on the edges to represent the indices running from $1$ to $N$ and arrows to distinguish complex numbers from their conjugates. 
We shall now treat the specific example of the flow \eqref{basicequa}.

\subsection{Tree expansion for the case under study }

We return to our pair of evolution equations \eqref{basicequa1}, which we write as
\bea 
\frac{d}{dt}\alpha_j (t) &=& -i \sum_{S = j}^\infty \sum_{k=0}^S C^S_{jk} \bar \alpha_{S-j}  (t)\alpha_k (t) \alpha_{S-k} (t),  \label{eq:evolution-equation1}\\
\frac{d}{dt} \bar \alpha_j (t)  &=&i  \sum_{S = j}^\infty \sum_{k=0}^S C^S_{jk} \alpha_{S-j} (t) \bar \alpha_k (t) \bar \alpha_{S-k} (t). \label{eq:evolution-equation2}
\eea
They are homogeneous of degree $q=3$, which we assume now in the rest of this paper. Like the simpler equation
\eqref{trivialq}, there is an iterative solution to these equations in terms of \emph{suitably oriented and indexed} trees $T$ (for \eqref{eq:evolution-equation1})
 and anti-trees $\bar T$  (for \eqref{eq:evolution-equation2}) in $\cT_{1,3}^h$ (they are heap-ordered 3-ary 1-rooted trees). We write these expansions as
\bea  
\alpha_\R (t) &=& \sum_{h \in {\mathbb N}} \frac{t^h}{h!} \sum_{T \in \cT_{1,3}^h}  \A_\R (T),  \label{eq:tree1} \\
\bar \alpha_\R (t) &=& \sum_{\bar h \in {\mathbb N}} \frac{t^{\bar h}}{\bar h!} \sum_{\bar T \in  \cT_{1,3}^{\bar h}}  \A_\R (\bar T)  \label{eq:tree2} .
\eea
where the amplitudes $\A_\R (T)$ and $\A_\R (\bar T)$ take into account the orientation and indexation of the trees and anti-trees $T$ and $\bar T$ in a way that we now explain.

\ 

\noindent{\bf Orientation of the trees. }
The $h$ 4-valent vertices of a tree in $\cT_{1,3}^h$ represent the way in which the $\alpha$ factors have been recursively differentiated. The heap-ordering precisely keeps track of the differentiation history: the vertex labeled $\sigma$ corresponds to the $\sigma$th differentiation step. The root of a tree (resp.~an anti-tree) initially represented an $\alpha$ (resp.~an $\bar \alpha$) factor, which we picture as out-going (resp.~in-going). The first true vertex resulted from the differentiation of this initial factor. Our graphical rule at a true vertex $v$ is to orient the children-edges which carried $\alpha$ factors at the  $\sigma(v)$th differentiation step as out-going and those which carried $\bar \alpha$ factors as in-going. The orientation of the full tree then results from recursively applying this ``parent" rule to the true vertices while following the heap-ordering of the vertices as follows. 
Around a vertex $v$ we denote the parent-edge $e_1(v)$.\footnote{Note that for the leaves it is the only incident (dashed) edge.} 
 The children-edges, which are ordered from 2 to 4, are denoted respectively by $e_2(v), e_3(v)$, and $e_4(v)$.  
Among children-edges at a vertex $v$,  the edge $e_2(v)$ is endowed with the same orientation (in-going or out-going) as the parent-edge $e_1(v)$, and the remaining two edges incident to $v$ are endowed with the opposite orientation.\footnote{It is convenient to draw trees with counterclockwise labeling of edges around the vertices and anti-trees	with the opposite clockwise ordering of edges around vertices, but this is not essential. It is the convention we adopt in the figures of the paper.}

\begin{figure}[t]
	\centering
	\includegraphics[scale=1]{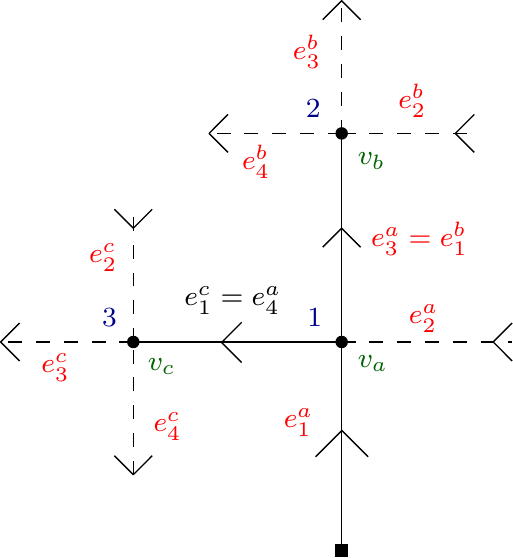}\hspace{2cm}\includegraphics[scale=1]{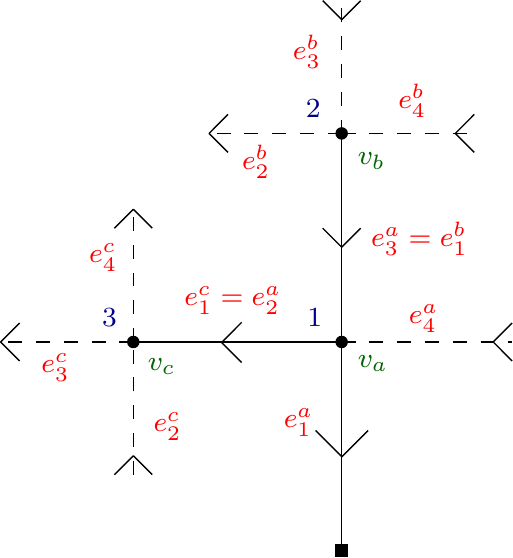}
	\caption{A heap-ordered tree oriented as a tree (left) or an anti-tree (right).}
	\label{fig:Orientations-trees}
\end{figure}

We remind the reader that, importantly, in a tree, only the leaves actually carry $\alpha$ or $\bar \alpha$ factors, the root and the solid edges do not.
In our amputated representation of Figures \ref{fig:Ordered-trees1}-\ref{fig:Ordered-trees2}, the leaves are half-edges and
now carry arrows: an arrow pointing out of the tree corresponds to a leaf and to an $\alpha$ factor whereas an arrow pointing 
into the tree corresponds to what we call an \emph{anti-leaf} and to an $\bar \alpha$ factor.\footnote{We stress however that in this amputated representation, the root (which also has valency one), is still represented as a vertex and does not bring any $\alpha$ or $\bar \alpha$ factor.}
Note that a tree  in $\cT_{1,3}^h$ with $h$ vertices has exactly $h+1$ leaves and $h$ anti-leaves; conversely
an anti-tree with $\bar h$ vertices has exactly $\bar h$ leaves and $\bar h+1$ anti-leaves. 
See some examples in Figure~\ref{fig:Orientations-trees}.

\

\noindent{\bf Momenta. }
In analogy with the Feynman graph terminology, let us call now the indices $j,S-j,k,S-k$ in (\ref{eq:evolution-equation1}-\ref{eq:evolution-equation2}) \emph{momenta}.\footnote{The resonance condition at each vertex is indeed reminiscent of energy-momentum conservation.} For a given 1-rooted tree with root index $\R$ entering the root, we now define its \emph{momentum attribution} $\cI_T$. 
It is a set of integers, defined first by a choice, for each 4-valent vertex $v$ of the tree, of three non-negative integers $S_v \in {\mathbb N}$, $j_v\le S_v$, and $k_v\le S_v$. The two momenta $j_v$ and $S_v- j_v$ are respectively attributed to the parent-edge $e_1(v)$ and the edge $e_2(v)$ and the two
momenta, $k_v$ and $S_v-k_v$, are respectively attributed to the edges $e_3(v)$ and $e_4(v)$. These choices  furthermore satisfy the constraints that if a vertex $v$ is incident to the root, the momentum of its parent-edge is the root momentum $j_v=\R$, and the momenta of the two half-edges forming any edge must be the same.

Therefore to each leaf $\ell$ is associated a momentum $j(\cI_T, \ell)$ and to each anti-leaf $\bar \ell$ is associated a momentum  $j(\cI_T, \bar \ell)$, namely those of their parent-vertex.

\ 

\noindent{\bf Amplitude of a tree. }
We then have the following ``Feynman rules":
\begin{itemize}
	\item to  each (4-valent) vertex $v$ of the tree or anti-tree, one associates a factor $C^{S_v}_{j_vk_v}$;
	\item to each leaf $\ell$  is associated a factor $\alpha_{j(\cI_T, \ell)} (0)$ 
	and to each anti-leaf $\bar\ell$, one associates a factor $\bar \alpha_{j(\cI_T, \bar \ell)} (0)$, where we stress again that the root vertex is not counted among leaves;
	\item each 4-valent vertex $v$ of the tree or anti-tree whose unique parent-edge $e_1(v)$ is in-going (resp.~out-going) is weighted by $(-i)$ (resp. $(+i)$).
\end{itemize}
The amplitude $\A_\R (T)$ is defined by multiplying all these factors and summing over all indices $\cI_T$:
\bee 
\A_{\R}(T) = \sum_{\cI_T}  \prod_v (\pm i) C^{S_v}_{j_vk_v} \prod_{\ell} \alpha_{j(\cI_T, \ell)} (0) \prod_{\bar \ell} \bar \alpha_{j(\cI_T, \bar \ell)}(0).
\ee
The summation over $\cI_T$ more precisely stands for the following summations and constraints
\be 
\biggl ( \sum_{S_{v_1} \ge\,\R} \sum_{ k_{v_1} =0 }^{S_{v_1}}\,\biggr)\biggl (\prod_{\substack{{v \text{ true} } \\ \text{vertex}}} \sum_{S_v \ge 0} \sum_{ j_v, k_v =0 }^{S_v}\,\biggr)\biggl (\prod_{ l\text{ leaf}} \sum_{j_l \ge 0}\,\biggr)  \prod_{e \text{ edge} }\, \delta_{a_e}^{b_e},
\ee
where for every edge $e$, $a_e$ and $b_e$ are the momenta of its two half-edges (including the leaves), and the vertex $v_1$ is the true vertex incident to the root ($\sigma(v_1)=1$). 
The amplitude $\A_{\R}(T)$  is a function of the entering momentum $\R$, of the couplings $C$ and of the initial data $\{\alpha_j (0), \bar{\alpha}_j(0)\}_{j \in {\mathbb N}}$.

Let us return to the Sobolev norms $S_\gamma(t) =\sum_{\R\ge 0} \R^\gamma \bar{\alpha}_\R(t)\alpha_\R(t)$.
Their time evolution, combining \eqref{eq:tree1}-\eqref{eq:tree2}, is written as
\bea 
S_\gamma(t) &=& \sum_{\R\ge 0} \frac{\R^\gamma}{N}\, G_2(\R , t) , \\
G_2(\R , t)  &=& N\bar{\alpha}_\R(t)\alpha_\R(t) = N\sum_{h \in {\mathbb N},   \bar h \in {\mathbb N}}  \frac{t^{h + \bar h}}{h ! \bar h !}  \sum_{T \in \cT_{1,3}^h,  \bar T \in \cT_{1,3}^{\bar h} }  
\A_\R (T) \A_\R (\bar T),
\label{eqref:AlphaBarAlpha}
\eea
where we included a factor $N$ in $G_2(\R,t)$ for more natural scaling properties.

\ 

\noindent{\bf 2-rooted tree. } 
It is possible to simplify the factorial factors in the expansion \eqref{eqref:AlphaBarAlpha} by using a slightly different notion of trees. By merging the roots of a tree $T$ with $h$ vertices and an anti-tree $\bar T$ with $\bar h$ vertices, we obtain a tree $U$ with $n = h + \bar h$ 4-valent (true) vertices, $2n+2$ leaves, and a single distinguished root-vertex of valency two. We call such trees \emph{2-rooted} trees. Note that in the case where the tree is trivial ($h=0$), the bivalent root is directly linked to a leaf (a dashed half-edge) and not to a true vertex, and similarly for the anti-tree. Most of what has been said for 1-rooted trees (ordering, parent-edge, heap-ordering) still holds for 2-rooted trees, and we denote $\cT_{2,3}^n$ the set of heap-ordered 3-ary 2-rooted trees with $n$ true vertices. The 2-rooted tree $U$ inherits the orientations of $T$ and $\bar T$: its bivalent root has one in-going edge and one out-going edge, and its $2n+2$ leaves divide into  $n+1$ leaves and $n+1$ anti-leaves. The momentum attribution $\cI_U$ of $U$ follows the exact same rules as the momentum attributions for $T$ and $\bar T$, the only difference being that there are now two vertices incident to the root.

Note however that when merging the roots of two heap-ordered 1-rooted trees, the resulting 2-rooted tree is not heap-ordered, and in order to heap-order it, we need to relabel its vertices. 
 There are several ways to define a heap-ordering on $U$ given the heap-orderings of $T$ and $\bar T$. Indeed, there is one such heap-ordering on $U$ for every set injection $\iota_T:\{1,\ldots,h\}\rightarrow \{1,\ldots, n=h+\bar h\}$ that preserves the natural order of integers. In fact, such an injection induces a relabeling of the vertices of $T$ seen as a subgraph of $U$. Meanwhile, the complement in $\{1,\ldots, n\}$ of the image $\textrm{Im} \ \iota_T$ induces a relabeling of the vertices of $\bar T$ seen as a subgraph of $U$. The above constructed relabelings thus indeed defines a heap-ordering of $U$. Therefore, for each pair of heap-ordered $T, \bar T$  there are as many heap-ordered 2-rooted trees as there are order-preserving injections $\iota_T$, namely $\frac{n!}{h! \bar h !}$. It follows that if we define the amplitude of $U$ as $\A_\R (U) := \A_\R (T) \A_\R (\bar T) $, we have:
\begin{equation}
	\sum_{U \in \cT_{2,3}^n }\A_\R (U)=\sum_{U \in \cT_{2,3}^n} \A_\R (T) \A_\R (\bar T) = \frac{n!}{h! \bar h !}\sum_{T \in \cT_{1,3}^h,  \bar T \in \cT_{1,3}^{\bar h} }\A_\R (T) \A_\R (\bar T).
\end{equation} 
From this we conclude, using \eqref{eqref:AlphaBarAlpha}, that $G_2(\R,t)$ is rewritten as a sum over heap-ordered $2$-rooted trees as   
\bea 
G_2(\R , t)  = N \sum_{n \in {\mathbb N}}  \frac{t^{n}}{n !}  \sum_{U \in \cT_{2,3}^n }  
\A_\R (U).
\label{eqref:AlphaBarAlphaU}
\eea

In the following lemma, we denote the indices of the leaves by $j_l$ instead of $j(\cI_T, \ell)$. Taking into account that at each true vertex $v$ there are four indices $j_v, S_v-j_v$ and $k_v, S_v-k_v$, we have the following very crude bound.

\begin{lemma} \label{flow}
Consider a tree $U\in \cT_{2,3}^n$ with a non-empty set $\cV$ of $n$ 4-valent vertices and a total set $\cL$ of $2n+2$ leaves (we do not distinguish leaves from anti-leaves here)
and a momentum attribution $\cI_U$. Then
\begin{equation} \sum_{v \in \cV} S_v  \le n \sum_{\ell \in \cL} j_\ell  
 \label{flow1}
\end{equation}
\end{lemma}
\proof The proof goes by induction. For $U$ with a single true vertex ($n=1$), there are four leaves. One of them is attached to the root, and therefore carries the index $\R$, and the three others are attached to the true vertex, with indices $S_v-\R, k_v, S_v - k_v$.  Hence, $\sum_{v \in \cV} S_v  =
\frac{1}{2}\sum_{\ell \in \cL} j_\ell  \le \sum_{\ell \in \cL} j_\ell   $ and the bound is true. Then, by induction, in a tree $U$  with $n$ true vertices, we consider a vertex $v_1$ with three incident leaves $l_2, l_3, l_4$ respectively linked to $v_1$ by $e_2(v_1), e_3(v_1), e_4(v_1)$. We denote  the leaf-set of $U$ by  $\cL=\{l_2, l_3, l_4\}\cup\cL''$. Removing $v_1$ and its three leaves, we obtain a tree $U'$ (the parent-edge  $e_1$ of $v_1$ is now a leaf $\ell_1$ of $U'$)  with $n-1$ true vertices and leaf-set $\cL' = \cL'' \cup \{\ell_1\}$.
Denote by $j_1$ the index of $e_1$ (which is also the index of $\ell_1$) and by $j_2, j_3, j_4$ those of the three leaves $l_2, l_3, l_4$. 
In our index convention, the pair of leaves $\ell_3, \ell_4$ at $v_1$ are such that their indices $j_3 , j_4$ add up to $S_{v_1}$; moreover we also have $j_1 + j_2 =S_{v_1}= j_3 + j_4$ for the other pair, hence $j_1 \le j_3 + j_4 $.
 Therefore  applying the induction hypothesis
	\bea 
	\nonumber \sum_{v \in \cV} S_v &=& S_{v_1} + \sum_{v \neq v_1} S_v \\\nonumber 
	&\le&  j_3 +j_4  + (n-1)\sum_{\ell \in \cL' } j_\ell   \\\nonumber 
	&\le&  j_3 + j_4  + (n-1) j_1+  (n-1)\sum_{\ell \in \cL'' } j_\ell   \\\nonumber 
	&\le& n ( j_3 + j_4 ) + n \sum_{\ell \in \cL'' } j_\ell  \\\nonumber 
	&\le& n( j_2 + j_3  + j_4 +\sum_{\ell \in \cL'' } j_\ell )  = n \sum_{\ell \in \cL} j_\ell  .
	\eea
\qed

The bound is of course far from optimal\footnote{It could be easily improved but there is little point in doing that until we get a better picture of the constructive
aspects of the full model (not just the melonic approximation) 
at finite $N$ (i.e., at $p$ bounded away from 1).} but will be enough to ensure convergence
of the sum over $\cI_U$.

\subsection{Averaged Sobolev norms}
\label{sec:Averaged-Sobolev}
The averagings over $C$ and $\alpha$ commute. It is quite convenient to first average over $\alpha$, then over $C$.

\

\begin{figure}[!ht]
	\centering
	\includegraphics[scale=0.8]{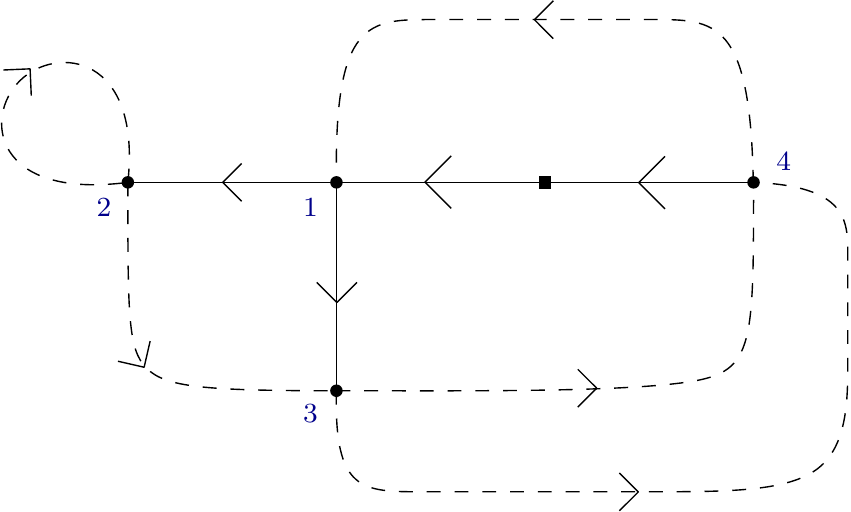}
	\caption{Oriented graph defined by a tree in $U\in\cT_{2,3}^n$ and a pairing $w \in \cW_\alpha  (U)$.}
	\label{fig:Example-alpha-graph}
\end{figure}

\noindent {\bf Averaging over $\alpha$.}  
We recall that the initial conditions are Gaussian distributed random variables of zero mean and covariance \eqref{initN}
\begin{equation} 
\langle \alpha_{j}(0) \bar \alpha_{j'}(0) \rangle_{\alpha}= \frac{\delta_{jj'}}{N} \chi_N (j). \label{remalphamean}
\end{equation}
A tree $U \in \cT_{2,3}^n$ has a binary root plus $n$ 4-valent true vertices forming a set $\cV (U)$, and $n+1$ leaves and $n+1$ anti-leaves. 
The averaging over $\alpha$ pairs together in all the $(n+1)!$ possible ways the $n+1$ leaves with the $n+1$ anti-leaves of $U$ into $n+1$ new $\alpha$-edges. We call $\cW_\alpha (U)$ the set of the $(n+1)!$ different pairings of leaves with anti-leaves and $\cE_\alpha (w)$ the set of $\alpha$-edges obtained for a given $w \in \cW_\alpha (U)$. Any pair $(U \in \cT_{2,3}^n, w \in \cW_\alpha (U))$  defines a new oriented graph with a bivalent root-vertex. Its $\alpha$-edges are naturally represented as \emph{dashed}, and oriented from the leaf to the anti-leaf. An example is shown in  Figure~\ref{fig:Example-alpha-graph} (note that the 2-rooted in this example tree is composed of the tree on the left of Fig.~\ref{fig:Orientations-trees}, and of the only anti-tree with one true vertex). The remaining $n$ edges in the graph are not dashed (they link rooted or true vertices of $U$), and are  depicted as  solid, to distinguish them from the dashed edges, because only the latter carry a $\frac{\chi_N}{N}$ factor. By \eqref{remalphamean}, any dashed edge $e$ also constrains the two indices $j_e, \bar j_e$ of the leaf and anti-leaf that it joins to be equal. 

 Note that any $w \in \cW_\alpha  (U)$ must connect the $T$ and $\bar T$ pieces of $U$
simply because the number of leaves and anti-leaves differ by one in $T$ and also in $\bar T$.

The $\alpha$-averaged $G_2$ function is therefore a sum over trees $U\in\cT_{2,3}^n$ and pairings $w\in\cW_\alpha(U)$
of an associated amplitude in which the leaf factor $\prod_{\ell} \alpha_{j(\cI_U, \ell)} (0) \prod_{\bar \ell} \bar \alpha_{j(\cI_U, \bar \ell)}(0)$
in the tree amplitudes have been replaced by a dashed edge factor $\prod_{e\in \cE_\alpha (w)} \frac{\chi_N (j_e) }{N}$.
Hence, remembering the scaling factor $N$, the fact that $\vert \cE_\alpha (w) \vert = n + 1$ and $ \chi_N (j_e) = p^{j_e}$, we have
\bee 
\label{eq:G2-Average-alpha}
\langle G_2(\R , t) \rangle_\alpha =  
\sum_{n \in {\mathbb N}}    \frac{1}{N^{n }} 
\frac{t^{n}}{n !}  \sum_{U \in \cT_{2,3}^h} 
 \sum_{\cI_U} 
\prod_{v \in \cV(U)} (\pm i ) C^{S_v}_{j_vk_v}  \sum_{w \in \cW_\alpha  (U)} \prod_{e\in \cE_\alpha (w)} \delta_{j_e \bar j_e}p^{j_e} ,
\ee
where $\cI_U$ is the momentum attribution of $U$.

\

\noindent {\bf Averaging over $C$. } As a reminder, the tensor coefficients $C$ are Gaussian distributed variables of zero mean and covariance \eqref{eq: CovarianceNew}
\bea
\langle C^S_{jk}C^{S'}_{j'k'}\rangle_C &=&\frac{\delta_{SS'}}{8} \Bigl( \delta_{jj'}\delta_{kk'} + \delta_{j,S-j'}\delta_{kk'} + \delta_{jj'}\delta_{k,S-k'}+ \delta_{j,S-j'}\delta_{k,S-k'} \nonumber \\ 
&&\ + \ \delta_{jk'}\delta_{kj'} + \delta_{j,S-k'}\delta_{kj'} + \delta_{jk'}\delta_{k,S-j'}+ \delta_{j,S-k'}\delta_{k,S-j'} \Bigr).
\label{eq: CovarianceNew1}
\eea
A first consequence is that when averaged over $C$, the terms in the expansion \eqref{eq:G2-Average-alpha} of $\langle G_2(\R , t) \rangle_\alpha$ that correspond to graphs $(U, w)$ 
with an odd number $n$ of true vertices vanish. 

Let us focus on the contribution to the expansion $\langle G_2(\R , t) \rangle_\alpha$ of a graph with an even number of true vertices. The averaging over $C$ of the corresponding term is expressed as a sum over all the possible ways of pairing the true vertices of the graph two-by-two.  For each such partition in pairs of vertices, the coefficients $C$ associated with the vertices of a given pair are replaced with the covariance \eqref{eq: CovarianceNew1} (the indices $S,j,k$ and $S',j',k'$ correspond to the indices $S_v,j_v,k_v$ and $S_{v'}, j_{v'}, k_{v'}$ associated with the two true vertices $v$ and $v'$). This is known as the Wick theorem, and it is common to call such a pairing of two $C$'s a \emph{Wick contraction}. 
For a graph $(U, w)$  with $n$ true vertices, there are $n!!:=n\cdot(n-1)\cdot(n-3)\cdot \ldots \cdot 3\cdot 1$ possible ways of performing the $n/2$ Wick contractions. 
\begin{figure}[!h]
	\centering
	\includegraphics[scale=1]{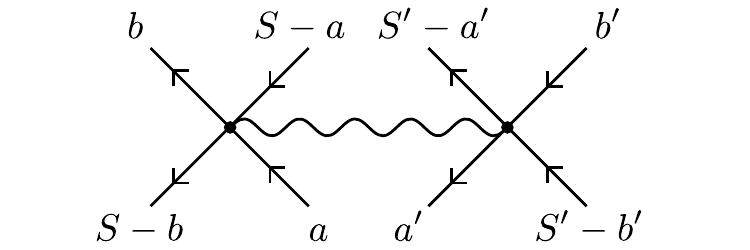}
	\caption{A wavy edge represents the averaging of two tensors $C$ at two different true vertices. }
	\label{fig:CWickPairing}
\end{figure}

We represent a Wick contraction between two tensors $C$ as a wavy line between the two corresponding true vertices, as depicted in Fig.~\ref{fig:CWickPairing} (the half-edges are solid in the figure, but up to three of them at each true vertex might be dashed). In Fig.~\ref{fig:CWickPairing}, depending on whether the parent-edge is in-going or out-going, the indices $a$ and $b$ take the value $j_v$  or $k_v$, and similarly for $a',b'$ and $j_{v'}, k_{v'}$.
The graphs obtained after averaging over $C$ thus have a new set of $n/2$ wavy edges. For each such edge, there is a sum implementing the eight different terms in \eqref{eq: CovarianceNew1}.

We denote by $\cW_C$ the set of Wick contractions of all the $C$ factors together with one of the eight different possibilities for each wavy line.
In the following, we call the eight terms in \eqref{eq: CovarianceNew1} \emph{propagators}. An element of $\cW_C$ is then a choice of a partition of all of the true vertices in pairs of vertices (represented by wavy lines), together with a choice of propagator, \textit{i.e.}~of one of the eight terms in \eqref{eq: CovarianceNew1}, for each wavy line. Each  $w' \in \cW_C$ gives a set of new momentum identifications, which we denote for the moment as $\delta_{w'} (\cI_U)$.

Note that 
$\vert \cW_C \vert = 8^{n/2} n!!$. Indeed, the number of pairings of all the true vertices is $n!!$, and it should be  multiplied by $8^{n/2}$, because there are eight choices of possible propagators for each wavy line. 

In this way, the expansion for the function $G_2$, when averaged over $\alpha$ and $C$, is expressed as a sum over graphs $G=(U,w,w')$ which have a set $\cV$ of $n$ true vertices (which are now \emph{five-valent} if we count the wavy edges) and one bivalent root, a set $\cE_s$ of $n$ solid edges, a set $\cE_\alpha$ of $n+1$ dashed edges,  
and a set $\cE_{C}$ of $n/2$ wavy edges.
The root constrains the momenta of the two edges attached to be $\R$. 
We write this expansion as follows:
\bea 
\langle G_2(\R , t) \rangle_{\alpha, C} &=& \sum_{n \ {\rm even}}
 \frac{t^n}{n !} 8^{-n/2} \sum_{U \in \cT_{2,3}^n  } \epsilon (U )   \sum_{w \in \cW_\alpha (U) \atop w' \in \cW_C (U)} \cA_\R (G),
\label{fullexpan1} \\
\cA_\R (G) &=&   \frac{1}{N^{n }} \sum_{\cI_U} \delta_{w'} (\cI_U) \prod_{e\in \cE_\alpha (w)} \delta_{j_e \bar j_e} p^{j_e} ,
 \label{fullexpans}
\eea
where $ \epsilon (U)$ is the sign obtained by collecting all the $n$ factors $\pm i$ in the previous formula  (since $n=h+\bar h$ is even, these factors 
must multiply to a real sign $\pm 1$), and $\cA_\R (G)$ is the \emph{amplitude} associated to the graph $G=(U, w, w')$, which is now obviously strictly positive. 
Indeed, at fixed root-momentum  $\R$, it evaluates the sum over all the $\{S_v,j_v,k_v\}$ integers using the delta constraints in $\delta_{w'} (\cI_U) \prod_{e\in \cE_\alpha (w)} \delta_{j_e \bar j_e} $, the exponential decays $ \prod_{e\in \cE_\alpha (w)}  p^{j_e}$ for the momenta of the dashed edges and the root constraint that the two incident edges have fixed momentum $\R$.

\ 

\noindent {\bf A first bound on the graph amplitudes. } We are in fact interested in understanding the scaling in $N$ of $\cA_\R (G)$ at fixed $n$, in order to identify the dominant amplitudes as $N \to \infty$. This will be done in Section~\ref{melodomi}. But before that, we can provide a very crude first bound on the graph amplitudes. 

In any graph $G$ with $n$ true vertices, since $p\in [0,1)$ and each $\alpha$-edge contracts two former leaves of $U$, using Lemma \ref{flow},  
\bee \prod_{e\in \cE_\alpha (w)}  p^{j_e} \le   \prod_{v \in \cV} p^{\frac{S_v}{2n}}.
\ee
Focusing on a wavy line between two true vertices of $G$, we see that the six sums involved are reduced to at most three sums $S,j,k$. Omitting the constraints from the solid and dashed edges, we thus bound the amplitude of a graph by  $$\frac{1}{N^{n }} 8^{-n/2} \Bigl( \sum_{S=0}^\infty \sum_{j=0}^{S} \sum_{k=0}^{S} p^{\frac{S}{n}} \Bigr)^{n/2}.$$ Computing these $3n/2$ sums, we find that 
\be 
\cA_\R (G) \le \frac{1}{N^{n }} 8^{-n/2} \Biggl( \frac{2n^3}{(1-p)^3} + O\left(\frac{n^3}{(1-p)^2}\right) \Biggr)^{n/2}.
\ee
Thus, there exists a constant independent of $p$, namely $K_{n}$ (thus possibly $n$-dependent), such that
\bee
\cA_\R (G)  \le (K_{n})^n n^{3n/2} N^{\frac{n}{2} }. \label{crudebound}
\ee
This bound gives us a first upper bound on the scaling in $N=\frac1{1-p}$ and allows one to check that each amplitude is finite, for $p<1$.

\

These are however very loose bounds. In Section~\ref{melodomi}, we will show that the scaling behavior in $N$ of the amplitudes $\cA_\R(G)$ is bounded from above by $N^{-d(G)}$ with $d(G) \ge 0$.
Also, the $n^{3n/2}$ factorial growth is only a very crude overestimate.
This is what the melonic analysis of Section~\ref{melodomi} will prove. Before that, let us however describe  the heuristic  asymptotic behavior that is expected for the series \eqref{fullexpan1} at finite $N$, but at large order $n$.

\subsection{Large order heuristic analysis}

 In this section, we perform a \emph{heuristic} analysis of the analytic properties of the series expansion in $t$ of the Sobolev norms around $t=0$. This analysis relies on the study of the large order behavior of the coefficients. This behavior is inferred using a graph counting argument. The result seems to indicate that the series cannot be fully summed and, consequently, that the underlying function of $t$ is not analytic at $t=0$. This gives an additional motivation to study the relevant (melonic) sub-series in the limit $p\rightarrow 1$, which is expected to be analytic at $t=0$.
(In our analysis of the coefficients of the series, we shall only be keeping factorials and $K^n$ factors,
as this is sufficient to infer, \textit{a priori}, the analytic behavior of the series at $t=0$.
Neglecting finite powers of $n$, we denote this large order analysis by the mathematical symbol $\simeq$.)

Consider a fixed even order $n =  h + \bar h$ of perturbation theory. The number of ordered trees in $\cT_{2,3}^n$ at order $n$, divided by the symmetry factor $\frac{1}{n!}$, is 
\bee \frac{\vert \cT_{2,3}^n \vert}{n!} = \frac{1}{n!} \sum_{h=0}^n  \frac{n!}{h! (n-h)!} \vert \cT_{1,3}^h \vert \vert \cT_{1,3}^{n-h} \vert \simeq 2^n ,
\ee
since $\vert \cT_{1,3}^h \vert = \prod _{k=1}^{h-1} [2k+1] \simeq 2^h h! $
(compare with \eqref{trivialsol}). The total number of pairings $w$ from leaves to anti-leaves is $(n+1 )! \simeq n!$. Finally the number of pairings of the
$C$ couplings is $8^{n/2} n!!  \simeq 8^{n/2}  \sqrt{ n! } $. 
Indeed, using Stirling formula, and the fact that since $n$ is even, $n=2k$, for some integer $k$, we have
\begin{align}
&n!!\sim 2^k \sqrt{2\pi}(k/e)^k\nonumber \\
&n!\sim 2^{2k} \sqrt{2\pi} (k/e)^{2k} \nonumber,
\end{align}
thus $n!!\simeq 
\, \sqrt{n!}$.
Since the total number of graphs to sum, when divided by the symmetry factor $\frac{1}{n!}$, is 
$$\frac{\vert \cT_{2,3}^n \vert}{n!}8^{n/2}(n+1)!n!!,$$ the coefficients of the series at large order scale in $n$ as $2^{5n/2}  (n!)^{3/2}$. 
At finite values of $N$ (that is when our parameter $p$
is bounded away from 1) we expect the various sums
over indices to be exponentially convergent, hence a generic exponential bound on any amplitude in the number of its vertices (the bound \eqref{flow1} being very far from sharp). 
In this case we therefore expect that the $n$-th order of the time perturbative series for $\bar S_\gamma$   should behave as
\bee \vert s_{\gamma, n} \vert  \le K^n  (n!)^{3/2}
\ee
for some value of $K$.
Of course amplitudes have various signs and can compensate (this happens after all for $\gamma = 0$ and $\gamma =1$).
However for $\gamma >1$, there are no apparent reasons for such cancellations, hence a lack of analyticity  of the averaged Sobolev norms $\bar S_\gamma (t)$ is suggested by the graph counting argument.\footnote{This is typical of QFT-like expansions, from which one typically expects at best some kind of Borel summability, depending on the stability properties at $t=0$ of the particular model considered \cite{RivConstruct}.} Such crude arguments are, of course, far from being water-tight, and the situation requires further study.

Now the $N \to \infty$ limit allows, like any $1/N$ expansion, to turn this (potential) problem around. 
In  the next section, we shall see indeed that this limit selects a much smaller family of dominant graphs, the \emph{melonic} graphs, 
whose number is exponentially bounded in the number of vertices, with also an exponential upper bound on individual amplitudes. 
The melonic graphs in fact are those with the maximal number of \emph{faces} in a stranded representation defined below. There are \textit{a priori}
three indices to sum over in an 8-valent node but at least one sum can be considered ``external" and is hence spared. So heuristically, we expect a
maximal scaling as $N^{2n/2} = N^{n}$ for the sums over indices in $\cI_T$ and $\cI_{\bar T}$, plus a 
remaining constraint similar to $\chi_N (\R)$ for the root index.
 Compensating  with the $1/N^{n}$ factor in \eqref{fullexpans} and taking into account that 
 $\sum_{\R \le N}  \R^\gamma  \chi_N (\R)= O(N^\gamma)$
 we expect the behavior \eqref{domiscaling1}-\eqref{domiscaling3} for the melonic approximation.
 Moreover, with the same reasoning, we expect 
 the sum of the melonic sub-series $ S^{melo}_\gamma (t) = \sum_n s^{melo}_{\gamma,n} t^n$, in contrast to the sum of the full series,
to straightforwardly \emph{exist and to define an analytic function in a neighborhood of $t=0$}.
This is in fact what we \emph{prove} in Section~\ref{melodomi}.

Physically, there are then two interesting regimes to consider for this melonic approximation $S^{melo}_\gamma$. The small $t$ behavior is governed
by the first non-trivial term $s^{melo}_{\gamma,2}$ and we show in Section \ref{sec:order_2} that it is strictly positive for $\gamma>1$, leading to 
a cascade towards higher modes at least during a finite time interval. Then, the asymptotic regime at large $n$
is also physically interesting. Specifying for instance $\gamma =2$, i.e., the simplest non trivial norm  
$S^{melo}_2 (t)$, and \emph{no longer neglecting finite powers of $n$}, we expect an asymptotics for $s^{melo}_{2,n}$ presumably of the form
\bee s^{melo}_{2,n} \sim K_2^n n^{\beta} .
\ee
Such asymptotics leads to a finite critical time $t_c = K_2^{-1}$, at which this Sobolev norm $S_2^{melo}$ would no longer be analytic. For $\beta > -1$
it would in fact blow up\footnote{This blow up behavior would hopefully be universal
for a large class of such models, being an analog of some susceptibility in the quantum gravity context.} with critical rate  
$(t - t_c)^{-1- \beta} $. \
However, such a critical time $t_c$ could turn out to be a negative number. 
In such a case, the positive time dynamics for that Sobolev norm may not blow-up in finite time. 
This is clearly an issue deserving future investigation

\section{Explicit computations at order $t^2$}\label{sec:order_2}

Before establishing general bounds for the graph amplitudes \eqref{fullexpans}, let us compute the first non-trivial order of perturbation theory, namely $n=2$. In that case, there is a single possible Wick contraction $w'$, hence, in the figures, the corresponding wavy edge will be omitted, but of course the indices identification that it implies will be included in the computations. 

\subsection{Amplitudes at order 2}\label{sec:order_2-amplitudes}

We shall now list these contributions at order two in $t$,  and compute the corresponding graph amplitudes. 
We do not represent the heap-orderings on the diagrams in the figures, as they simply provide a counting factor which we will indicate in each case. 
We arrange the contributions into four different groups.

\begin{figure}[!ht]
	\centering
	\includegraphics[scale=1]{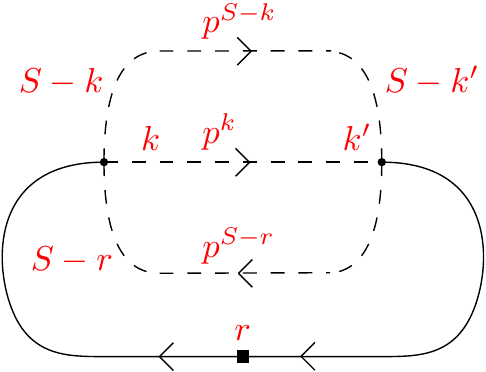} 
	\hspace{1cm}
	\includegraphics[scale=1]{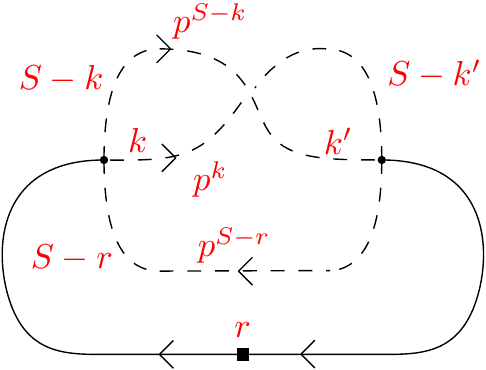}
	\caption{Type-$\I$ diagrams.}
	\label{fig:Dipole1}
\end{figure}

\

\noindent {\bf Graphs of type $\I$.}  
Each one of the diagrams in Fig.~\ref{fig:Dipole1} has two heap-orderings (the root is labeled 1 and there are  two ways of labeling the two other true vertices). As we shall see, these four diagrams give the same total contribution to $\langle G_2(\R ,t) \rangle_{\alpha, C}$ at order 2, which can be understood from the symmetries of $C$. However, for a given choice of propagator for the wavy edge, the amplitudes of the corresponding graphs for the diagrams on the left and on the right of Fig.~\ref{fig:Dipole1} actually differ. In total, there are $2\times2\times8$ graphs corresponding to the diagrams shown in Fig.~\ref{fig:Dipole1}. We call them graphs of type $\I$. In the following, we provide step-by-step details for the computation of the amplitude associated with any one of the $2\times8$ graphs $G=(U, w, w')$ corresponding to the diagram on the left of Fig.~\ref{fig:Dipole1}. Then, we give the results for the amplitude of the other graphs.

Using \eqref{fullexpans}, the amplitude of a graph corresponding to the left diagram of Fig.~\ref{fig:Dipole1} reads
\be 
\cA_\R(G)= \frac 1 {N^2} \sum_{S\ge \R
}\sum_{j,k=0}^{S}\sum_{S'\ge \R
}\sum_{j',k'=0}^{S'} \delta_{w'}(\cI_U) p^{S-k} \delta_{S-k}^{S'-k'} p^{k} \delta_{k}^{k'} p^{S-j} \delta_{S-j}^{S'-j'} \delta_{j}^{r} \delta_{r}^{j'},
\ee 
where $\delta_{w'}(\cI_U)$ is one of the eight propagators in \eqref{eq: CovarianceNew1}. 
As a first step, we use the identification between $S$ and $S'$ in $\delta_{w'}(\cI_U)$ to sum over $S'$, 
and sum over $j$ and $j'$, which are fixed to $r$, so that we obtain
\be 
\cA_\R(G)= \frac 1 {N^2} \sum_{S\ge r}\sum_{k,k'=0}^{S}  \tilde\delta_{w'}(\cI_U) p^{2S-r} \delta_{k}^{k'},
\ee
where $\tilde\delta_{w'}(\cI_U)$ is now one of the eight propagators  $1$,  $\delta_k^{S-k'}$,  $\delta_\R^{S-\R}$, $\delta_k^{S-k'}\delta_\R^{S-\R}$, $\delta_k^{\R}$ , $\delta_k^{S-\R}\delta_\R^{k'}$, $\delta_k^{\R}\delta_\R^{S-k'}$, or $\delta_k^{S-\R}$. The contribution from the first trivial propagator $\tilde \delta_{w'}(\cI_U)=1$ (originally $ \delta_{w'}(\cI_U)=\delta_S^{S'}\delta_j^{j'}\delta_k^{k'}$) is
\be 
\label{eq:AI-1}
\cA_\R(G_m^{\I})= p^\R  \Bigl(\frac {p^2}{(1+p)^2} + \frac 1 N \frac {\R+1}{1+p}\Bigr).
\ee
The sum of the contributions from the other seven propagators is
\be 
\label{eq:AI-2}
\frac 1 N \Bigl[ \frac{ p^\R}{(1+p)(1+p^2)} 
\bigl(\delta_{\R [2]}^0 +p^2\delta_{\R[2]}^1 \bigr) + \frac{2p^\R}{1+p}\Bigr]  +\frac 1 {N^2}2 p^{3\R}(\R+2), 
\ee
where $\delta_{\R [2]}^0$ vanishes if $\R$ is even, and conversely for $\delta_{\R [2]}^1$. In particular, as will be clarified in the following, the contributions of these seven propagators for the wavy line are subdominant when $p\rightarrow 1$.

Using the same reasoning, the amplitude of a graph corresponding to the diagram on the right of Fig.~\ref{fig:Dipole1} is 
\be 
\cA_\R(G')= \frac 1 {N^2} \sum_{S\ge \R}\sum_{k, k'=0}^{S} \tilde\delta_{w'}(\cI_U) p^{2S-\R} \delta_{k}^{S-k'},
\ee 
where $\tilde\delta_{w'}(\cI_U)$ is one of the eight propagators
$\delta_k^{k'}$, $\delta_k^{k'}\delta_\R^{S-\R}$, $1$, $\delta_\R^{S-\R}$, $\delta_k^{\R}$ , $\delta_k^{S-\R}\delta_\R^{k'}$, $\delta_k^{\R}\delta_\R^{S-k'}$, or $\delta_\R^{S-\R}$. Now, the dominant contribution only comes from the third propagator $\delta_{w'}(\cI_U)=\delta_S^{S'}\delta_j^{j'}\delta_k^{S-k'}$ and gives the same result as \eqref{eq:AI-1}. The same holds for the seven other propagators and \eqref{eq:AI-2}.

Therefore, we observe that the total sum of the contributions of the graphs from the left and from the right of Fig.~\ref{fig:Dipole1} is the same. This result can actually be traced back to the symmetries of $C$. Indeed, using these symmetries, one can untwist the dashed edges of the graphs from the right of Fig.~\ref{fig:Dipole1}. Then, by a local relabelling $k' \rightarrow S'-k'$, we directly obtain the graphs from the left of the figure. Besides, this also explains why it is the first propagator $\delta_S^{S'}\delta_j^{j'}\delta_k^{k'}$ that gives the dominant contribution for the graphs from the left of the figure whereas it is the third propagator $\delta_S^{S'}\delta_j^{j'}\delta_k^{S-k'}$ for the graphs from the right. In both cases, it is obtained for the trivial propagator $\tilde\delta_{w'}(\cI_U)=1$.

In total, we see that the sum of the amplitudes of the 32 graphs of type~$\I$, denoted $\cA_r^{\I}$, is four times \eqref{eq:AI-1} plus four times \eqref{eq:AI-2}. The total contribution of these graphs to $\langle G_2(\R , t) \rangle_{\alpha, C}$ is $\frac{t^2}{16}\cA_r^{\I}$.

\begin{figure}[ht!]
	\centering
	\includegraphics[scale=1]{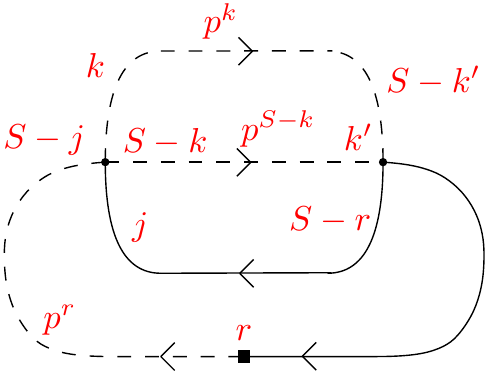} 
	\caption{Type-$\II$ diagram.}
	\label{fig:Dipole2}
\end{figure}

\noindent {\bf Graphs of type $\II$.} The second kind of graphs is shown in Fig.~\ref{fig:Dipole2}. We only draw one example, however there are four diagrams which all give the same contribution due to the symmetries of $C$: they are obtained by exchanging the role of the tree and the anti-tree, and by crossing the $k$ and $S-k$ edges as in Fig.~\ref{fig:Dipole1}. Each one of these four diagrams gives 8 graphs, thus a total of 32 graphs (here, the trees have a unique heap-ordering).

Taking into account the symmetries of $C$, the total contribution from these 32 graphs to $\langle G_2(\R , t) \rangle_{\alpha, C}$ is  
\be 
 \frac {2t^2}{N^2} \sum_{S\ge \R}\sum_{k,j=0}^{S}
 \sum_{S'\ge \R}
 \sum_{k'=0}^{S'}\langle C_{jk}^S C_{\R k'}^{S'}\rangle p^{S+\R}
\delta_{j}^{S'-\R}\delta_{k}^{S'-k'}\delta_{S-k}^{k'}\delta_{S-j}^{\R},
\ee 
and it is enough to compute the 8 contributions for the example shown in Fig.~\ref{fig:Dipole2} and multiply it by 4.
The amplitude of any one of the $8$ graphs $G=(U, w, w')$ in the example of Fig.~\ref{fig:Dipole2} is 
\be 
\cA_\R(G)= \frac 1 {N^2} \sum_{S\ge \R}\sum_{k, k'=0}^{S} \tilde\delta_{w'}(\cI_U) p^{S + \R}\delta_{k}^{S-k'},
\ee 
where $\tilde\delta_{w'}(\cI_U)$ is one of the 8 propagators
in  \eqref{eq: CovarianceNew1}:
$\delta^\R_{S-\R}\delta_k^{k'}$,  $\delta_k^{k'}$,  $\delta^\R_{S-\R}$, $1$, $\delta_{S-\R}^{k'}\delta_k^{\R}$ , $\delta_\R^{k'}\delta_k^{\R}$, $\delta_{S-\R}^{k'}\delta_k^{S-\R}$, or $\delta_\R^{k'}\delta^{S-\R}_{k}$. 
The contribution from the fourth trivial propagator $\tilde\delta_{w'}(\cI_U)=1$ (originally $\delta_S^{S'}\delta_j^{S-\R}\delta_k^{S-k'}$) is 
\be 
\label{eq:AII-1}
\cA_\R(G_m^{\II})= p^{2\R}\Bigl(p + \frac {\R+1}{N}\Bigr).
\ee
The contribution for the other seven propagators is
\be 
\label{eq:AII-2}
\frac 1 {N}\Bigl[\frac{ p^{2\R}}{1 + p} 
\bigl(\delta_{\R [2]}^0 +p\delta_{\R[2]}^1 \bigr) +  2p^{2\R}\Bigr]  + \frac 1 {N^2}2 p^{3\R}(\R+2).
\ee
As in the type-$\I$ case above, the sum of the contributions when crossing the upper edges or when exchanging the role of the tree and the anti-tree is the same. In the latter case, the dominant term is still obtained for the propagator $\delta_S^{S'}\delta_j^{S-\R}\delta_k^{S-k'}$. 
When crossing the upper edges, the dominant term is obtained for the second propagator $\delta_S^{S'}\delta_j^{S-\R}\delta_k^{k'}$.
Again, in both cases, the dominant term is obtained for the trivial propagator $\tilde\delta_{w'}(\cI_U)=1$.
In total,  the sum of the amplitudes of the 32 graphs of type~$\II$, denoted $\cA_r^{\II}$, is four times  \eqref{eq:AII-1} plus four times  \eqref{eq:AII-2}, and the contribution of these graphs to $\langle G_2(\R , t) \rangle_{\alpha, C}$ is $\frac{t^2}{16}\cA_r^{\II}$.

\begin{figure}[ht!]
	\centering
	\includegraphics[scale=1]{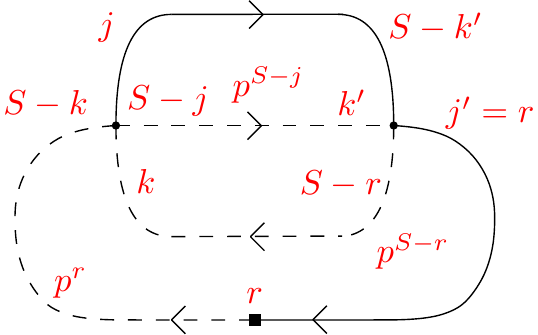} 
	\caption{Type-$\III$ diagram.}
	\label{fig:Dipole3}
\end{figure}

\noindent {\bf Graphs of type $\III$.} The third kind of graph is shown in Fig.~\ref{fig:Dipole3}. Again, we only draw one of them, however there are now eight diagrams which all give the same contribution due to the symmetries of $C$, and which we obtain  by exchanging the role of the tree and the anti-tree, by crossing the two upper edges as on the right of Fig.~\ref{fig:Dipole1}, or by choosing which one of the two upper edges is solid and which one is dashed. Each one of these 8 diagrams gives 8 graphs, thus a total of 64 graphs (again, the tree has a unique heap-ordering).

Importantly, here the total contribution from these 64 graphs to $\langle G_2(\R , t) \rangle_{\alpha, C}$  comes with a \emph{minus sign} (\textit{i.e.} $\epsilon(U)=-1$), because the two true vertices have parent-edges with the same orientation: both in-going or both out-going.
The amplitude of any one of the $8$ graphs $G=(U, w, w')$ for the diagram of Fig.~\ref{fig:Dipole3} is 
\be 
\cA_\R(G)= \frac 1 {N^2} \sum_{S\ge \R}\sum_{k, j, k'=0}^{S} \tilde\delta_{w'}(\cI_U)  p^{2S-j} \delta_{j}^{S-k'}\delta_{k}^{S-\R},
\ee 
where $\tilde\delta_{w'}(\cI_U)$ is one of the 8
propagators in  \eqref{eq: CovarianceNew1}:
$\delta_j^{\R}\delta_k^{k'}$,  $\delta_j^{S-\R}\delta_k^{k'}$,  $\delta_{j}^{\R}\delta_k^{S-k'}$, $\delta_j^{S-\R}\delta_k^{S-k'}$, $\delta_j^{k'}\delta_{k}^{\R}$, $\delta_k^{\R}$ , $\delta_j^{k'}$,  or $1$. 
The contribution from the eighth (trivial) propagator is 
\be 
\label{eq:AIII-1}
\cA_\R(G_m^{\III})= p^\R - \frac {p^{2\R+1}}{1+p}.
\ee
The contribution for the other seven propagators is
\be 
\label{eq:AIII-2}
\frac 1 {N}\Bigl[ \frac{ p^{3\R/2}}{1+p+p^2} 
\bigl(\delta_{\R[2]}^0 +p^{3/2}\delta_{\R[2]}^1 \bigr) +  2p^{2\R}-p^{4\R+1}+ \frac{p^\R }{1+p}\Bigr]  + \frac 1 {N^2}3 p^{3\R} .
\ee
As in the two cases above, we find that a single one of the 8 possibilities for the wavy line provides a dominant contribution. It is obtained for the eighth propagator $\delta_S^{S'}\delta_j^{S-k'}\delta_k^{S-\R}$
when the upper edges are not crossed, and for the fifth propagator $\delta_S^{S'}\delta_j^{k'}\delta_k^{\R}$
when the upper edges are crossed (note that because of the convention that $j, S-j, k, S-k$ are attributed clockwise starting from the parent-edge for the anti-tree, we have to modify the indices when crossing the two upper edges). 

In total,  the sum of the amplitudes of the 64 graphs of type~$\III$, denoted $\cA_r^{\III}$, is eight times  \eqref{eq:AII-1} plus eight times \eqref{eq:AII-2}, and the contribution of these graphs to $\langle G_2(\R , t) \rangle_{\alpha, C}$ is $-\frac{t^2}{16}\cA_r^{\III}$.

\

In the following, we will call \emph{leading propagator} $\delta_\text{lead}$ the particular choice of propagator for the wavy edge of a graph of type $\I$, $\II$ or $\III$, which leads to a dominant contribution. For each one of the diagrams presented in Fig.~\ref{fig:Dipole1}, \ref{fig:Dipole2}, and \ref{fig:Dipole3}, and similar diagrams, we showed that there is a unique leading propagator. This unique leading propagator always corresponds to the trivial propagator $\tilde \delta_{w'}(\cI_U)= 1$, i.e.~to the propagator which does not add additional constraints to the constraints imposed by the edges. This is quite intuitive, since constraints lower the number of free sums thus also lowering the number of potential $N$ factors. 

\

\begin{remark}
\label{rmk:Change-variable-El-melons}
Note that the computations above have been done in accordance with the conventions adopted earlier in the paper, regarding the assignment of $j_v, S_v - j_v, k_v, S_v-k_v$. These conventions have been adopted in order to have well-defined combinatorial objects, and render the counting transparent. Note that this counting is essential to  compute exactly the Sobolev norms. However, we would like to emphasize that in practice, it would have been simpler to change variables locally for each diagram above, to have matching indices on every edge: $j$ and $j'$ on the edges incident to the root, and respectively $S-j$ and $S-j'$, $k$ and $k'$, and $S-k$ and $S-k'$ for each one of the remaining edges. This way,  the constraints imposed by the edges are the same for all of the graphs above, and the only difference between two graphs is the exponent of $p$. In particular, the leading propagator is always $\delta_S^{S'}\delta_j^{j'}\delta_k^{k'}$. Indeed, these constraints are already imposed by the edges, and thus this propagator is the only one which does not impose further constraints. This remark will be useful in Section~\ref{melodomi}.\end{remark}

\begin{figure}[!ht]
	\centering
	\includegraphics[scale=1]{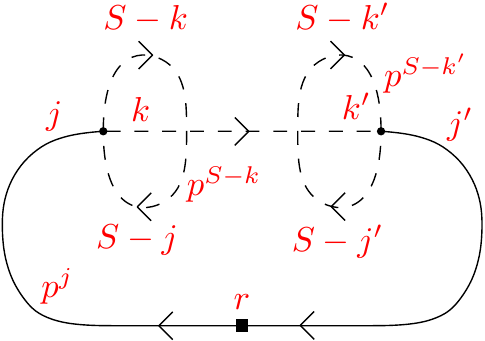} \hspace{1cm }\includegraphics[scale=1]{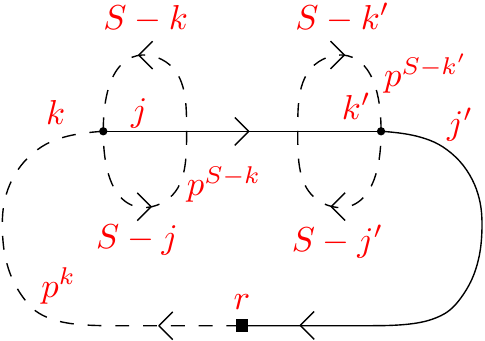} 
	\caption{Remaining diagrams.}
	\label{fig:Remaining-graphs}
\end{figure}

\noindent {\bf Remaining graphs.} The remaining graphs split into two categories. To the diagram on the left of Fig.~\ref{fig:Remaining-graphs} correspond $2\times 4 \times 8$ graphs: there are two heap-orderings, eight propagators, and the graphs obtained by exchanging the $k$ and $S-k$ edges or the $k'$ and $S-k'$ edges give the same contribution. Using the symmetries of $C$, we find that the total contribution to $\langle G_2(\R , t) \rangle_{\alpha, C}$ of these 64 graphs is
\be 
2 \times 4 \times \frac {t^2} {2 N^2} \sum_{S\ge \R} \sum_{j,k,j',k'=0}^S \langle C_{jk}^S C_{j'k'}^{S}\rangle  \delta_j^k\delta_{j'}^{k'}\delta_{k}^{k'}\delta_{j}^{\R}\delta_{j'}^{\R} p^{2S-\R}.
\ee
On the other hand, there are also 64 graphs corresponding to the diagram on the right of Fig.~\ref{fig:Remaining-graphs}. There is a single heap-ordering,  eight propagators, the graphs obtained by exchanging the $k$ and $S-k$ edges or the $k'$ and $S-k'$ edges, or exchanging the role of the tree and the anti-tree give the same contribution. Using the symmetries of $C$, we find that the total contribution to $\langle G_2(\R , t) \rangle_{\alpha, C}$ of these 64 graphs is
\be 
 - 2 \times 4 \times \frac {t^2} {2 N^2} \sum_{S\ge \R} \sum_{j,k,j',k'=0}^S \langle C_{jk}^S C_{j'k'}^{S}\rangle  \delta_j^k\delta_{j'}^{k'}\delta_{k}^{k'}\delta_{j}^{\R}\delta_{j'}^{\R} p^{2S-\R},
\ee
where the minus sign comes from the fact that the parent-edges at the two true vertices are both in-going or both out-going, thus giving $\epsilon(U)=-1$.
In summary, the total contribution of the remaining graphs to  $\langle G_2(\R , t) \rangle_{\alpha, C}$ vanishes, due to the symmetries of $C$.

\subsection{Sobolev norms at order 2}\label{sec:Sobolev-order2}

Let us first compute $\langle G_2(\R , t) \rangle_{\alpha, C}$ at order two, and then the Sobolev norms at order 2.

\

\noindent {\bf Total contribution of order 2 graphs.} The total contribution to $\langle G_2(\R , t) \rangle_{\alpha, C} = \frac {t^2} {16}(\cA_\R^{\I} + \cA_\R^{\II} -  \cA_\R^{\III})+O(t^4)$ at order 2 is given by 
\be 
\label{eq:melons_firstpart}
\langle G_2(\R , t) \rangle_{\alpha, C}^{(2)} = \frac{t^2 p^\R}{4}\biggl[ p^{\R+1} \frac {p+3}{p+1} + \frac {p^2}{(p+1)^2 }-2\biggr] + \frac{(\R+1)t^2p^\R}{4N} \Bigl[\frac 1{1+p}  + p^\R \Bigr] + R_\R(t),
\ee 
where the dominant contribution is always obtained when the leading propagator $\delta_\text{lead}$ is chosen for the wavy line, and where $R_\R(t)$ gathers the contributions of all the other propagators for $\cA_\R^{\I}$,  $\cA_\R^{\II}$, and  $\cA_\R^{\III}$, and the (vanishing) contribution of the tadpole graphs:
\begin{align} 
\label{eq:rest}
 R_\R(t) = \frac{t^2}{4N}\biggl[\frac {p^{4\Ceil{\frac \R 2} - \R}}{(1+p)(1+p^2)} + \frac{p^{\Ceil{\frac \R 2} + \R}}{1+p}- \frac{2p^{3\Ceil{\frac \R 2}}}{1+p+p^2}
+ 2(p^{4\R+1} - p^{2\R})\biggr]+ \frac{t^2(2\R+1)p^{3\R}}{2N^2}  ,
\end{align} 
where $\Ceil{\frac \R 2}$ is the ceiling of $\Ceil{\frac \R 2}$, which is $\R/2$ if $\R$ is even, and $(\R+1)/2$ if $\R$ is odd.

\

We immediately see that only the terms given by the leading propagator for $\cA_\R^{\I}$,  $\cA_\R^{\II}$, and  $\cA_\R^{\III}$ give dominant contributions (because when summed over $\R$,  a typical term of the form $\sum \R^{\gamma} p^\R$ behaves as $(1-p)^{-1-\gamma}$ as $p\rightarrow 1$), while the terms in $R_\R(t)$ are all sub-dominant. We make more precise statements in the following paragraph.

\

\noindent {\bf Sobolev norms at order 2.} We are interested in the averaged Sobolev norms at order 2,
\be 
\bar S_\gamma^{(2)}(t) := \langle S_\gamma(t)\rangle_{\alpha, C}^{(2)} = \sum_{\R\ge 0} \frac{\R^\gamma}{N}\, \langle G_2(\R , t)\rangle_{\alpha, C}^{(2)}  .
\ee 
For  $\gamma=0$, one verifies that both $\bar R(0,t)=\sum_{\R\ge 0}R_\R(t)=0$ and $\bar S_0^{(2)}(t) - \bar R(0,t)=0$, so that $\bar S_0^{(2)}(t) =0$, as expected. The same happens for $\gamma=1$. 

In general, for $\gamma>0$, we express the various  terms involved in $\bar S_\gamma^{(2)}(t)$ using the series $\Li_{\gamma}(z) =\sum_{\R\ge 1} \R^{\gamma} z^\R $ (they are polylogarithm functions). We have for $\gamma>0$
\begin{align} 
\label{eq:melons_firstpart-sobolev}
\bar S_\gamma^{(2)}(t) &= \frac{t^2}{4N}\biggl[ p \frac {p+3}{p+1}\Li_{\gamma}(p^2) + \Bigl(\frac {p^2}{(p+1)^2 }-2\Bigr)\Li_{\gamma}(p)\biggr] \\ \nonumber  &\qquad + \frac{t^2}{4N^2} \Bigl[\frac {\Li_{\gamma+1}(p) + \Li_{\gamma}(p)}{1+p}  + \Li_{\gamma+1}(p^2) + \Li_{\gamma}(p^2) \Bigr] + \bar R(\gamma,t),
\end{align} 
where making use of the fact that 
$$ 
\sum_{\R\ge 0} (2\R)^{\gamma}z^{2\R} = 2^\gamma \Li_{\gamma}(z^2),\text{ and } \sum_{\R\ge 0} (2\R+1)^{\gamma}z^{2\R+1} = \Li_{\gamma}(z)- 2^\gamma \Li_{\gamma}(z^2),
$$
the residual term is expressed as
\begin{align} 
\label{eq:rest-sobolev}
 \bar R(\gamma,t) &= \frac{t^2}{4N^2}\biggl[\frac {p^2\Li_{\gamma}(p)}{(1+p)(1+p^2)} + \frac p{1+p}\Li_{\gamma}(p^2)- \frac{2p^{3/2}\Li_{\gamma}(p^{3/2})}{1+p + p^2} + 2\bigl(p\Li_{\gamma}(p^4) - \Li_{\gamma}(p^2)\bigr)\biggr] \nonumber\\ 
  &\hspace{-8mm}
 + \frac{t^2}{4N^3}\biggl[ \frac {2^\gamma}{1+p^2}\Li_{\gamma}(p^2) + \frac{2^\gamma}{1+p}\Li_{\gamma}(p^4)- \frac{2^{\gamma+1}}{1+p^{3/2}}\Li_{\gamma}(p^3)+ 4\Li_{\gamma+1}(p^3) + 2\Li_{\gamma}(p^3) \biggr] 
\end{align}

\

\noindent {\bf Asymptotic behavior of the order 2 Sobolev norms.} 
Let us take a closer look at the behavior of $\Li_\gamma$ near 1, when $\gamma$ is a positive integer. In that case, 
\be 
\Li_{\gamma}(z)= \frac 1 {(1-z)^{\gamma + 1}}\sum_{k=0}^{\gamma-1}A(\gamma, k) z^{\gamma - k},
\ee
where the $A(\gamma, k)$ are the Eulerian numbers, which satisfy the identity 
\be 
\sum_{k=0}^{\gamma-1}A(\gamma, k)= \gamma!,
\ee
so that when approaching 1,
\be 
\Li_{\gamma}(z)= \frac {\gamma!} {(1-z)^{\gamma + 1}} + o\Bigl(\frac 1 {(1-z)^{\gamma + 1}}\Bigr).
\ee 
We find that when $N$ goes to infinity ($p$ goes to 1),
\be 
\bar  R(\gamma,t) = o\bigl(N^{\gamma}\bigr)
\ee 
and 
\begin{align} 
\label{eq:Dominant-sobolev-order2}
\bar S_\gamma^{(2)}(t) &= \frac{t^2N^{\gamma} \gamma! }{4}\biggl[ \frac 1 {2^\gamma} - \frac 7 4 + \frac{\gamma+1}2\bigl(1+\frac 1 {2^{\gamma+1}}\bigr) \biggr] +  o\bigl(N^{\gamma}\bigr),
\end{align} 
which we can rewrite as
\begin{align} 
\label{eq:Order2-Sobolev-asympt}
\bar S_\gamma^{(2)}(t) &= \frac{t^2 N^{\gamma}\gamma! }{16}\biggl[ \frac {5+\gamma} {2^{\gamma}} + 2\gamma-5 \biggr] +  o\bigl(N^{\gamma}\bigr),
\end{align} 
In particular, we see that since $\frac {5+\gamma} {2^{\gamma}} + 2\gamma-5 $ does not vanish for $\gamma>1$, the contribution of the rest term $ \bar  R(\gamma,t)$ is sub-dominant: for $\gamma>1$, the only dominant contributions are obtained for the leading propagators. Furthermore, very importantly, \emph{the order 2 averaged Sobolev norms are positive when $p$ is close to 1}.

\section{Melonic dominance} \label{melodomi}

In this section, we define a class of graphs called melonic graphs (Section~\ref{sub:Melo}), and show that the averaged Sobolev norms admit a $1/N$ expansion (Section~\ref{sub:1N-exp}), whose first leading term is given exactly by the restriction of the averaged Sobolev norms to the melonic graphs (Section~\ref{sub:Mel-Dom}). We then show in Section~\ref{sub:Exp-bounds} that this melonic approximation of the averaged Sobolev norms is analytic in a finite disc around 0. In Section~\ref{sub:stranded}, we introduce the stranded representation for the graphs involved, which is then used in the proofs found in the next sections.

We recall that the graphs are denoted $G=(U, w, w')$ where $U$ is a tree in $\cT_{2,3}^n$ with $n$ true vertices and one bivalent root. It has dashed edges (which result from the averaging over $\alpha$ that pairs the leaves and anti-leaves of $U$), solid edges (the edges of $U$ that are not incident to leaves), and wavy edges between true vertices, which result from the averaging over $C$ and carry a propagator (one of the eight products of deltas in the covariance \eqref{eq: CovarianceNew1}). 

In this section, we bound the graph amplitudes \eqref{fullexpans}. As these amplitudes do not depend on the heap-ordering of the tree $U$, we forget this heap-ordering. We stress however that it is essential to include these combinatorial factors when summing the graph amplitudes, as was done in the previous section at order 2.

\subsection{Melonic graphs}
\label{sub:Melo}

Among the four categories of order-2 graphs described in the previous section, only three give dominant contributions, the graphs of type $\I$, $\II$ and $\III$. The dominant terms are obtained for the leading propagators, as detailed above. Forgetting the heap-orderings, there are respectively 2, 4, and 8 dominant graphs of type $\I$, $\II$ and $\III$. We call these 14 graphs elementary melons of type $\I$, $\II$ and $\III$.

\

\noindent {\bf Melonic moves. } If we remove the bivalent root in one of these elementary melons, we obtain a graph with two pending half-edges, such as on the right of Fig.~\ref{fig:Move1}, \ref{fig:Move2} or \ref{fig:Move3}. We call such graphs elementary two-point melons of type $\I$, $\II$, and $\III$. To construct the melonic graphs of higher orders, we define the following operations on the dashed edges of the graphs. 
\begin{itemize}
\item The melonic insertion of type $\I$ consists in replacing a dashed edge of $\cE_\alpha$ by one of the 2 elementary two-point melons of type $\I$.

\begin{figure}[!h]
	\centering
	\includegraphics[scale=1.4]{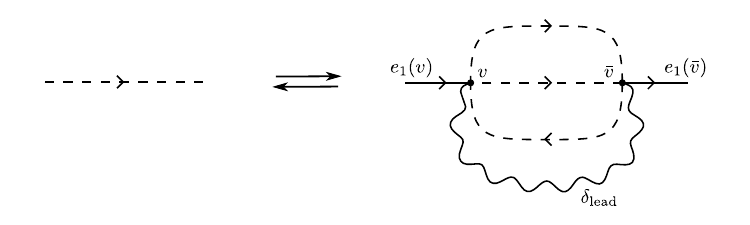}
	\vspace{-0.3cm}
	\caption{Melonic move of type $\I$}
	\label{fig:Move1}
\end{figure}
\item The melonic insertion of type $\II$ consists in replacing a dashed edge by one of the 4 elementary two-point melons of type $\II$.
\begin{figure}[!h]
	\centering
	\includegraphics[scale=1.4]{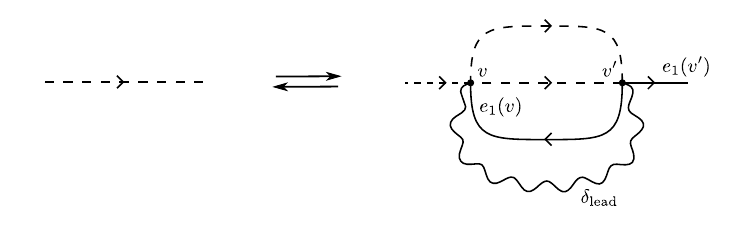}
		\vspace{-0.3cm}
	\caption{Melonic move of type $\II$}
	\label{fig:Move2}
\end{figure}
\item The melonic insertion of type $\III$ consists in replacing a dashed edge by one of the 8 elementary two-point melons of type $\III$.
\begin{figure}[!h]
	\centering
	\includegraphics[scale=1.4]{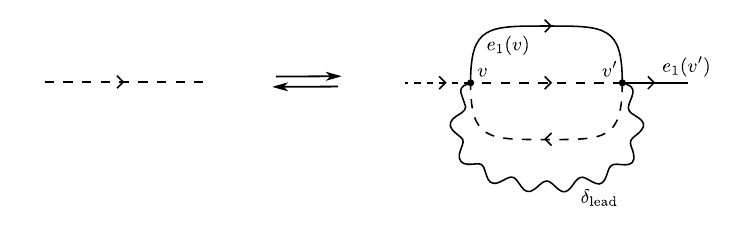}
		\vspace{-0.3cm}
	\caption{Melonic move of type $\III$}
	\label{fig:Move3}
\end{figure}
\end{itemize}
This is done so that the orientation remains coherent\footnote{Note that in the figures, with the present convention, the ordering of the half-edges around the vertices might need to be inverted, depending on whether the vertices belong to the tree or anti-tree part of $U$.}. Importantly, a melonic insertion imposes that we chose the leading propagator $\delta_{\text{lead}}$ for the wavy edge that links the two new vertices.  The inverse operations are called melonic reductions of type $\I$, $\II$ and $\III$.

\

To define similar operations on the solid edges, we also need to introduce the elementary two-point melons of type $\II s$, and $\III s$: they are simply the elementary two-point melons of type $\II$ and $\III$, for which the dashed pending half-edges have been changed for solid half-edges. There are now respectively 2 and 4 elementary two-point melons of type $\II s$, and $\III s$. We will need the following moves:
\begin{itemize}
\item The melonic insertion of type $\II s$ consists in replacing a solid edge of $\cE_s$ by one of the 2 elementary two-point melons of type $\II s$.
\begin{figure}[!h]
	\centering
	\includegraphics[scale=1.4]{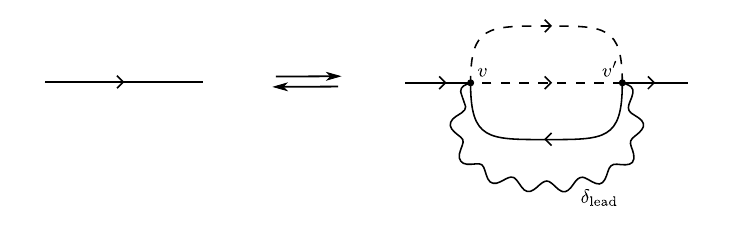}
		\vspace{-0.3cm}
	\caption{Melonic move of type $\II s$}
	\label{fig:Move4}
\end{figure}
\item The melonic insertion of type $\III s$ consists in replacing a solid edge  by one of the 4 elementary two-point melons of type $\III s$.
\begin{figure}[!h]
	\centering
	\includegraphics[scale=1.4]{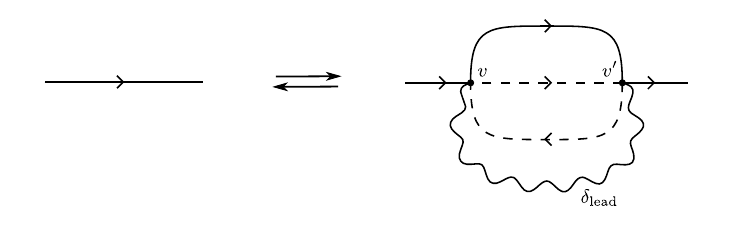}
		\vspace{-0.3cm}
	\caption{Melonic move of type $\III s$}
	\label{fig:Move5}
\end{figure}
\end{itemize}

\noindent {\bf Melonic graphs. }The trivial tree in $\cT_{2,3}^n$ has one bivalent root and one leaf and one anti-leaf. It is obtained for the terms of order 0 in the Taylor expansion of $\alpha(t)$ and $\bar \alpha(t)$. After the averaging, the corresponding graph has a single dashed edges and only one vertex: its bivalent root.

\begin{definition}
We say that $G= (U, w, w')$ is a melonic graph if it can be reduced to the trivial $n=0$ graph 
by a sequence of melonic reductions of any of the five types defined above (Figs.~\ref{fig:Move1}, \ref{fig:Move2}, \ref{fig:Move3}, \ref{fig:Move4}, \ref{fig:Move5}). 
\end{definition}

\begin{figure}[!ht]
	\centering
	\includegraphics[scale=0.8]{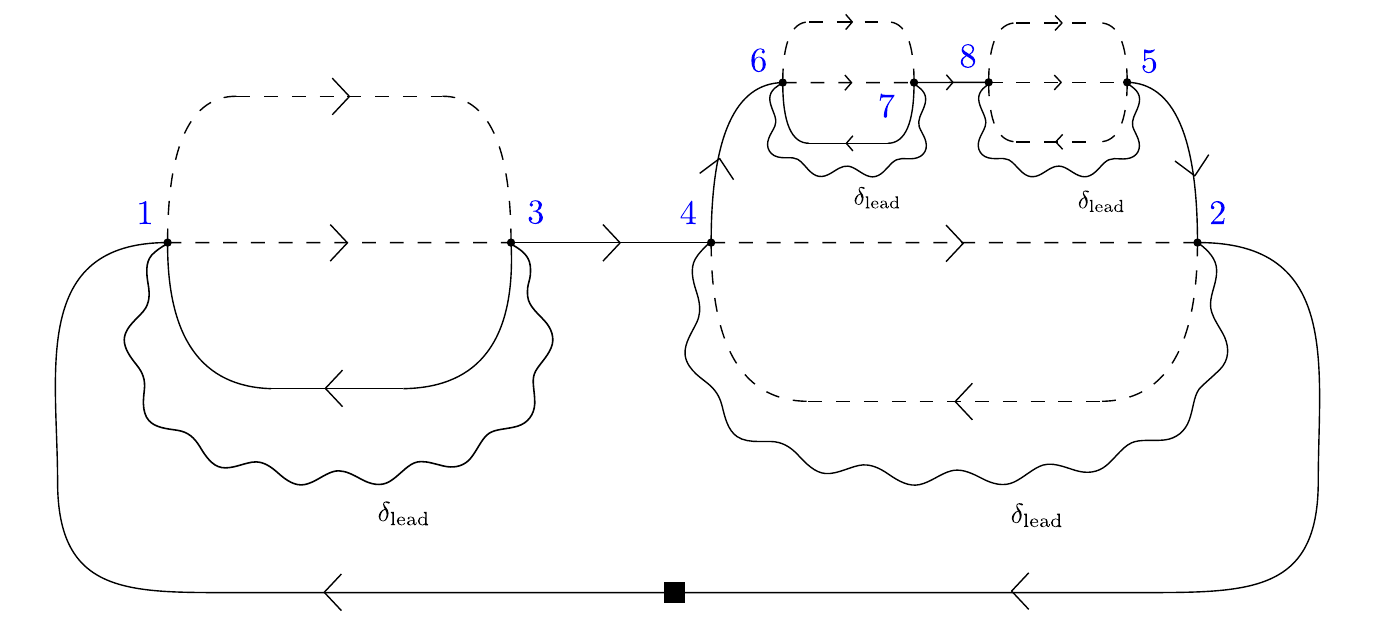}
	\caption{A melonic graph with eight true vertices.}
	\label{fig:Example-melonic}
\end{figure}

An example of melonic graph with eight true vertices is shown in Fig.~\ref{fig:Example-melonic}.

Of course, melonic graphs are quite special and most graphs are not melonic. Note that when we forget the arrows, the root, the wavy edges and the distinction between dashed and solid edges, then these melonic graphs become melons in the ordinary sense of rank-3 random tensor theory \cite{Bonzom:2011zz,Gurauetal}.
Note also that the last reduction in this sequence cannot be of type $\II s$ or $\III s$ since the trivial graph only has a single dashed edge.

\subsection{Stranded representation}
\label{sub:stranded}

We are interested in proving the existence of a $1/N$ expansion for our model and identifying the dominant family of graphs at each order $n$ as the melonic graphs. The orientation and the presence of both dashed and solid edges with different associated factors make the analysis a bit tedious.
To simplify it, we find it convenient to introduce still another representation, called \emph{stranded}.
The power of $N$ of any amplitude will then be related to the number of independent closed strand loops, in analogy with standard $1/N$ limits or power counting theorems
in quantum field theory. 

\begin{figure}[h]
	\centering
	\vspace{-0.8cm}
	\includegraphics[scale=1.8]{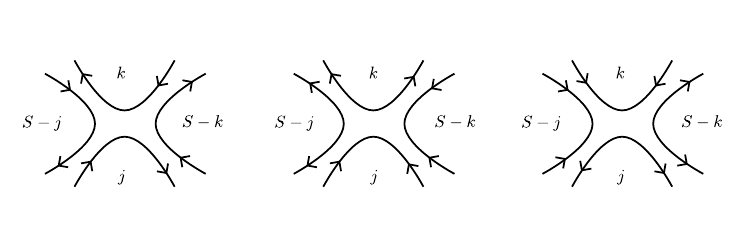}
		\vspace{-0.8cm}
	\caption{The three types of 8-valent stranded nodes.}
	\label{fig:8ValentVertices}
\end{figure}

Consider a wavy edge and its associated product of $\delta$'s in the global factor $\delta_{w'} (\cI_U)$ in \eqref{fullexpans}.
Deleting the wavy edge and the incident four-valent vertices, then identifying the half-edges according to the $\delta$'s of the wavy edge,  we obtain a representation of the pair of vertices as an 8-valent stranded node made of these two 4-valent vertices. 
There are a priori eight possible types for such stranded nodes because the $C$-covariance has eight terms. But first of all, note that there is a single integer $S$ associated to each 8-valent node.
Then there remain two $\delta$'s, which correspond to the identification of the $j $ and $k$ indices of one 4-valent vertex with the $j$ or $S-j$ and $k$ or $S-k$ of the other one.
In the end, it means that each index $j$, $k$, $S-j$, $S-k$ must occur exactly for two strands out of the eight strands attached to the 8-valent node. Hence, it is natural to pair the strands
into four matching pairs, and the 8-valent node becomes similar to a vector-model 8-valent node with four corners\footnote{By corner, we refer to the arc which links two paired strands inside an 8-valent node.}. However, there is a subtlety:
orientations may not agree. A moment of contemplation leads to the conclusion that we 
can obtain only three kinds of vector-like 8-valent nodes, 
which are represented in Fig.\ \ref{fig:8ValentVertices}. 

In the figure, the half-edges are represented as solid although some of  the full edges to which they belong may in fact be dashed.
Note that the half-edges around the 8-valent nodes are still ordered: one can distinguish which half-edge comes from which true vertex and the half-edges around each true vertex are ordered. For instance, two half-edges that share a corner may not have the same nature, dashed or solid, so they are not exchangeable. This is important when computing exact combinatorial weights, however this ordering is not so important when computing bounds for the graph amplitudes. 

We call $\tilde G$ the graph associated to $G$ in the stranded representation. It has a set of 8-valent nodes $\tilde \cV$ with $\vert \tilde \cV \vert = n/2$, since each node in $\tilde \cV$
is made of a pair of true vertices of the initial graph $G$. 

\begin{figure}[h]
	\centering
	\includegraphics[scale=1.3]{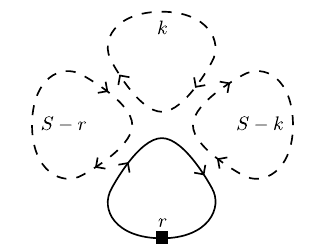}
	\hspace{0.5cm}
	\includegraphics[scale=1.3]{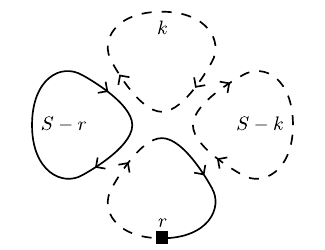}
	\hspace{0.5cm}
	\includegraphics[scale=1.3]{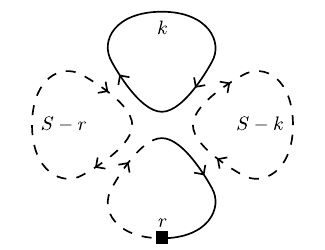}
	\caption{Elementary melons of type $\I$, $\II$ and $\III$ in the stranded representation.}
	\label{fig:ElementaryMelons}
\end{figure}

In Fig.~\ref{fig:ElementaryMelons}, we show three examples of stranded graphs, one for each kind of elementary melon. In this representation, we see explicitly that these graphs all have four closed loops (called faces, see below). In fact, the elementary melons are the only graphs with $n=2$ that have four faces, and each of these faces are of length one.

\subsection{Existence of the $1/N$ expansion
}
\label{sub:1N-exp}

In this section, we prove the following result, which is a consequence of Prop.~\ref{prop:globalbound} and Lemmas \ref{lemmamelo2}-\ref{lemmamelo3}, which are proven below.

\begin{proposition}
\label{keylemma1}
For any graph $G$ with $n$ true vertices, the scaling behavior in $N$ of the amplitudes $\cA_\R(G)$ at fixed order $n$ and as $N\rightarrow \infty$ is bounded from above by $N^{-d(G)}$ with $d(G)\ge 0$ a non-negative integer, which will be defined in this section.
\end{proposition}

The above result ensures that the averaged Sobolev norms admit\footnote{In the sense that  we can classify the graphs by grouping those whose amplitudes have the same dependence in $N$, and because the dependence in $N$ is bounded from above, this classification is in non-negative powers of $1/N$, up to a global rescaling. Note that to have a formal expansion in $1/N$, one should also show that the sums $\bar S_{\gamma, \omega}(t)$ of the amplitudes for the graphs of any given group (i.e.~graphs for which the dependence in $N$ is the same) is finite in a certain interval of time.} \emph{$1/N$ expansions} of the form 
\be 
\label{eq:1Nexp-def}
\bar S_\gamma(t) = N^\gamma \sum_{\omega \ge 0 } \frac 1 {N^\omega} \bar S_{\gamma,\omega}(t),
\ee 
where $\omega$ is non-negative and takes discrete values (that is, it takes values in a set in bijection with $\bN$) and $\bar S_{\gamma,\omega}(t)$ is the sub-series corresponding to the graphs whose amplitudes behave as $N^{-\omega}$.\footnote{For a graph $G$, we thus have $\omega(G) \ge d(G)$.} In the above equation, the dependence in $N$ is made explicit (i.e.~we expect $\bar S_{\gamma,\omega}(t)$ to scale as $N^0$). The dominant term when $N\rightarrow \infty$ ($p\rightarrow 1$) is then obtained by restraining the series to the graphs for which $\omega=0$. 
In the following section, we show that these graphs are precisely the melonic graphs. Note that the crude bound of Eq.~\eqref{crudebound} (Sec.~\ref{sec:Averaged-Sobolev}) does not allow one to define such an expansion, because it does not rule out the existence of an infinite family of graphs with unbounded behavior in $N$. 

\

\noindent{\bf Faces of a graph. }The graphs in the stranded representation are collections of closed loops, which meet around 8-valent nodes. These closed loops are called \emph{faces}. The \emph{length} of a face is defined as the number of corners of $8$-valent nodes that the face visits (the root vertex does not contribute to the length of the faces).

Our next lemma bounds the number of faces of any graph $\tilde G$ associated to some $G= (U,w,w')$. In the coming part of the text, we denote by $V=\lvert\tilde{\cV}\rvert=n/2$ the number of $8$-valent nodes and by $F$ the number of faces in $\tilde G$. We also denote by $F_l$ the number of faces of length $l$.
\begin{lemma} \label{lem:boundfaces}
The number $F$ of faces of any graph $\tilde G$ is bounded from above by $\frac{3n}{2}+1$.
\end{lemma}
\proof In $\tilde G$, we have
\begin{equation}
F= \sum_{l\geq 1} F_l , \quad 2n = \sum_{l\geq 1} l F_l \quad
\Rightarrow \quad F = 2n - \sum_{l\ge 2} (l-1) F_l \, .
\label{faceleng}
\end{equation}
Now, we use the fact that the graph $\tilde G$ is connected. Therefore, starting with the $n/2$ isolated 8-valent nodes of $\tilde \cV$, when we add the edges, we make up faces and obtain in the end a single connected component. A face of length $l$ connects at most $l$ 8-valent nodes into a single connected component, hence it decreases the number of connected components by at most $l-1$. Therefore 
\be
\label{facesat}
n/2 - \sum_{l\ge 2} (l-1)F_l \le 1,
\ee
where the $1$ on the right-hand side stands for the number of connected components of $\tilde G$.
\qed \\

\begin{remark}
Another way of proving the lemma is as follows. The graph $\tilde G$ is connected. If we break down the 8-valent nodes into four corners, the graph $\tilde G$ becomes a graph with 
exactly $F$ disconnected components. Then joining four corners in an 8-valent node can connect at most four faces hence decrease the number of connected components by at most 3. After $n/2$
such moves we have a single connected component, hence $F-3 n/2 \leq 1$.
\end{remark}

\

\noindent{\bf Amplitudes in terms of the face momenta. }
We call $\cC$ the set of the  $2n$ corners of $\tilde G$, and $i_c$ the momentum at corner $c$, so that $i_c \in \{j_v, S_v -j_v, k_v ,S_v - k_v\}$ if $c$ is a corner of $v$. The edges (dashed or solid) identify the corner momenta at their extremities. We see that all the corner momenta encountered along a face are ultimately identified.  For a given face $f$, we will call $i_f\in {\mathbb N}$ the (non-negative) face momentum common to all the corners of $f$. Intuitively, if for every face we sum over all the corner momenta in the face but one, we should reduce the graph amplitudes to sums over face momenta. 
At each vertex $v$, we respectively denote $i^{(1)}_{f,v}$,  $i^{(2)}_{f,v}$, $i^{(3)}_{f,v}$ and  $i^{(4)}_{f,v}$ the face momenta of the faces which respectively pass by the corners with momenta $j_v$, $S_v - j_v$, $k_v$, and $S_v-k_v$. Note that they might not be distinct as a face might visit several corners around the same vertex. We denote by $L_f^\alpha$ the number of dashed edges visited by the face $f$, and by $\fr$ the face that visits the root. The resonance constraints at every node are now expressed as $i^{(1)}_{f,v} + i^{(2)}_{f,v} = i^{(3)}_{f,v} + i^{(4)}_{f,v}$. 
We therefore obtain the following expression Lemma.

\begin{lemma} 
\label{lemma:ChangeVariable1}
The amplitudes of the graphs can be expressed in terms of the face momenta, 
\be 
\label{eq:Expression-Amp-Faces}
\cA_\R(G)=\frac {p^{\R L^\alpha _ {\fr}}} {N^n} \Bigl(\prod_{f\neq \fr} \sum_{i_f\ge 0}p^{i_f L^\alpha_f}\Bigr)\prod_{v\in\tilde \cV} \delta_{i^{(1)}_{f,v} + i^{(2)}_{f,v} \;,\;i^{(3)}_{f,v} +  i^{(4)}_{f,v} }.
\ee
\end{lemma}
This result is quite intuitive in the initial variable $j, j', k , k'$, but because the nodes constraints should be handled carefully, we provide a detailed proof in Appendix~\ref{Appendix:ChangeVariable1}.

\

At each vertex $v\in \tilde \cV$ of $\tilde G$ visited by $f$, the face-momentum $i_f$  must be among the four numbers $j_v, S_v-j_v ,k_v, S_v - k_v$. 
Among the $F$ faces, a unique one, say $\fr$, visits the root vertex, hence has momentum fixed to $i_{\fr} =\R$.
However the $F-1$ face momenta for $f \ne \fr$ are still not \emph{independent}. Indeed, the 
momentum conservation rule $j_v + (S_v -j_v) = k_v + (S_v - k_v)$ at each vertex of $\tilde G$ 
has to be taken into account, since it can lead some face momenta to be expressed in terms of other face momenta. To find out the true set of independent face momenta we introduce some incidence matrices.
We recall that $\cC$ is the set of the  $2n$ corners of $\tilde G$, and $i_c$ the momentum at corner $c$, 
so that  $i_c \in \{j_v, S_v -j_v, k_v ,S_v - k_v\}$ if $c$ is a corner of $v$. For each one
of the two corners of $v$ with momenta $j_v, S_v - j_v$ we define $\zeta_{vc} = +1$ and for the other two we define $\zeta_{vc} = -1$; for the other corners not belonging to $v$ we put $\zeta_{vc}=0$. The conservation rule at each vertex $v$ can then be written in terms of the $i_c$ as
 \bee 
 \forall v, \quad \sum_{c\in \cC} \zeta_{vc}   i_c = 0  
\ee 
Now, to rewrite it in terms of the face momenta $i_f$, we introduce the matrix $\eta_{cf}$ which is $1$ if the face $f$ goes through the corner $c$ and 0 otherwise.
The linear system of the $n/2$ vertex momentum conservations is then represented as
\bee  
 \forall v, \quad \sum_{c, f} \zeta_{v c}  \eta_{cf} i_f = 0,
\ee 
or, more compactly, $E\cdot \vec i = 0$, where $E=  \zeta . \eta$ is a  $V\times F$ incidence matrix between vertices and faces with elements in $\{ -2, -1, 0, 1, 2\}$ and $\vec i$ is the vector of the face momenta $i_f$. 
Thus, the contraints in \eqref{eq:Expression-Amp-Faces} can be expressed as
\be 
\label{eq:new-constr-1}
\prod_{v\in\tilde \cV}\delta_{i^{(1)}_{f,v} + i^{(2)}_{f,v} \;,\;i^{(3)}_{f,v} +  i^{(4)}_{f,v} } = \delta\bigl(E\cdot \vec i \bigr).
\ee

Let us compute these matrices in a simple example such as the elementary melon on the left of Fig.~\ref{fig:ElementaryMelons}. In that case, there are four faces, and each one of them visits a single corner. We label the faces respectively $f_0, f_1, f_2, f_3$ 
corresponding to the face (and corner) momenta $r, S-r, k, S-k$. We have $\zeta=(1,1,-1,-1)$ and $\eta=\un$, so that $E=\zeta$. 

\

\noindent{\bf Amplitudes in terms of the independent face momenta. }
We call $R\le V=n/2$ the rank of this matrix $E$. 
We can select a subset $F_{(R)}$ of $R$ independent columns of $E$, and consider the $V\times R$ matrix $E_{(R)}$ obtained from $E$ by keeping only these $R$ columns, as well as the matrix $E_{(I)}$ of the remaining columns. Similarly, the vector $\vec i$ splits into two vectors $\vec i_{(R)}$ and $\vec i_{(I)}$, and the equation $E\cdot \vec i = 0$ can be rewritten as $E_{(R)}\cdot \vec i_{(R)} +  E_{(I)}\cdot \vec i_{(I)}= 0$. As the columns of $E_{(R)}$ are linearly independent, the rectangular matrix $E_{(R)}$ has a left inverse $E_{(R)}^+$, given by the Moore-Penrose inverse $E_{(R)}^+=(E_{(R)}^TE_{(R)})^{-1} E_{(R)}^T$, so that $\vec i_{(R)} =-  E_{(R)}^{+}E_{(I)}\cdot \vec i_{(I)}$. 

If $F=R$, we have $\vec i =-  E^{+}0=0$, 
so that this case does not occur as long as $\R>0$. 

If not, $F>R$, and we can always include $\fr$ in $F_{(I)}$ ($i_{\fr}$ is an element of $\vec i_{(I)}$). We define 
\be \label{eq:ItildeG}
I(\tilde G) = F-R -1, 
\ee
and call $F_{(R)} = \{\fR_1, \ldots, \fR_R\}$ and the $F-R$  remaining  face momenta (including the root face momentum $f_0$) $F_{(I)} = \{\fr, \fI_1, \ldots, \fI_I\}$. Writing $a_{j,k} := (-  E_{(R)}^{+}E_{(I)})_{j,k}$, we can express any face momentum $ i_{\fR_j}$ for $\fR_j \in F_{(R)}$ as a linear combination
$ i_{\fR_j} = \sum_{k=0}^{I(\tilde G)} a_{j,k} i_{\fI_k}$ of the elements of $ \vec i_{(I)}$.

Then the $V$ \emph{discrete} constraints $\delta( E\cdot \vec i)$ can  be replaced in the expression \eqref{eq:Expression-Amp-Faces} of $\cA_\R (G) $ by the smaller equivalent set of $R$ constraints 
$
 \prod_{j=1}^R \delta\bigl(i_{\fR_j}  - \sum_{k=0}^{I(\tilde G)}  a_{j,k} i_{\fI_k}\bigr).
$
The amplitude of a graph is therefore 
\be 
\label{eq:Expression-Amp-Faces-2}
\cA_\R(\tilde G)=\frac {p^{\R L^\alpha _ {\fr}}} {N^n}
\prod_{\substack{{\fI\in F_I}\\{f\neq \fr}}} \sum_{i_{\fI}\ge 0} p^{i_{\fI} L^\alpha_{\fI}} \prod_{\fR\in F_R}\left( \sum_{i_{\fR}\ge 0} p^{i_{\fR} L^\alpha_{\fR}}\prod_{j=1}^R \delta\Bigl(i_{\fR_{j}}  - \sum_{k=0}^{I(\tilde G)}  a_{j,k} i_{\fI_k}\Bigr)\right).
\ee
We can perform the sums over the face momenta $i_{\fR_j}$.
However we must be careful: the linear combinations $\sum_{k=0}^{I(\tilde G)}  a_{j,k} i_{\fR_k}$ are not necessarily non-negative, while  the face momenta $i_{\fR_j}$ run over $\bN$ and thus are constrained to be  non-negative. Therefore, we need to implement the condition that the $i_{\fR_j}\ge 0$ in the resulting summand. 
We write these conditions using Heaviside functions $\Theta(x)$ which vanish for all $x<0$. 
\begin{lemma}
The amplitudes of the graphs are expressed in terms of the independent face momenta $\fI_k \in F_{(I)}$ only:
\be 
\label{eq:Expression-Amp-Faces-Ind}
\cA_\R(\tilde G)=\frac {p^{\R \tilde L _ {\fr}}} {N^n}  \sum_{i_{\fI_1} , \ldots, i_{\fI_{I(\tilde G)}}\ge 0 
}\;\prod_{k = 1}^{I(\tilde G)}p^{i_{\fI_k} \tilde L_{k}}\; \prod_{j=1}^R\;\Theta\left(\sum_{l=0}^{I(\tilde G)}  a_{j,l} i_{\fI_l} \right) ,
\ee
 where $\tilde L_k = L^\alpha_{\fI_k} + \sum_{j=1}^R a_{j,k}L^\alpha_{\fR_j}$. 
 The Heaviside functions restrict the sums over the $i_{\fI_k} \in \bN$ to smaller summation intervals. 
 \end{lemma}

Let us comment on the importance of the positivity conditions, implemented by the Heaviside functions.  Consider the example on the left of Fig.~\ref{fig:ElementaryMelons}. As detailed previously in the present section, for this example, $R=1$. We choose the third face (corresponding to $k$) as the face in $F_{(R)}$ (so that $f_2 = \fR_1$, $f_1=\fI_1$, and $f_3=\fI_2$). We can rewrite $i_{f_2} = i_{f_0} + i_{f_1} - i_{f_3}$, which is the linear combination $\sum_{k=0}^{I(\tilde G)} a_{j,k} i_{\fI_k}$ for $j=1$. Eq.~\eqref{eq:Expression-Amp-Faces-Ind} translates as  
\be 
\cA_\R (\tilde G) = \frac 1 {N^2} \sum_{i_{f_1}, i_{f_3} \ge0} p^{i_{f_1}}p^{i_{f_3}} p^{i_{f_0} + i_{f_1} - i_{f_3}} \Theta(i_{f_0} + i_{f_1} - i_{f_3}).
\ee
Now if we suppress the $\Theta$ constraint, the expression diverges because of the sum over $i_{f_3}$. 

\

\noindent{\bf Existence of the $1/N$ expansion. }
Using the above lemma, we can now find an upper bound on the graph amplitudes that improves the one found in Section \ref{sec:Averaged-Sobolev} and then show the existence of a $1/N$ expansion for the model. 
\begin{proposition}\label{prop:globalbound}
The amplitude of any graph $G$ with $n$ true vertices is bounded from above as
\be
\cA_\R(G)\le (4n)^{3n/2-1}p^{\frac{\R }{4n}} N^{-d(G)},
\ee
where $d(G):=n-I(\tilde G)$ is called the degree of $G$.
\end{proposition}

\proof The first thing to remark is that because each face momentum touching $v$ is bounded by $S_v$, each vertex touches at
most four faces and each face touches at least a vertex,
\bee \sum_{v \in \tilde \cV}  S_v \ge \sum_{f \ni v} \frac{i_f }{4}  \ge \sum_{f} \frac{i_f }{4}  .
\ee 
Then using Lemma~\ref{flow}, 
\bee 
\label{eq:bound-on-ps}
\prod_{e\in \cE_\alpha (w)} p^{j_e} \le \prod_{v \in  \cV} p^{\frac{S_v}{2n}}
\le \prod_{v \in \tilde \cV} p^{\frac{S_v}{n}} \le \prod_{f \in F} p^{\frac{i_f}{4n}} \le \prod_{k=0}^{I(\tilde G)} p^{\frac{i_{\fI_k}}{4n}} , \ee 
where the expression of $I(\tilde G)$ is given in \eqref{eq:ItildeG}. We can apply this bound on the original expression of the graph amplitudes \eqref{fullexpans} thus obtaining  
\be 
\cA_\R (G)\le    \frac{1}{N^{n }} \sum_{\cI_U} \delta_{w'} (\cI_U) \prod_{e\in \cE_\alpha (w)} \delta_{j_e \bar j_e} \prod_{k=0}^{I(\tilde G)} p^{\frac{i_{\fI_k}}{4n}}.
\ee
We can now rewrite this bound in terms of the face momenta, 
\be 
\label{eq:Upp-Bound-1}
\cA_\R (G)\le    \frac {p^{\frac{\R }{4n}}} {N^n} \Bigl(\prod_{f\neq \fr} \sum_{i_f\ge 0}\Bigr)\delta(E\cdot\vec i) \prod_{k=1}^{I(\tilde G)} p^{\frac{i_{\fI_k}}{4n}},
\ee
and then of the face momenta in $i_{\fI_j}$, exactly as was done above for the graph amplitudes themselves,
\be 
\cA_\R( G)\le\frac {p^{\frac{\R }{4n}}} {N^n}  \sum_{i_{\fI_1} , \ldots, i_{\fI_{I(\tilde G)}}\ge 0 
}\; \prod_{j=1}^R\;\Theta\left(\sum_{l=0}^{I(\tilde G)}  a_{j,l} i_{\fI_l} \right) \prod_{k=1}^{I(\tilde G)} p^{\frac{i_{\fI_k}}{4n}},
\ee
with the difference that now, removing the positivity constraints from the $\Theta$, we still have a finite quantity. 
\be 
\cA_\R(G)\le\frac {p^{\frac{\R }{4n}}} {N^n}  \sum_{i_{\fI_1} , \ldots, i_{\fI_{I(\tilde G)}}\ge 0 
}\;\prod_{k=1}^{I(\tilde G)} p^{\frac{i_{\fI_k}}{4n}}=\frac {p^{\frac{\R }{4n}}} {N^n}  \Biggl( \frac 1 {1- p^{\frac{1}{4n}}}\Biggr)^{I(\tilde G)}.
\ee
Factorizing the dependence on $N$, we get
\be 
\cA_\R(G)\le\frac {p^{\frac{\R }{4n}}} {N^{n-I(\tilde G)}} h(p)^{I(\tilde G)},
\ee
with $h(p)$ a smooth increasing positive function on $(0,\infty)$, thus bounded on $(0,1)$ by its value at $p=1$. Since $h(1)=4n$, we have
\be
\cA_\R(G)\le (4n)^{I(\tilde G)}p^{\frac{\R }{4n}} N^{-d( G)}.
\ee
Finally, using the definition of $I(\tilde G)= F-R-1$ and $F\le 3n/2+1$ (from Lemma \ref{lem:boundfaces}) and $R\ge 1$, we get 
\be
\cA_\R(G)\le (4n)^{3n/2-1}p^{\frac{\R }{4n}} N^{-d( G)}.
\ee 
\qed

\

The existence of the $1/N$ expansion is then guaranteed by the following two lemmas.

\begin{lemma} \label{lemmamelo2}
If $G $ is melonic, then $d(G)= 0$.
\end{lemma} 
\proof 

\begin{figure}
	\centering
	\includegraphics[scale=0.6]{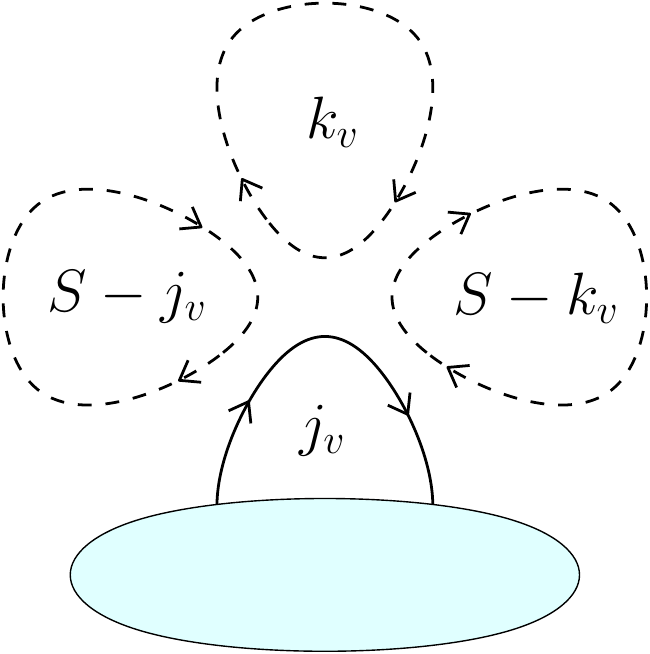}
	\hspace{1cm}
	\raisebox{1cm}{$\longrightarrow$}
	\hspace{1cm}
	\includegraphics[scale=0.6]{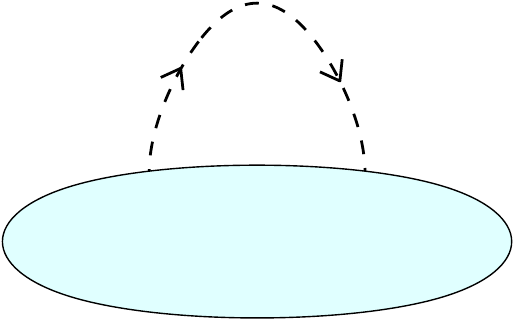}
	\caption{Melonic reduction of type $\I$ in the stranded representation.}
	\label{fig:move-stranded}
\end{figure}

Let $G$ be a melonic graph. By definition, it can be reduced to the trivial ($n=0$) graph by a sequence of melonic reductions, where at each step, the reduced elementary 2-point melon does not contain the root. If we represent the melonic reduction moves of Figs.~\ref{fig:Move1}, \ref{fig:Move2}, \ref{fig:Move3}, \ref{fig:Move4}, \ref{fig:Move5} in the stranded representation using the  representation of Fig.~\ref{fig:ElementaryMelons} for the elementary melons, it is straightforward to see that each reduction removes three faces (see Fig.~\ref{fig:move-stranded} for instance for a reduction of type $\I$). 

The recursive sequence of melonic reductions thus shows that the number of faces in a melonic graph is $F=3n/2+1$, as each melonic reduction removes three faces and two 4-valent vertices (or equivalently one 8-valent node). Melonic graphs thus  saturate the bound of Lemma \ref{lem:boundfaces}.  

On the other hand, at each melonic reduction step, the three faces that are removed have length one. In terms of the incidence matrix $E$, they each correspond to a column with zeros everywhere except a $+1$ or a $-1$ on the line of the corresponding 8-valent node. Therefore, there is only one independent column (and line) in $E$ associated with these three faces. As the incidence matrix of the graph after the melonic reduction is just $E$ without the columns and line corresponding to the three faces of length one and the reduced 8-valent node,  the melonic reduction reduces the rank $R$ of $E$ by one. As a consequence, $R$ is maximal for a melonic graph, i.e.\ $R=V=n/2$, since it requires $n/2$ melonic reductions to reduce it to the trivial graph, which is characterized by $n=R=0$. This result can also be understood in terms of the face momenta. Recall that the dependent face momenta correspond to the independent columns in $E$. When performing a melonic reduction, the three faces that are removed possess a distinct face momentum; but one of these three face momenta depends on the other two because of the momentum conservation at the corresponding 8-valent node.

Hence, in the case of a melonic graph $G$, we have $I(G)=F-R-1=3n/2+1-n/2-1=n$, so that $d(G)=n-I(G)=0$. 
\qed

\

\begin{figure}[h!]
	\centering
	\includegraphics[scale=1.4]{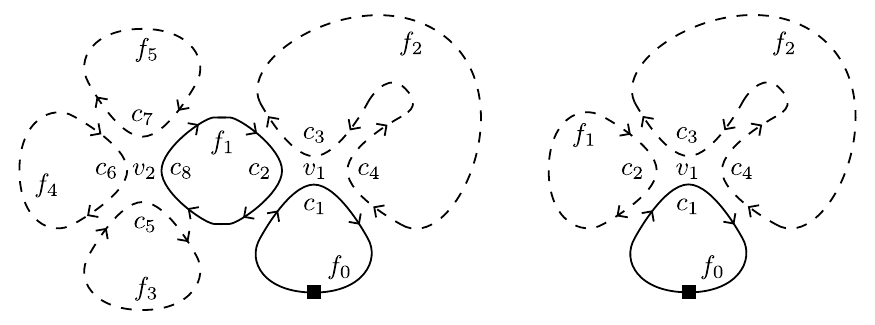}
	\caption{Example of a melonic reduction of type $\I$ (the 8-valent node $v_2$ is removed), in the stranded representation, on a non-melonic graph $\tilde{G}$ (left-hand side), which gives a non-trivial graph $\tilde{G}'$ (right-hand side). For illustration purposes, we added the 8-valent node labels $v_1, v_2$, the face labels $f_0$ to $f_5$ as well as the corner labels $c_1$ to $c_8$.}
	\label{fig:example-melonic-reduction}
\end{figure}

We provide in Fig.~\ref{fig:example-melonic-reduction} an explicit example of a reduction of a 2-point melon in a non-melonic graph $G$ (given in the stranded representation on the left-hand side), yielding another non-melonic graph $G'$ (given on the right-hand side). We then compute the incidence matrices to illustrate the change of rank during a melonic reduction, as performed in the proofs of Lemma~\ref{lemmamelo2} and \ref{lemmamelo3}.

In the figure, we labelled in $\tilde{G}$  (resp.\ in $\tilde{G}'$) the two 8-valent nodes $v_1$ and $v_2$ (resp.\ $v_1$), the five faces $f_0$ to $f_5$ (resp.\ $f_0$ to $f_2$) and the corners $c_1$ to $c_8$ (resp.\ $c_1$ to $c_4$). $G$ has $n/2=2$ 8-valent nodes and $F=6$ faces. The melonic reduction removes the 2-point melon associated with the 8-valent node $v_2$. The new graph $G'$ then has $n'/2=n/2-1=1$ remaining 8-valent node and  $F'=F-3=3$ remaining faces. The incidence matrix $E$ of $G$ is given by the $2\times 6$ matrix $E=\bigl(\begin{smallmatrix} 1&1&-2&0&0&0 \\ 0&-1&0&1&1&-1 \end{smallmatrix} \bigr)$. It has rank $R=2$. One can verify that out of the three faces of length one associated with the 2-point melon, namely $f_3, f_4$ and $f_5$, only one of them is independent. On the other hand, the incidence matrix $E'$ of $G'$ is obtained by removing in $E$ the last three columns associated with $f_3, f_4$ and $f_5$, as well as the second line associated with $v_2$. It is thus given by the following $1\times 3$ matrix $E'=(\begin{smallmatrix} 1&1&-2 \end{smallmatrix})$. As expected, its rank is $R'=R-1=1$. Finally, one can check explicitly that $d(G)=d(G')=1$.

\begin{lemma}\label{lemmamelo3}
If $ G$ is not melonic, then $d (G ) >0$.
\end{lemma}
\proof
Let $G$ be a non-melonic graph. We first reduce recursively all the elementary 2-point melons in $G$, if there are any, which gives a non-trivial graph $G'$. Using the same reasoning as in the proof of the previous lemma, applying $0\leq m<n/2$ melonic reduction moves eliminates $m$ 8-valent node, $3m$ faces and $m$ independent columns or lines of the incidence matrix $E$. Therefore, the number of 8-valent nodes and faces of $\tilde G'$, and  the rank $R'$ of the associated incidence matrix $E'$, are respectively $n'/2=n/2-m$, $F' = F- 3m$, and $R'= R-m$.  Hence, $d(G')=d(G)$,  and the proof reduces to the case of a graph $G'$ which has $n'$ true vertices and which does not contain any elementary 2-point melons.

Let us define $V_l'$ the number of 8-valent nodes in $\tilde{G}'$ adjacent to exactly $l$ faces of length one (we recall that the stranded graph corresponding to $G'$ is denoted by $\tilde G'$). On the one hand, because $G$, and therefore $G'$, are not melonic, $V_4'=0$. Indeed, the only graphs with $V_4'\neq 0$ are the elementary melons of Fig.~\ref{fig:ElementaryMelons}. On the other hand, since $G'$ doesn't contain any elementary 2-point melon, $V_3'=0$, because 
by definition, the latter are the only subgraphs with a single 8-valent node and three adjacent faces of length one. Hence, there are at most two faces of length one adjacent to a given 8-valent node in $\tilde{G}'$.

Now, we remark that the rank $R'$ of the incidence matrix $E'$  associated with $\tilde{G}'$ is at least $V_1' + V_2'$. Indeed, for each 8-valent node $v\in\tilde{\cV}(\tilde{G}')$ that contains at least one face of length one, we can express the face momentum of this face of length one in terms of the other adjacent face momenta using  the momentum conservation at $v$. Therefore, $R'
\geq V_1' + V_2'$.

Starting from the definition of $I'=I(\tilde{G}')$ and using $F'= \sum_{l\geq 1} F'_l$, where  $F'_l$ is the number of faces of length $l$ in $\tilde{G}'$,
we can thus write
\begin{equation*}
    I'
    = F'
    -R'
    -1 \leq \sum_{l\geq1}F'_l - V_1' - V_2' -1.
\end{equation*}
The number of faces of length one in $\tilde{G}'$ is given by $F'_1=V_1' + 2V_2'$. Hence, we find that
\begin{equation}\label{upperboundI}
    I'
    \leq \sum_{l\geq2}F'_l + V_2' -1 \leq \sum_{l\geq2}\frac{l}{2}F'_l + V_2' -1.
\end{equation}
We now use the fact that $2n' = \sum_{l\geq 1} l F'_l$
to obtain the relation
\begin{align*}
    \sum_{l\geq2}\frac{l}{2}F'_l &= n'-\frac{1}{2}F'_1 \\
        &= n'-\frac{1}{2}(V_1' + 2V_2') \\
        &= 2V_0' +\frac{3}{2}V_1' + V_2',
\end{align*}
where the last equality comes from the fact that the total number of 8-valent nodes in $\tilde{G}'$ is given by $n'/2=V_0' +V_1' + V_2'$. Replacing the above relation in the RHS of \eqref{upperboundI} finally yields
\begin{align*}
    I'
    & \leq 2V_0' +\frac{3}{2}V_1' + 2V_2'-1 \\
        &< 2V_0' +2V_1' + 2V_2' = n'.
\end{align*}
This eventually shows that $d(G)=d(G')=n'-I'
>0$.
\qed

\

We make a few remarks. First, Prop.~\ref{prop:globalbound} only provides an upper bound for the $N$ scaling of the graph amplitudes. Though it is sufficient for proving the existence of a $1/N$ expansion (in the sense detailed at the beginning of Sec.~\ref{sub:1N-exp}), it doesn't give the exact $N$ scaling of a given graph $G$. Second, Lemma \ref{lemmamelo2} strongly suggests that the melonic graphs should be the dominant graphs in the $1/N$ expansion. Indeed, they are the only graphs that can be part of the leading sector $d(G)=0$ of the expansion. However, this is not enough for proving that all the melonic graphs are part of the leading sector since Prop.~\ref{prop:globalbound} only provides an upper bound. 

Given the expression of the graph amplitudes in term of the independent faces \eqref{eq:Expression-Amp-Faces-Ind}, it is likely that the behavior of a graph $G$ is truly in $N^{-d(G)}$, thus restricting the $1/N$ expansion of the averaged Sobolev norms to non-negative integer powers of $1/N$. We however leave this to future studies.

\subsection{Melonic dominance
}
\label{sub:Mel-Dom}

To have a stronger statement for melonic graphs, we now prove that all the melonic graphs are part of the leading sector $d(G)=0$. To do this, we find a lower bound on the amplitude of melonic graphs in Lemma~\ref{lemmamelobounds}, with the right scaling in $N$. Together with Lemma~\ref{lemmamelo3}, this establishes the following result.
\begin{proposition} 
\label{prop:Mel-Dom}
Melonic graphs are all dominant and are the only dominant graphs.
\end{proposition}
\proof Indeed, from Lemma~\ref{lemmamelo3}, we know that the behavior in $N$ of the amplitudes of melonic graphs is bounded from above by 1, and in Lemma~\ref{lemmamelobounds} below, we show that it is bounded from below by 1. \qed

\begin{lemma} \label{lemmamelobounds}
If $G= (U,w,w')$ is a melonic graph, then
\begin{equation} 
\frac{1}{(2n+1)^n} p^{(2n+1)r} \le   \cA_r (G).
\label{melobound}
\end{equation}
\end{lemma}
\proof
Since $G$ is a melonic graph, there exists a way of constructing it by recursively inserting $n/2$ elementary two-point melons, starting from the trivial graph. We consider one particular way of doing so.  In the following, for $0\le q \le n$ ($q$ even), we denote by $G_q$ the melonic graph obtained in this particular process, but after $q/2$ melonic insertions only, so that $G_0$ is the trivial graph, and $G_n=G$. In $G_q$, the last elementary two-point melon that has been inserted is refered to in the following as the $q/2^{\textrm{th}}$ two-point melon.

We start from the expression of the amplitude of 
the graph $G_n=G$ with $n$ vertices given in terms of $3n$ sums
\be 
\cA_\R(G_n)=\frac 1 {N^n} \Bigl(\prod_{v\in
\cV} \sum_{S_v}\sum_{j_v, k_v \le S_v}\Bigr) \Bigl(\prod_{\substack{e \text{ dashed } \\ \text{or solid}}} \delta_{j_e, j'_e}
\Bigr) \prod_{e\text{ dashed} } p^{j_e}\prod_{e\in\cE_C} \delta_\text{lead}(e).
\ee
First, we notice that this expression of the amplitude can be bounded from below by weighting \emph{every} edge (dashed or solid) with momentum $j_e$ by a factor $p^{j_e}$, thanks to the fact that $p<1$. In the following however, we will need to bound the amplitude by the more general quantity which we now define. We introduce an edge-weight $a_e\in \bN$ for every edge $e$, and we define a quantity such that every edge of given momentum $j_e$ and edge-weight $a_e$ is weighted by a factor $p^{a_ej_e}$, 
\be
\cB_\R(G_{n},\vec a):=\frac 1 {N^n} \Bigl(\prod_{v\in
\cV} \sum_{S_v}\sum_{j_v, k_v \le S_v}\Bigr) \Bigl(\prod_{\substack{e \text{ dashed } \\ \text{or solid}}} \delta_{j_e, j'_e} p^{a_e j_e}
\Bigr)\prod_{e\in\cE_C} \delta_\text{lead}(e),
\ee 
where $\vec a$ is the vector of all edge-weights. We obviously have, for all $\vec a$,
\be 
\cA_\R(G_{n})\ge\cB_\R(G_{n}, \vec a).
\ee
Let us consider the melonic graphs $G_{q}$ and $G_{q-2}$ as defined above, with $q=n-2m$. These graphs are respectively obtained from $G_n$ by performing $m$ and $m+1$ melonic reductions, so that they have respectively $n-2m$ and $n-2m -2$ vertices. In the following, we bound $\cB_\R(G_{n-2m}, \vec a_m)$ in terms of $\cB_\R(G_{n-2m -2}, \vec a_{m+1})$, where we assume that the edge-weights $\vec a_m$ have been computed from the initial weights $\vec a_0$ after the $m$ melonic reductions, and we compute the new edge-weights $\vec a_{m+1}$. 
\begin{figure}[!ht]
	\centering
	\includegraphics[scale=1]{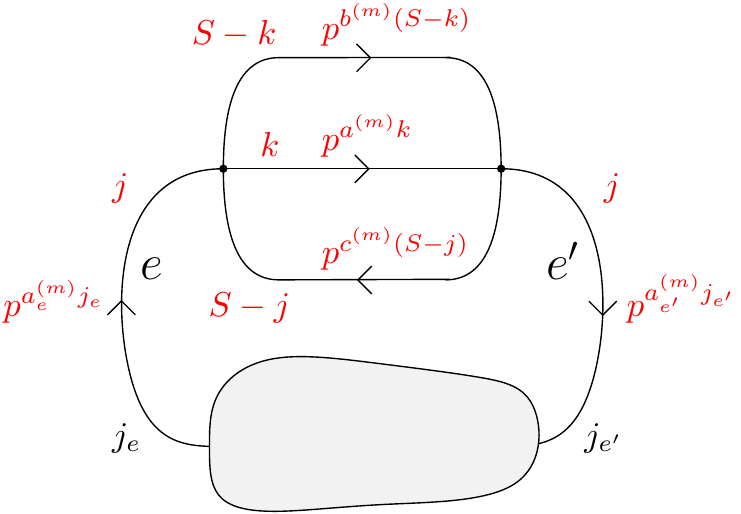} 
	\caption{Local assignment of momenta on the $(n/2-m)^\text{th}$ two-point melon. We represent all edges as solid since in $\cB_r$, every edge $e$ carries a weight $p^{a_e}$.}
	\label{fig:proof-lower-bound}
\end{figure}

We consider the $(n/2-m)^{\textrm{th}}$  two-point melon. As explained in Remark~\ref{rmk:Change-variable-El-melons} of Section~\ref{sec:order_2-amplitudes}, in an elementary two-point melon, we can always locally relabel the indices  (within the sums in the amplitude) so that the indices on the external legs of the two-point melon are $j$ and $j'$, and so that for the remaining three edges that link the two true vertices of the 2-point melon, the indices are respectively $k$ and $k'$, $S-k$ and $S-k'$, and $S-j$ and $S-j'$. This way, the leading propagator is always the propagator $\delta_S^{S'}\delta_j^{j'}\delta_k^{k'} \sim 1$. Summing over $j' , k', S'$, the momenta are locally as in Fig.~\ref{fig:proof-lower-bound}, where we denoted by $j_e$ and $j_{e'}$ the indices on the other end of the edges $e$ and $e'$ that also carry the index $j$, and by $a^{(m)}, b^{(m)}, c^{(m)}$ the $a^{(m)}_e$ edge-weights after $m$ melonic reductions associated with the internal lines of the $(n/2-m)^{\textrm{th}}$ two-point melon.
Focusing on the one hand on the sums over $\{S,j,k\}$ involving the indices associated with the $(n/2-m)^{\textrm{th}}$ two-point melon and on the other hand on the indices $j_e$ and $j_{e'}$, we have
\begin{align}
\cB_\R(G_{n-2m},\vec a_m)\! &= \!\frac { B^{(n-2m)}_{j_e, j_{e'}}(p, \vec a_m)}{N^{n-2m}} p^{a_e^{(m)}j_e+a_{e'}^{(m)}j_{e'}} \sum_{S=0}^\infty\sum_{j,k = 0}^S p^{a^{(m)}k+b^{(m)}(S-k)+c^{(m)}(S-j)}\delta_{j_e, j}\delta_{j_{e'}, j}\nonumber \\
&= \frac { B^{(n-2m)}_{j_e, j_{e'}}(p,\vec a_m)} {N^{n-2m}}  p^{(a^{(m)}_e+a^{(m)}_{e'})j_e}\delta_{j_e, j_{e'}} \sum_{S=j_e}^\infty\sum_{k = 0}^S p^{a^{(m)}k+b^{(m)}(S-k)+c^{(m)}(S-j_e)},
\end{align}
where $B^{(n-2m)}_{j_e, j_{e'}}(p, \vec a_m)$ represents the remaining sums and weights of the expression of $\cB_\R(G_{n-2m},\vec a_m)$. We have 
\bea
\cB_\R(G_{n-2m},\vec a_m) &\ge&\frac{B^{(n-2m)}_{j_e, j_{e'}}(p,\vec a_m)}{N^{n-2m}}    p^{(a^{(m)}_e+a^{(m)}_{e'})j_e}\delta_{j_e, j_{e'}} \sum_{S=j_e}^\infty\sum_{k = 0}^S p^{(a^{(m)}+b^{(m)}+c^{(m)})S},
\eea
where we used that $k, S-k,S-j_e\le S$. The above sums can be straightforwardly computed:
\bea
\sum_{S=j_e}^\infty\sum_{k = 0}^S p^{(a^{(m)}+b^{(m)}+c^{(m)})S} &=& \sum_{S=j_e}^\infty(S+1) p^{(a^{(m)}+b^{(m)}+c^{(m)})S} \\
&=& \frac{p^{(a^{(m)}+b^{(m)}+c^{(m)})j_e}(1+j_e(1-p^{a^{(m)}+b^{(m)}+c^{(m)}}))}{(1-p^{(a^{(m)}+b^{(m)}+c^{(m)})})^2}\\
&\ge& \frac{p^{(a^{(m)}+b^{(m)}+c^{(m)})j_e}}{(1-p^{(a^{(m)}+b^{(m)}+c^{(m)})})^2}\\
&\ge& \frac{p^{(a^{(m)}+b^{(m)}+c^{(m)})j_e}}{(a^{(m)}+b^{(m)}+c^{(m)})^2}N^2.
\eea
Denoting $\lvert \vec a_m \rvert$ the $L^1$ norm of $\vec a_m$ we have 
\bea
\cB_\R(G_{n-2m},\vec a_m) &\ge& \frac{B^{(n-2m)}_{j_e, j_{e'}}(p,\vec a_m)}{N^{n-2m-2}} \frac{\delta_{j_e, j_{e'}}}{\lvert \vec a_m \rvert^2} p^{(a^{(m)}_e+a^{(m)}_{e'}+a^{(m)}+b^{(m)}+c^{(m)})j_e}.
\eea
Thus we can reduce the $(n/2-m)^{\textrm{th}}$ two-point melon and replace it by a new edge $e_{\textrm{new}}$ in $G_{n-2m-2}$ with edge-weight $a_{e_{\textrm{new}}}=a^{(m)}_e+a^{(m)}_{e'}+a^{(m)}+b^{(m)}+c^{(m)}$. The graph $G_{n-2m-2}$ has four edges less  than $G_{n-2m}$ and consequently four edge-weights less. We define the edge-weights $\vec a_{m+1}$ of $G_{n-2m-2}$ such that all the edge-weights of the edges untouched under the melonic reduction are the ones of the corresponding edges of $G_{n-2m}$, while the edge-weight of the new edge is defined to be $a_{e_{\textrm{new}}}$,
$$\vec a_{m+1} = \bigl(\vec a_m \setminus (a^{(m)}_e, a^{(m)}_{e'}, a^{(m)}, b^{(m)}, c^{(m))}\bigr) \cup a_{e_{\textrm{new}}}.$$
Doing so we notice that since the weights $a^{(m)}_e, a^{(m)}_{e'}, a^{(m)}, b^{(m)}, c^{(m)}$ are non-negative, then $a_{\textrm{new}}$ is too, and $\lvert \vec a_m \rvert =\lvert \vec a_{m+1}\rvert$, thus the $L^1$ norm of the edge-weights vector stays constant under a melonic reduction. We can further rewrite
\bea
\cB_\R(G_{n-2m},\vec a_m) &\ge& \frac{\cB_\R(G_{n-2m-2},\vec a_{m+1})}{\lvert \vec a_m \rvert^2} = \frac{\cB_\R(G_{n-2m-2},\vec a_{m+1})}{\lvert \vec a_{m+1} \rvert^2}.
\eea
This formula induces a sequence of nested lower bounds in terms of $\cB_\R$ evaluated on smaller graphs 
\be
\ldots \ge\frac{ \cB_\R(G_{n-2m},\vec a_m)}{\lvert \vec a_m\rvert^{2m}}\ge \frac{\cB_\R(G_{n-2m-2},\vec a_{m+1})}{\lvert \vec a_{m+1} \rvert^{2m+2}} \ge \ldots.
\ee
Consequently, if we start the process of melonic reductions with a melonic graph $G_n$ with $n$ vertices and initial edge-weights $\vec a_0$, we can perform $n/2-1$ melonic reductions to sequentially bound $\cB_\R(G_{n},\vec a_0)$ in term of $\cB_\R(G_{2},\vec a_{n/2-1})$, where $\vec a_{n/2-1}$ represents the remaining five edge-weights after the $n/2-1$ melonic reductions. We have
\be
\cB_\R(G_{n},\vec a_0)\ge \frac{\cB_\R(G_{2},\vec a_{n/2-1})}{\lvert \vec a_0\rvert^{n-2}},
\ee
and
\begin{align}
\cB_\R(G_{2},\vec a_{n/2-1})&=\frac1{N^2}p^{j_{e_r}a_{e_r}+j_{e'_r}a_{e'_r}}\sum_{S=j_{e_\R}}^{\infty}\sum_{j,k=0}^{S}p^{a^{(n/2-1)}k}p^{b^{(n/2-1)}(S-k)}p^{c^{(n/2-1)}(S-j)}\delta_{j_{e_r},j}\delta_{j_{e'_r},j} \nonumber\\
&\ge\frac1{N^2}p^{j_{e_r}a_{e_r}+j_{e_r}a_{e'_r}}\delta_{j_{e'_r},j_{e_r}}\sum_{S=j_{e_\R}}^{\infty}p^{(a^{(n/2-1)}+b^{(n/2-1)}+c^{(n/2-1)})S} \nonumber\\
&\ge\frac{1}{\lvert \vec a_0 \rvert^2}p^{j_{e_r}(a_{e_r}+a_{e'_r}+a^{(n/2-1)}+b^{(n/2-1)}+c^{(n/2-1)})}=\frac{p^{j_{e_r}\lvert \vec a_0 \rvert}}{\lvert \vec a_0 \rvert^2},
\end{align}
thus
\be
\cA_\R(G_{n})\ge \cB_\R(G_{n},\vec a_0)\ge \frac{p^{j_{e_r}\lvert \vec a_0 \rvert}}{\lvert \vec a_0 \rvert^n}.
\ee
Choosing $\vec a_0$ such that all initial edge-weights are set to one, we have $$\lvert \vec a_0 \rvert=\sum_{e\in G_n}a^{(0)}_e=2n+1$$ which is the number of edges of $G_n$. Since $j_{e_r}=r$, we end up with the desired bound $\cA_\R(G_{n})\ge \frac{p^{\R(2n+1)}}{(2n+1)^n}$. In particular, we have the desired melonic scaling in $N$, that is, $d(G)=0$.  \qed

\subsection{Exponential bound for melonic graphs and analyticity of the melonic Sobolev norms }
\label{sub:Exp-bounds}

Proposition~\ref{prop:Mel-Dom} shows that all the melonic graphs are part of the leading order in the $1/N$ expansion. This key fact allows one to define the melonic approximation
\bee
G_2^{melo} (\R,t) := \sum_{n \ {\rm even}}  \frac{t^n}{n !} 8^{-n/2} \sum_{G \ {\rm melonic \ of\ order}\ n} \epsilon (U ) \cA_\R (G) \label{meloapp}
\ee
to $\langle G_2 (\R,t) \rangle_{\alpha , C}$. It remains to prove that this melonic approximation defines an analytic function of $t$. This is the object of this section: we will prove the following proposition.
\begin{proposition} 
\label{prop:Analyc-Melo}
There exists an interval around 0 on which the  melonic approximation $S_\gamma^{melo}(t) = \sum_{\R\ge 0} \frac{\R^\gamma}{N} G_2^{melo} (\R,t) $ to the averaged Sobolev norms
$\bar S_\gamma$ is  analytic in $t$. 
\end{proposition}

\

The following lemma gives an exponential upper bound on melonic graphs. This is the first step to prove the analyticity of the restriction of the Sobolev norms to the melonic regime. 
\begin{lemma}\label{keylemma2}
If $G= (U,w,w')$ is a melonic graph, then
\begin{equation} 
\cA_r (G) \le {4}^{n } p^{r/2} .
\label{melobound2}
\end{equation}
\end{lemma}

\proof 
In this proof, we first bound from above the amplitude of melonic graphs using a quantity that depends on some weights, similarly as in the proof of Lemma~\ref{lemmamelobounds}. We then compute this quantity inductively, by performing melonic reductions. 

To start with, we notice that we have the following simple bound on the amplitude of a graph $G$
\be\label{eq:simple_bound}
\cA_\R(G)\le \frac 1 {N^n} \Bigl(\prod_{v\in\cV} \sum_{S_v, j_v, k_v \in \mathbb Z }\Bigr) \Bigl(\prod_{\substack{e \text{ dashed } \\ \text{or solid}}} \delta_{j_e, j'_e}
\Bigr) \prod_{e\text{ dashed} } p^{\vert j_e \vert}\prod_{e\in\cE_C} \delta_\text{lead}(e).
\ee
The right hand side bounds $\cA_r (G)$ simply because it extends the sum over all $\mathbb{Z}$ while ensuring convergence by changing all the factors $p^k$ into $p^{\vert k\vert }$. Since $G$ is a melonic graph, there exists a way of constructing it by recursively inserting $n/2$ elementary two-point melons, starting from the trivial graph. We pick one way of doing so. This induces an order on the two-point melons of $G$. In particular, we can distinguish the last melonic insertion.
\
We introduce the length-$L$ ($L\in\mathbb N$) chain functions, which are tool functions for our proof, 
\begin{eqnarray}
F_0(j)=p^{\vert j \vert}, \quad 
F_L(j) = \sum_{j_1, \cdots,  j_L} p^{\vert j-j_1 \vert + 
\vert j_1-j_2 \vert + \cdots+ \vert j_{L-1}-j_L \vert + \vert j_L\vert}.
\end{eqnarray}
\
Note  that these functions are even ($F_L(j) = F_L(-j) $). We have the sum and concatenation rules 
\begin{align}
&\sum_j F_L(j-k) = K(p)^{L+1} \;\;\forall k\in\mathbb Z, \label{sum1} \\ 
&\sum_k F_L(j-k) F_M(k) = F_{L+M+1}(j).\label{concat}
\end{align}
where $K(p)=\frac{1+p}{1-p} = 2N + O(1)$. The properties of these chain functions are used to bound the amplitude recursively.
\\

The right hand side of \eqref{eq:simple_bound} can be expressed using chain functions of length 0 attached to every dashed edge, it is written as
\be\label{eq:simple_with_lengths}
\frac 1 {N^n} \Bigl(\prod_{v\in \cV} \sum_{S_v, j_v, k_v \in \mathbb Z }\Bigr) \Bigl(\prod_{\substack{e \text{ dashed } \\ \text{or solid}}} \delta_{j_e, j'_e}
\Bigr) \prod_{e\text{ dashed} } F_0(j_e)\prod_{e\in\cE_C} \delta_\text{lead}(e).
\ee
For initializing the recursion and for illustratory purposes, we here compute the contribution of the last melonic insertion  
of $G$ to the right hand side of \eqref{eq:simple_bound}. The last melonic insertion has to be of one of the types $\I$, $\II$, $\III$, $\II s$ or $\III s$ as depicted in Figures \ref{fig:Move1}-\ref{fig:Move5}. As in the proof of Lemma~\ref{lemmamelobounds} (see also Remark~\ref{rmk:Change-variable-El-melons}), we can do a local change of variables and sum over the indices of one of the true vertices of the 2-point melon, so that the momenta are displayed as in Fig.~\ref{fig:proof-lower-bound}, i.e.~the two external edges of the 2-point melon have the same momentum $j$, and the three internal edges linking its true vertices have respectively the momenta $S-j$, $k$, and $S-k$.  Depending on the type of melonic insertion, we have different possible results. In order to keep the proof concise, we show the initialization cases for two different melonic insertion types only,
\begin{itemize}
    \item Type $\I$:
    \be
    \sum_{S,j,k\in \mathbb{Z}}F_0(k)F_0(S-k)F_0(S-j)\delta_{j_e,j}\delta_{j_{e'},j}=F_2(j_e)\delta_{j_e,j_{e'}} \label{eq:typeI-init}
    \ee
    which is computed using the concatenation rule \eqref{concat} and the evenness of the chain functions. Notice that in this case, the result does not involve the function $K(p)$. From this result, we can now perform a melonic reduction of type $\I$ on the last melonic insertion at the expense of weighting the newly created edge by $F_2(j_e)$. Notice that the chain function weighting this new edge is of greater length $L$.
    \item Type $\II$:
    \be
    \sum_{S,j,k\in \mathbb{Z}} F_0(j)F_0(S-k)F_0(k)\delta_{j_e,j}\delta_{j_{e'},j}=\delta_{j_e,j_{e'}}F_0(j_e)K(p)^2, \label{eq:typeII-init}
    \ee
    which is obtained using the sum rule first over $S$ and then over $k$ (the sum over $j$ is trivial thanks to the Kronecker delta). Again we can perform a melonic reduction of type $\II$ on the last melonic insertion at the expense of weigthing the newly created edge with $F_0(j_e)K(p)^2$. Notice that the length of the chain function weighting the new edge remains the same in this case.
\end{itemize}
In order to keep track of the length of the different chain functions weighting the dashed edges of a graph, we define length labels $L_e \in \bN_0$ associated to each dashed edge $e$. We write $\vec L_e$ the vector of all edge-lengths. We then define the \emph{extended amplitude} $\cD_\R(G, \vec L_e)$, which depends on a graph and its edge-lengths, as
\be
\cD_\R(G, \vec L_e):=\frac 1 {N^n} \Bigl(\prod_{v\in\cV} \sum_{S_v, j_v, k_v \in \mathbb Z }\Bigr) \Bigl(\prod_{\substack{e \text{ dashed } \\ \text{or solid}}} \delta_{j_e, j'_e}
\Bigr) \prod_{e\text{ dashed} } F_{L_e}(j_e)\prod_{e\in\cE_C} \delta_\text{lead}(e).
\ee
For $\vec L_e = 0$, it reduces to \eqref{eq:simple_with_lengths}, thus we have
\be
\cA_\R(G)\le \cD_\R(G,0).
\ee
We now want to evaluate the extended amplitude $\cD_\R(G,0)$ inductively, by performing successive melonic reductions starting from the initial graph $G$ and ending with the trivial graph. As illustrated above, reducing two-point melons can introduce chain functions with increasing lengths and factors of $K(p)$. Therefore, one needs to understand the effect of reducing a two-point melon, for each type, in a graph with \emph{a priori} unknown edge-length labels. This is what is done below. In the case of an elementary 2-point melon of type $\I$, we use twice the concatenation rule and the evenness of $F$ to write
\begin{equation}
\sum_j \delta_{j, j_{e}} \delta_{j, j_{e'}}
\sum_S  \sum_k  F_L(S-k) F_M(k) F_P(S-j) 
=  \delta_{j_e, j_{e'}} F_{L+M+P +2}(j_e).
\end{equation}
In the case of type $\II$, we use twice the sum rule to write 
\begin{equation}
\sum_j \delta_{j, j_{e}} \delta_{j, j_{e'}}
\sum_S \sum_k F_L(j) F_M(S-k) F_P(k)
= \delta_{j_e, j_{e'}} K(p)^{M+P+2} F_L(j_e) .
\end{equation}
In the case of type $\III$, we use twice the sum rule to write 
\begin{equation}
\sum_{j} \delta_{j, j_{e}} \delta_{j, j_{e'}}
\sum_{S} \sum_{k} F_L(j) F_M(S-k) F_P(S-j) 
= \delta_{j_e, j_{e'}} K(p)^{M+P+2} F_L(j_e) .
\end{equation}
In the case of type $\II s$, we use twice the sum rule to write
\begin{equation}
\sum_{j} \delta_{j, j_{e}} \delta_{j, j_{e'}}
\sum_{S} \sum_{k} F_L(S-k) F_M(k) 
= \delta_{j_e, j_{e'}} K(p)^{L+M+2} .
\end{equation}
Finally, in the case of type $\III s$, we use twice the sum rule to write
\begin{equation}
\sum_{j} \delta_{j, j_{e}} \delta_{j, j_{e'}}
\sum_{S} \sum_{k} F_L(S-k) F_M(S-j) 
= \delta_{j_e, j_{e'}} K(p)^{L+M+2} .
\end{equation}
As is apparent already in the computations of equations \eqref{eq:typeI-init}, \eqref{eq:typeII-init} and more generally in the above results, only type-$\I$ melonic reductions can create new edges with associated edge-length label greater (by $2$) than the sum of the edge-length labels associated to the reduced $2$-point melon. The other types of melonic reductions produce factors of $K(p)^2$ for each reduction, as well as additional factors of powers of $K(p)$ that depend on the length of the internal edges (see above results for the details). They also produce new edges, either with edge-lengths smaller than the sum of the internal edge-lengths or with no edge-length (if the resulting created edge is a plain edge). 

Let us compute $\cD_\R(G,0)$ in terms of powers of $K$ and factors $F_{L_e}$. There is an obvious factor $K(p)^{2\cM_{\hspace{-0.5mm}\not \hspace{0.5mm} \I}}$ from the reduction of 2-point melons which are not of type $\I$, where $\cM_{\hspace{-0.5mm}\not \hspace{0.5mm} \I}$ denotes the number of such 2-point melons. Let us then focus on the contribution of the dipoles of type $\I$. Although the type-$\I$ two-point melons do not produce any factor of $K$, they do add a factor of 2 to the length of the new edge. We can trace down what happens to this particular factor of 2 during the melonic reductions that follow. 
Every time a 2-point melon which is not of type $\I$ is reduced, the total length of the internal dashed edges of this 2-point melon is converted into powers of $K$. If the factor of 2 resulting from the reduction of a 2-point melon of type $\I$ ends up on an internal dashed edge of some other 2-point melon, then it becomes a factor $K^2$. We denote by $\phi$ the number of 2-point melons of type $\I$ for which this happens. Consequently, the only 2-point melons which do not contribute with a factor of $K^2$ are those of type $\I$ for which the factors of 2 end up on the dashed edge of the trivial graph after the last melonic reduction. We denote by $\nu$ the number of such type-$\I$ two-point melons. From the above discussion, we deduce that   
\be
\cD_\R(G,0)=\frac 1 {N^n}K(p)^{2\cM_{\hspace{-0.5mm}\not \hspace{0.5mm} \I}}K(p)^{2\phi}F_{2\nu}(r),
\ee
with 
\be 
\phi+\nu=\cM_{\I},
\ee 
$\cM_{\I}$ being the total number of 2-point melons of type $\I$. Now using the fact that $$F_{2\nu}(\R)\le p^{ \R/2}N^{2\nu} 4^{2\nu},$$ as will be proved in Lemma~\ref{bound-tool-function} below, we obtain that 
\be
\cD_\R(G,0)\le \frac 1 {N^n}N^{2(\cM_{\hspace{-0.5mm}\not \hspace{0.5mm} \I}+\phi + \nu)}4^{\cM_{\hspace{-0.5mm}\not \hspace{0.5mm} \I}+\phi + 2\nu}p^{ \R/2} \le 4^{n}p^{ \R/2},
\ee
where we used that $\cM_{\hspace{-0.5mm}\not \hspace{0.5mm} \I}+\phi + \nu=\cM_{\hspace{-0.5mm}\not \hspace{0.5mm} \I}+\cM_{\I}=n/2$.
\qed 

\

\begin{lemma} 
\label{bound-tool-function}
For $s\ge 0$, we have the following bound on the tool functions,
\bea
F_s(r)\le p^{ \R /2}N^s 4^s.
\eea
\end{lemma}
\proof Let us first change variable as follows,
\be
F_s(r) = \sum_{j_1, \ldots, j_s} p^{\vert r-j_1\vert} \cdots p^{\vert j_{s-1}-j_s\vert}p^{\vert j_s\vert} 
= \sum_{j_1', \ldots, j_s'} p^{\vert j_1'\vert} \cdots p^{\vert j_s'\vert} p^{\vert r-j_1'-j_2' -\cdots - j_s'\vert}.
\ee
Denoting $J=\sum_{k=1}^s j_k'$, we therefore have
\bea
F_s(r) &=& \sum_{j_1', \ldots, j_s'} p^{\sum_{k=1}^s\vert j_k'\vert}  p^{\vert r-J\vert} \\
&=& \sum_{j_1', \ldots, j_s'} p^{\sum_{k=1}^s\vert j_k'\vert/2} p^{\sum_{k=1}^s\vert j_k'\vert/2}  p^{\vert r-J\vert} \\
&\le& \sum_{j_1', \ldots, j_s'} p^{\vert J\vert/2}p^{\vert r-J\vert} p^{\sum_{k=1}^s\vert j_k'\vert/2}  \\
&\le& \max_{j_k'\in \mathbb{Z}} \left( p^{\vert J\vert/2}  p^{\vert r-J\vert}\right) \sum_{j_1', \ldots, j_s'} p^{\sum_{k=1}^s\vert j_k'\vert/2} \\
&=& p^{ r/2} \sum_{j_1', \ldots, j_s'} p^{\sum_{k=1}^s\vert j_k'\vert/2}=p^{ r/2}\left( \frac{1+p^{1/2}}{1-p^{1/2}}\right)^s .
\eea
Now, there exists a smooth increasing positive function $h(p)$ on $(0,\infty)$, thus bounded on $(0,1)$ by its value at $p=1$, such that
\bea
\frac{1+p^{1/2}}{1-p^{1/2}}=\frac{h(p)}{1-p} .
\eea
Thus 
\bea
F_s(r)\le p^{ \R/2}N^sh(1)^s.
\eea
\qed

\

We must now bound the number of melonic graphs at order $n$.

\begin{lemma}\label{keylemma3}
The number of melonic graphs $G=(U,w,w')$ of order $n$ is bounded from above by $K^n n!$ for some constant $K$.
\end{lemma}
\proof 
Let us bound the number of melonic graphs $G=(U,w,w')$ of order $n$ from above. By forgetting arrows, wavy edges and the difference between solid and dashed edges, one can associate to any melonic graph $G= (U, w, w')$ a simpler 4-regular melonic graph $\bar G$ with labeled vertices from 1 to $n$ and one bivalent root. We call $\bar G$ the projected graph of $G$. The vertices of $\bar G$ are labeled simply because they inherit the heap-ordering labels of $U$. The number of melonic graphs $G$ of order $n$ is given by $\sum_{\bar G} \cN(\bar G)$, where $\cN(\bar G)$ is the number of graphs $G$ that have the same projected graph $\bar G$, and the sum is taken over all labeled projected 4-regular melonic graphs $\bar G$. Let us bound $\cN(\bar G)$ from above. 
\\
Given a labeled projected melonic graph $\bar G$, we obtain all the graphs $G$ which project to $\bar G$ (if there exist any) as follows. We first choose a spanning tree so that the labels of $\bar G$ define a heap ordering of the spanning tree (leaves included). Of course, this might not be possible, however we bound the number of ways of doing so by the number of non-necessarily connected subgraphs of $\bar G$, which is $2^{2n+1}$ (for every edge, we decide whether it is included in the subgraph or not, and there are $2n+1$ edges). The edges not included in the spanning tree are the dashed edges. 
\\
Then, we must choose the orientation of the edges, so that there are two in-going and two out-going edges at every true vertex, and one in-going and one out-going edges at the root. Again, this might not be doable, but we bound the number of ways of doing so by $2^{2n}$, which is the number of ways of orienting the edges of the graph with only the condition for the edges incident to the root (i.e.~when forgetting the conditions at the true vertices).
\\
We must then choose an ordering of the children half-edges around every true vertex, so that the orientation of the parent-edge matches that of the edge $e_2$, and so that the orientation of the edges $e_3$ and $e_4$ is opposite to that of $e_1$ and $e_2$ at every vertex. This might not be possible, but we bound the number of ways of doing so by $3^n$.
\\
Finally, any  melonic graph can be constructed by recursively inserting elementary 2-point melons, so that the vertices of the graphs are naturally associated in pairs. There is a unique way of adding the wavy edges between these pairs of vertices. 
\\
This way, we see that the number of melonic graphs $G=(U,w,w')$ of order $n$ is bounded from above by $2n!C^4_n  {48}^n$, where $C^4_n$ is the order $4$ Fuss-Catalan number $\frac 1 {4n+1} \binom {4n+1} n $, which is the number of 4-regular projected melonic graph with $n$ unlabeled vertices and  one bivalent root \cite{Bonzom:2011zz},  and the labeling of the vertices corresponds to the $n!$ factor. The Fuss-Catalan number  $C^4_n$ behaves asymptotically as 
$ \frac 2{3\sqrt {6\pi}} \Bigl(\frac {4^4}{3^3}\Bigr)^n n^{-3/2} $, in particular it is bounded by $\Bigl(\frac {4^4}{3^3}\Bigr)^n$. 
We may set for instance $K=96(\frac {4^4}{3^3})$. This concludes the proof. \qed 

\

\noindent{\bf Proof of Prop.~\ref{prop:Analyc-Melo}. }
Taking into account the $1/n!$ symmetry factor in \eqref{fullexpan1}, Lemmas \ref{keylemma2}-\ref{keylemma3} 
prove that the melonic approximation \eqref{meloapp} to $\langle G_2 (\R,t) \rangle_{\alpha , C}$ is  analytic in $t$ at least in a finite disk, and for $t$ sufficiently close to 0,
\be 
\lvert G_2^{melo} (\R,t) \rvert \le \frac{p^{r/2}} {1-2K^2t^2}.
\ee 
Therefore,
the melonic approximation $S_\gamma^{melo} = \sum_{\R\ge 0} \frac{\R^\gamma}{N} G_2^{melo} (\R,t) $ to the averaged Sobolev norms
$S_\gamma$ is also analytic in $t$ at least in a finite disk. This completes the proof of Prop.~\ref{prop:Analyc-Melo}, and therefore of Theorem \ref{theo1}. \qed

\ 

This leads to a bound on the averaged Sobolev norms $S^{melo}_\gamma$, namely $$\sum_{\R\ge 0} \frac{\R^\gamma}{N} G_2^{melo} (\R,t).$$
This bound is written as,
\be 
\lvert S^{melo}_\gamma \rvert \le \frac 1 N \frac{L_\gamma(\sqrt p)}{1-2K^2t^2} \le \frac{K_1N^{\gamma} }{1-2K^2t^2}. 
\ee 
Thus we expect the averaged melonic Sobolev norms to scale in $N^\gamma$ (so that $S^{melo}_\gamma = N^\gamma \bar S_{\gamma, \omega}$ in \eqref{eq:1Nexp-def}), but proving it requires to take into account the possible cancellations in the series, due to the minus signs. 

\medskip
Since the averaged melonic Sobolev norms are analytic in a finite disk, we know that the second order computation of Section~\ref{sec:Sobolev-order2} corresponds to the Taylor expansion of the averaged melonic Sobolev norm $S^{melo}_\gamma$ around $t=0$. Since the coefficient of $t^2$ is positive for $\gamma>1$ (see Eq.~\eqref{eq:Order2-Sobolev-asympt}), the melonic Sobolev norm $S^{melo}_\gamma$ increases over a time interval $[0,\delta]$ for some $\delta>0$ thus proving Theorem \ref{theo2}. This growth phenomenon is called \emph{melonic turbulence}. 

\newpage

\begin{appendix}

\section{Proof of Lemma~\ref{lemma:ChangeVariable1}}
\label{Appendix:ChangeVariable1}

 The graph amplitudes can be expressed as  
\be 
\label{prflem3eq1}
\cA_\R(G)=\frac 1 {N^n} \Bigl(\prod_{v\in\tilde\cV} \sum_{S_v}\sum_{j_v, k_v \le S_v}\Bigr) \Bigl(\prod_{\substack{e \text{ dashed } \\ \text{or solid}}} \delta_{j_e j'_e}
\Bigr) \prod_{e\text{ dashed} } p^{j_e}.
\ee
We know that for every corner, the corner momentum $i_c$ (which is $j_v, S_v-j_v, k_v, S_v - k_v$ for the corresponding node $v$) is equal to all of the indices of the other corners in the face it belongs to. Indeed, a face consists of edges linking corners, and the constraint on a dashed or solid edge $e$ between two corners $c$ and $c'$  identifies the corner momenta $j_e=i_c$ and $j_e' = i_{c'}$.  So if we introduce a new set of indices $\{i_f\}_{f}$, we can rewrite the constraints on the edges as 
\be
\prod_{\substack{e \text{ dashed } \\ \text{or solid}}} \delta_{j_e j'_e} = \Bigl(\prod_{f} \sum_{i_f\ge0}\Bigr)\Bigl(\prod_f \prod_{c\in f} \delta_{i_f}^{i_c}\Bigr).
\ee
Indeed, to recover the left hand side, we perform the sums over the $\{i_f\}$ on the right hand side. For each face, it identifies all of the indices of the visited corners and gives back the constraints on the dashed and solid edges that the face visits. We rewrite \eqref{prflem3eq1}  as 
\bea
\cA_\R(G)&=&\frac 1 {N^n} \Bigl(\prod_{f} \sum_{i_f\ge0}\Bigr) \Bigl(\prod_{v\in\tilde\cV} \sum_{S_v}\sum_{j_v, k_v \le S_v}\Bigr) \Bigl(\prod_f \prod_{c\in f} \delta_{i_f}^{i_c}\Bigr) \prod_{e\text{ dashed} } p^{j_e}\\
\label{eq:ChDeVr2}
&=&\frac 1 {N^n} \Bigl(\prod_f \sum_{i_f}p^{i_f L^\alpha_f}\Bigr)\Biggl[ \Bigl(\prod_{v\in\tilde\cV} \sum_{S_v}\sum_{j_v, k_v \le S_v}\Bigr) \Bigl(\prod_f \prod_{c\in f} \delta_{i_f}^{i_c}\Bigr)\Biggr],
\eea
where $L^\alpha_f$ is the number of dashed edges in the face $f$. We would now like to perform the sums over the $\{S,j,k\}$, to reduce the term between brackets into a product of Kronecker deltas. This would succeed if we use one delta for each one of the $3n/2$ sums. There is exactly one delta per corner. For a given vertex $v_0$, these four deltas depend only on the three indices $j_v, k_v$ and $S_v$, and on some face momenta which are fixed. Therefore, we should be able to reduce the node constraints to a product of deltas.

More precisely, we consider a  node $v_0$ and perform the sums for the two corners $c_1$ and $c_3$ for which the corner momenta are $i_{c_1} = j_{v_0}$ and $i_{c_3}=k_{v_0}$. This uses one delta each, $\delta_{i_{f_1}}^{i_{c_1}}$ and $\delta_{i_{f_3}}^{i_{c_3}}$ (where we shortened the notation $i^{(a)}_{f,v_0} = i_{f_a}$), and it means that for the two remaining corners $c_2$  and $c_4$ of the node, respectively corresponding to the face momenta $S_{v_0} - j_{v_0}$ and $S_{v_0} - k_{v_0}$, the remaining deltas are  $\delta_{i_{f_2}}^{S_{v_0} - i_{f_1}}$ and $\delta_{i_{f_4}}^{S_{v_0} - i_{f_3}}$. 
We stress that the condition $S_{v_0}\ge i_{f_1}, i_{f_3}$ is implemented in the deltas because $i_{f_2}$ and $i_{f_4}$ are non-negative. We perform the sum over $S_{v_0}$, leaving us with a $\delta_{i_{f_2}}^{i_{f_3} + i_{f_4} - i_{f_1}}$ which we re-arrange as a $\delta_{i_{f_1} + i_{f_2}}^{i_{f_3} + i_{f_4}}$. We have to be sure that the constraint $S_{v_0}\ge \max(i_{f_1},i_{f_3})$ is implemented, and this is the case, as $i_{f_3} + i_{f_4} - i_{f_1}$ is non-negative since $i_{f_2}\ge0$ and $i_{f_3} + i_{f_4}\ge i_{f_3}$ so $i_{f_3} + i_{f_4}\ge \max(i_{f_1},i_{f_3}) $ and similarly for $i_{f_1} + i_{f_2}$.
The sums corresponding to the vertex $v_0$ in \eqref{eq:ChDeVr2} have been taken care of and we proceed with another vertex.
 \qed 

\end{appendix}


\begin{thebibliography}{99}

\bibitem{Wishart}
 J. Wishart, ``Generalized product moment distribution in samples,'' Biometrika. 20A (1928) 32. 

\bibitem{Wigner}
E. Wigner,  ``Characteristic vectors of bordered matrices with infinite dimensions,''  Annals of Mathematics. 
62 (1955) 548. 

\bibitem{Dyson} F. Dyson, ``Correlations between eigenvalues of a random matrix,''
Comm. Math. Phys. 19 (1970) 235.

\bibitem{MajumdarSchehr1} 
S Majumdar and G. Schehr,
``Top eigenvalue of a random matrix: large deviations and third order phase transition,''
J. Stat. Mech. (2014) P01012, arXiv:1311.0580.

\bibitem{Hooft}
  G.~'t Hooft,
  ``A Planar Diagram Theory for Strong Interactions,''
  Nucl.\ Phys.\ B {\bf 72}, 461 (1974).

\bibitem{2dgrav}
  P.~Di Francesco, P.~H.~Ginsparg and J.~Zinn-Justin,
  ``2-D Gravity and random matrices,''
  Phys.\ Rept.\  {\bf 254}, 1 (1995),
  arXiv:hep-th/9306153.

\bibitem{May}
R.~M.~May, ``Will a large complex system be stable?'' 
Nature 238 (1972) 413; \\
Stability and Complexity in Model Ecosystems, Princeton (1973).

\bibitem{Allesina}
S.~Allesina and S.~Tang,
``The stability/complexity relationship at age 40: a random matrix
perspective,'' Popul. Ecol. 57 (2015) 63.

\bibitem{Dine}
X.~Chen, G.~Shiu, Y.~Sumitomo and S.~H.~H.~Tye,
  ``A Global View on The Search for de-Sitter Vacua in (type IIA) String Theory,''
  JHEP {\bf 1204} (2012) 026,
  arXiv:1112.3338 [hep-th].\\
M.~Dine,
  ``Classical and Quantum Stability in Putative Landscapes,''
  JHEP {\bf 1701} (2017) 082,
  arXiv:1512.08125 [hep-th].

\bibitem{Ambjorn} 
J. Ambj\o rn, B. Durhuus and T. Jonsson, 
``Three-Dimensional Simplicial Quantum Gravity 
And Generalized Matrix Models,''
Mod. Phys. Lett. A 6 (1991) 1133; \\
N. Sasakura, 
``Tensor model for gravity and orientability of manifold, 
Mod. Phys. Lett. A 6 (1991) 2613;
\\
M. Gross, 
``Tensor models and simplicial quantum gravity in $>$ 2-D,''
Nucl. Phys. Proc. Suppl. 25A (1992)
144;\\
J. Ambj\o rn, 
``Simplicial Euclidean and Lorentzian Quantum Gravity", 
arXiv:gr-qc/0201028.

\bibitem{1/N}{R.~Gurau,  ``The $1/N$ expansion of colored tensor models,'' Ann. Henri Poincar\'e \textbf{12} (2011) 829, arXiv:1011.2726;\\
R.~Gurau and V.~Rivasseau, ``The $1/N$ expansion of colored tensor models in arbitrary dimension,'' Europhys.\ Lett.\ \textbf{95} (2011) 50004, arXiv:1101.4182;\\
R.~Gurau, ``The complete $1/N$ expansion of colored tensor models in arbitrary dimension,'' Ann. Henri Poincar\'e \textbf{13} (2012) 399, arXiv:1102.5759.}


\bibitem{Bonzom:2011zz} 
  V.~Bonzom, R.~Gurau, A.~Riello and V.~Rivasseau,
  ``Critical behavior of colored tensor models in the large N limit,''
  Nucl.\ Phys.\ B {\bf 853}, 174 (2011), arXiv:1105.3122 [hep-th].

\bibitem{Gurauetal}
R.~Gurau,  ``Colored Group Field Theory,''
Commun.\ Math.\ Phys.\  {\bf 304} (2011) 69;\\
R.~Gurau and J.~P.~Ryan,  ``Colored Tensor Models - a review,''
SIGMA {\bf 8} (2012) 020; \\
R.~Gurau, ``Random Tensors", Oxford University Press (2016),
 "Tensor Models, Formalism and Applications", SIGMA special issue, 2016,

\bibitem{TensUniv} 
  R.~Gurau,
 ``Universality for Random Tensors,''
  Ann.\ Inst.\ H.\ Poincare Probab.\ Statist.\  {\bf 50} (2014) 1474,  arXiv:1111.0519 [math.PR].

\bibitem{MelonUniv} 
 D.~Benedetti, S.~Carrozza, R.~Gurau and M.~Kolanowski,
  ``The $1/N$ expansion of the symmetric traceless and the antisymmetric tensor models in rank three,''
  arXiv:1712.00249 [hep-th];\\
  S.~Carrozza,
  ``Large $N$ limit of irreducible tensor models: $O(N)$ rank-$3$ tensors with mixed permutation symmetry,''
  JHEP {\bf 1806}, 039 (2018),
  arXiv:1803.02496 [hep-th];\\
 S.~Carrozza and V.~Pozsgay,
 ``SYK-like tensor quantum mechanics with $\mathrm{Sp}(N)$ symmetry,''
  arXiv:1809.07753 [hep-th].
  
  \bibitem{EnhScal}
  V.~Bonzom, L.~Lionni and V.~Rivasseau, ``Colored Triangulations of Arbitrary Dimensions are Stuffed Walsh Maps,'' Electr. J. Comb. 24(1): P1.56 (2017), 	arXiv:1508.03805 [math.CO];\\
  L. Lionni and J. Th\" urigen, ``Multi-critical behaviour of 4-dimensional tensor models up to order 6,'' 	arXiv:1707.08931 [hep-th];\\
  F.~Ferrari, V.~Rivasseau and G.~Valette,
  ``A New Large N Expansion for General Matrix-Tensor Models,'' arXiv:1709.07366 [hep-th].

 
\bibitem{TensRen}
  J.~Ben Geloun and V.~Rivasseau,
  ``A Renormalizable 4-Dimensional Tensor Field Theory,''
  Commun.\ Math.\ Phys.\  {\bf 318} (2013) 69,
  arXiv:1111.4997 [hep-th];\\
  J.~Ben Geloun,
  ``Renormalizable Models in Rank $d\geq 2$ Tensorial Group Field Theory,''
  Commun.\ Math.\ Phys.\  {\bf 332} (2014) 117,
  arXiv:1306.1201 [hep-th];\\
  S.~Carrozza, D.~Oriti and V.~Rivasseau,
  ``Renormalization of a SU(2) Tensorial Group Field Theory in Three Dimensions,''
  Commun.\ Math.\ Phys.\  {\bf 330} (2014) 581,
  arXiv:1303.6772 [hep-th];\\
  S.~Carrozza, D.~Oriti and V.~Rivasseau,
  ``Renormalization of Tensorial Group Field Theories: Abelian U(1) Models in Four Dimensions,''
  Commun.\ Math.\ Phys.\  {\bf 327} (2014) 603,
  arXiv:1207.6734 [hep-th];\\
 A.~Tanasa, ``Multi-orientable Group Field Theory'', J. Phys. A {\bf 45} (2012) 165401, arXiv:1109.0694.

\bibitem{SigmaF}
S. Carrozza, ``Flowing in Group Field Theory Space: a Review",
SIGMA {\bf 12} (2016) 070, arXiv:1603.01902 [gr-qc];\\
T. Krajewski and R. Toriumi 
``Exact Renormalisation Group Equations and Loop Equations for Tensor Models",
SIGMA {\bf 12} (2016) 068, arXiv:1603.00172 [gr-qc].

\bibitem{SigmaC}
  R.~Gurau,
  ``The 1/N Expansion of Tensor Models Beyond Perturbation Theory,''
  Commun.\ Math.\ Phys.\  {\bf 330} (2014) 973,
  arXiv:1304.2666 [math-ph];\\
V. Rivasseau, ``Constructive Tensor Field Theory," SIGMA {\bf 12} (2016) 085,  arXiv:1603.07312 [math-ph].

\bibitem{Kitaevetal} A.~Kitaev, ``A Simple Model of Quantum Holography,'' KITP Program ``Entanglement in Strongly-Correlated Quantum Matter,'' unpublished, see http://online.kitp.ucsb.edu/online/entangled15;\\
J.~Polchinski and V.~Rosenhaus,
  ``The Spectrum in the Sachdev-Ye-Kitaev Model,''
  JHEP {\bf 1604} (2016) 001,
  arXiv:1601.06768 [hep-th].\\
J.~Maldacena and D.~Stanford,
  ``Remarks on the Sachdev-Ye-Kitaev model,''
  Phys.\ Rev.\ D {\bf 94} (2016) 106002
  arXiv:1604.07818 [hep-th].

\bibitem{wittenetal} E.~Witten, ``An SYK-Like Model Without Disorder,'' arXiv:1610.09758 [hep-th];\\
R.~Gurau, ``The complete $1/N$ expansion of a SYK-like tensor model,'' Nucl.\ Phys.\ B {\bf 916} (2017) 386, arXiv:1611.04032 [hep-th];\\
  S.~Carrozza and A.~Tanasa,
  ``$O(N)$ Random Tensor Models,''
  Lett.\ Math.\ Phys.\  {\bf 106} (2016) 1531,
  arXiv:1512.06718 [math-ph];\\
S.~Dartois, V.~Rivasseau and A.~Tanasa, ``The 1/N expansion of multi-orientable random tensor models'', Ann. Henri Poincare {\bf 15} (2014) 965, arXiv:1301.1535 [hep-th]; \\
I.~R.~Klebanov and G.~Tarnopolsky,
  ``Uncolored random tensors, melon diagrams, and the Sachdev-Ye-Kitaev models,''
  Phys.\ Rev.\ D {\bf 95} (2017) 046004,
  arXiv:1611.08915 [hep-th];
  
  \bibitem{LargeD}
  F.~Ferrari, ``The Large D Limit of Planar Diagrams,'' arXiv:1701.01171 [hep-th];
  
  T.~Azeyanagi, F.~Ferrari and F.~I.~Schaposnik Massolo, ``Phase Diagram of Planar Matrix Quantum Mechanics, Tensor, and Sachdev-Ye-Kitaev Models,''
  Phys.\ Rev.\ Lett.\  {\bf 120} (2018) 061602, arXiv:1707.03431 [hep-th];
  
  T.~Azeyanagi, F.~Ferrari, P.~Gregori, L.~Leduc and G.~Valette, ``More on the New Large $D$ Limit of Matrix Models,''
  Annals Phys.\  {\bf 393} (2018) 308, arXiv:1710.07263 [hep-th].

\bibitem{BohigasFlores} O.~Bohigas and J.~Flores, ``Two-body random Hamiltonian and level density,'' Phys. Lett. 34B (1971) 261;
``Spacing  and  individual  eigenvalue  distributions  of  two-body  random
Hamiltonians,'' Phys. Lett. 35B (1971) 383.

\bibitem{FrenchWong}J. B.~French and S. S. M.~Wong, ``Validity of random matrix theories for many-particle
systems,'' Phys. Lett. 33B (1970) 449;
``Some  random-matrix  level  and  spacing  distributions for fixed-particle-rank interactions,'' Phys. Lett. 35B (1971) 5.

\bibitem{Kota}V. K. B.~Kota, ``Embedded random matrix ensembles in quantum physics,'' Springer (2014).

\bibitem{quantres}
  O.~Evnin and W.~Piensuk,
  ``Quantum resonant systems, integrable and chaotic,''
  arXiv:1808.09173 [math-ph].

\bibitem{AdSres}
  V.~Balasubramanian, A.~Buchel, S.~R.~Green, L.~Lehner and S.~L.~Liebling,
  ``Holographic thermalization, stability of anti-de Sitter space, and the Fermi-Pasta-Ulam paradox,''
  Phys.\ Rev.\ Lett.\  {\bf 113} (2014) 071601,
  arXiv:{1403.6471} [hep-th];\\
B.~Craps, O.~Evnin and J.~Vanhoof,
 ``Renormalization group, secular term resummation and AdS (in)stability,''
 JHEP {\bf 1410} (2014) 48,
 arXiv:{1407.6273} [gr-qc];\\
B.~Craps, O.~Evnin and J.~Vanhoof,
``Renormalization, averaging, conservation laws and AdS (in)stability,''
JHEP {\bf 1501} (2015) 108,
arXiv:{arXiv:1412.3249} [gr-qc];\\
 P.~Bizo\'n, M.~Maliborski, A.~Rostworowski, ``Resonant dynamics and the instability of anti-de Sitter spacetime,'' Phys.\ Rev.\ Lett. {\bf 115} (2015) 081103,
    arXiv:{1506.03519} [gr-qc];\\
P.~Bizo\'n, B.~Craps, O.~Evnin, D.~Hunik, V.~Luyten and M.~Maliborski,
 ``Conformal flow on $S^3$ and weak field integrability in AdS$_4$,'' Comm.\ Math.\ Phys. {\bf 353} (2017) 1179, arXiv:{1608.07227} [math.AP]\\
P.~Bizo\'n, D.~Hunik-Kostyra and D.~Pelinovsky,
  ``Ground state of the conformal flow on $\mathbb{S}^3$,''
    arXiv:{1706.07726} [math.AP];\\
B.~Craps, O.~Evnin and V.~Luyten,
  ``Maximally rotating waves in AdS and on spheres,'' JHEP {\bf 1709} (2017) 059,
   arXiv:{1707.08501} [hep-th];\\
P.~Bizo\'n, D.~Hunik-Kostyra and D.~Pelinovsky,
``Stationary states of the cubic conformal flow on $\mathbb{S}^3$,'' arXiv:1807.00426 [math-ph].

\bibitem{BR} P.~Bizo\'n and A.~Rostworowski,
 ``On weakly turbulent instability of anti-de Sitter space,''
 Phys.\ Rev.\ Lett.\ {\bf 107} (2011) 031102,
 arXiv:{1104.3702} [gr-qc].

\bibitem{GPres} P. Germain, Z. Hani and L. Thomann,
    ``On the continuous resonant equation for NLS: I. Deterministic analysis,'' J. Math. Pur. App. {\bf 105} (2016) 131,
    arXiv:{1501.03760} [math.AP];\\
P.~Germain and L.~Thomann,  ``On the high frequency limit of the LLL equation,'' Quart. Appl. Math. {\bf 74} (2016) 633, arXiv:{1509.09080} [math.AP];\\
A.~F.~Biasi, J.~Mas and A.~Paredes,
  ``Delayed collapses of Bose-Einstein condensates in relation to anti-de Sitter gravity,''
  Phys.\ Rev.\ E {\bf 95} (2017) 032216,
  arXiv:1610.04866 [nlin.PS];\\
A.~Biasi, P.~Bizo\'n, B.~Craps and O.~Evnin,
  ``Exact lowest-Landau-level solutions for vortex precession in Bose-Einstein condensates,''
  Phys.\ Rev.\ A {\bf 96} (2017) 053615,
  arXiv:{1705.00867} [cond-mat.quant-gas];\\
 P.~G\'erard, P.~Germain and L.~Thomann,
  ``On the cubic Lowest Landau Level equation,''
  arXiv:{1709.04276} [math.AP];\\
J.~Fennell, ``Resonant Hamiltonian systems associated to the one-dimensional non-linear Schr\"odinger equation with harmonic trapping,'' arXiv:1804.08190 [math.AP];\\
A.~Biasi, P.~Bizo\'n, B.~Craps and O.~Evnin,
  ``Two infinite families of resonant solutions for the Gross-Pitaevskii equation,'' Phys.Rev. E {\bf 98} (2018) 032222,
arXiv:1805.01775 [cond-mat.quant-gas].

\bibitem{murdock}  J.~A.~Murdock, ``Perturbations: Theory and Methods,'' SIAM (1987).

\bibitem{KM} S.~Kuksin and A.~Maiocchi, ``The effective equation method,'' in New Approaches to Non-linear Waves, Springer (2016), arXiv:{1501.04175} [math-ph].

\bibitem{GG}P.~G\'erard and S.~Grellier,
``The cubic Szeg\H o equation,'' Ann. Scient. \'Ec. Norm. Sup. {\bf 43} (2010) 761,
arXiv:{0906.4540} [math.CV];\\
``Effective integrable dynamics for a certain nonlinear wave equation,'' Anal. PDE {\bf 5} (2012) 1139, arXiv:{1110.5719} [math.AP];\\
``An explicit formula for the cubic Szeg\H o equation,'' Trans. Amer. Math. Soc. {\bf 367} (2015) 2979, arXiv:{1304.2619} [math.AP];\\
``The cubic Szeg\H{o} equation and Hankel operators,'' Ast\'erisque {\bf 389} (2017),
 arXiv:{1508.06814} [math.AP].

\bibitem{AO}A.~Biasi, P.~Bizo\'n and O.~Evnin,
 ``Solvable cubic resonant systems,''
  arXiv:1805.03634 [nlin.SI].
  
\bibitem{Kraichnan}
R.~H.~Kraichnan, ``Dynamics of nonlinear stochastic systems,'' J. Math. Phys. {\bf 2} (1961) 124.

\bibitem{Bouchaud:1995wx} 
  J.~P.~Bouchaud, L.~Cugliandolo, J.~Kurchan and M.~Mezard,
  ``Mode coupling approximations, glass theory and disordered systems,''
  Physica A {\bf 226} (1996) 243,
  [cond-mat/9511042].
  
\bibitem{NLStorus} J.~Colliander, M.~Keel, G.~Staffilani, H.~Takaoka and T.~Tao,
``Transfer of energy to high frequencies in the cubic defocusing
nonlinear Schr\"odinger equation,'' Invent. Math. {\bf 181} (2010) 39,
arXiv:0808.1742 [math.AP].

\bibitem{TensScal}
R.~Gurau and G.~Schaeffer, ``Regular colored graphs of positive degree,'' Ann. Henri Poincar\'e Prob. Stat.  D {\bf 3} (2016) 257, arXiv:1307.5279 [math.CO];\\
  S.~Dartois, R.~Gurau and V.~Rivasseau,
  ``Double Scaling in Tensor Models with a Quartic Interaction,''
  JHEP {\bf 1309} (2013) 088,
  arXiv:1307.5281 [hep-th];\\
  V.~Bonzom, R.~Gurau, J.~P.~Ryan and A.~Tanasa,
  ``The double scaling limit of random tensor models,''
  JHEP {\bf 1409} (2014) 051,
  arXiv:1404.7517 [hep-th].

\bibitem{FussCatalan}
W.~M\l otkowski and
K.~A.~Penson,
``Probability distributions with binomial moments", Inf. Dim. Analys. Quant. Prob. 17 (2014) 1450014, arXiv:1309.0595 [math.PR].

\bibitem{RivConstruct} V. Rivasseau, ``From Perturbative to Constructive Renormalization," Princeton (1991).

\end{thebibliography}
\end{document}